\documentclass[aps,rmp,twocolumn]{revtex4-1} 
\usepackage{graphicx}
\usepackage{amsmath}

\renewcommand{\ol}[1]{\overline{#1}}
\newcommand{\E}{{\rm E}}
\newcommand{\Em}{\tilde{\mu}_m}
\newcommand{\x}{\mathbf{x}}

\def\eq#1{Eq.\ \eqref{eq:#1}}

\def\fig#1{Fig.\ \ref{fig:#1}}

\newcommand{\FIG}[3]{
\begin{figure}
\centering
\includegraphics[width=0.8\linewidth]{#1}
\caption{\label{#2}#3}
\end{figure}}

\begin{document}

\title{Physics of adherent cells}
\author{Ulrich S. Schwarz}
\affiliation{BioQuant and Institute for Theoretical Physics, Heidelberg University, Philosophenweg 19, 69120 Heidelberg, Germany}
\author{Samuel A. Safran}
\affiliation{Department of Materials and Interfaces, Weizmann Institute, Rehovot 76100, Israel}

\date{\today}

\begin{abstract}
One of the most unique physical features of cell adhesion to external
surfaces is the active generation of mechanical force at the
cell-material interface. This includes pulling forces generated by
contractile polymer bundles and networks, and pushing forces generated
by the polymerization of polymer networks. These forces are
transmitted to the substrate mainly by focal adhesions, which
are large, yet highly dynamic adhesion clusters. Tissue cells use
these forces to sense the physical properties of their environment and
to communicate with each other. The effect of forces is intricately
linked to the material properties of cells and their physical
environment. Here we review recent progress in our understanding of
the role of forces in cell adhesion from the viewpoint of theoretical
soft matter physics and in close relation to the relevant experiments.
\end{abstract}

\maketitle
\tableofcontents

\section{Introduction}

Observations of swimming bacteria, crawling animal cells or developing
organisms dramatically indicate that physical force and movement are
central to the behavior of biological systems
\cite{thompson_growth_1992,Ingber99,lecuit_cell_2007,kollmannsberger_physics_2011}. The functions of cells have evolved in
the context of very specific physical environments leading to a close
coupling between cells and their surroundings \cite{alberts_molecular_2007,phillips_physical_2008}. This is especially true
for animal cells, which have evolved in the controlled environment
provided by multi-cellular organisms and therefore appear to be more
sensitive to environmental cues than {\it{e.g.}},\ uni-cellular organisms that
can sometimes live in very harsh surroundings. Therefore it is an essential element of understanding animal
cells to consider their physical interactions
with the environment.

During recent years, it has become increasingly clear that the
cell-material interface determines the behaviour and fate of
biological cells to a much larger extent than was formerly appreciated
\cite{discher_tissue_2005,uss:schw05,vogel_local_2006,janmey_hard_2009,
discher_growth_2009,mitragotri_physical_2009,geiger_environmental_2009,
de_theoretical_2010,hoffman_dynamic_2011,uss:schw12b,ladoux_physically_2012}. For example, it
has been shown that the differentiation of stem cells can be guided by
the mechanical or adhesive properties of the substrate
\cite{engler_matrix_2006,kilian_geometric_2010,fu_mechanical_2010}.
Such observations can lead the way to exciting new applications for
regenerative medicine and tissue engineering, because physical signals
are easier to control and can be more permanent than biochemical or genetic
manipulations. On the scientific side, however, the fundamentals of
these processes are puzzling and not yet well understood,
despite their importance in development, health and disease
\cite{janmey_mechanisms_2011,dufort_balancing_2011}.

From a physical point of view, the most important aspect of the
cellular response to the physical properties of the environment is the
observation that cells show a controlled response only if they are
able to actively generate force and to transmit this force to the
surroundings. This finding makes sense because cells must actively
sense the passive properties of their environment. For rigidity
sensing, for example, cells must actively strain their surroundings
to probe their elastic properties (similar to a bat
that senses the geometry of its environment by sending out ultrasound).
Cells have evolved special sensory systems for this purpose.
For example, it has been found that the size of the contacts between cells
and their environment grow with physical force
\cite{chrzanowska-wodnicka_rho-stimulated_1996,choquet_extracellular_1997,riveline_focal_2001,uss:bala01,tan_cells_2003,uss:paul08a,uss:colo09,trichet_evidence_2012}.
Although this finding makes sense from a biological viewpoint,
it is at the same time puzzling to the physicist, since in materials science,
force usually disrupts adhesion contacts. 

Physicists have traditionally been reluctant to study living systems
due to their molecular complexity. However, this has recently changed
in many ways. An important development that has led to physics
approaches to describe cells and tissue is the maturation of soft matter physics
into an independent and very active field of research. Soft matter
physics traditionally has focused on the properties of liquid
crystals, colloidal dispersions, emulsions, fluid membranes, polymer
gels and other complex fluids
\cite{pggnobel_1992,chaikin_principles_2000,safran_statistical_2003,jones_soft_2002}.
These systems are \textit{soft} since the interaction energies are of
the same order as the thermal energy. They are thus very sensitive to
thermal fluctuations and concepts from both mechanics and statistical
mechanics must be employed to understand phenomena such as, {\it{e.g.}},
conformational changes of membranes \cite{seifert_configurations_1997,powers_dynamics_2010}
or deformations of polymer networks \cite{bausch_bottom-up_2006,chen_rheology_2010}. 
While soft matter physics has established a firm physical basis of the building blocks of
biological cells, their behavior critically depends on additional elements, 
most prominently active remodeling controlled by genetic and
signaling networks. Meeting the challenge of combining the physics of soft
matter physics with active processes to describe \emph{active matter} will enable insight into many
biological processes, guide the design of new types of materials and further extend the range of phenomena
that can be analyzed by concepts and methods from physics
\cite{ramaswamy_mechanics_2010,fletcher_active_2009,mackintosh_active_2010,marchetti_soft_2012,gonzalez-rodriguez_soft_2012,huber_emergent_2013}.

To understand the physical aspects of cell adhesion, soft matter
physics provides useful model reference systems, such as the wetting
of substrates by droplets \cite{de_gennes_wetting:_1985}, adhesion of
vesicles made of fluid membranes \cite{seifert_configurations_1997} or
the adhesion of capsules that comprise thin polymer shells
\cite{pozrikidis_modeling_2003}. The challenge is to combine these
reference systems with the molecularly specific and active processes
that they support at the cell-material interface, such as force
generation by polymerization \cite{mogilner_edge:_2006} or the binding
and unbinding of transmembrane adhesion receptors
\cite{evans_forces_2007}. Over the last decade, several soft matter
systems have been revisited with a focus on this particular point of
view. The physical understanding of the properties of active materials
is rapidly growing; particular attention has been paid to active
membranes \cite{manneville_active_2001,gov_membrane_2004} and active
gels
\cite{liverpool_instabilities_2003,kruse_asters_2004,julicher_active_2007,marchetti_soft_2012}.

Cell adhesion is a multi-scale problem because the molecular processes
at the cell-material interface are dramatically amplified on the scale
of cells. Cellular processes such as spreading, adhesion, migration
and proliferation are in turn dramatically amplified on the scale of
tissues \cite{gonzalez-rodriguez_soft_2012}. Interestingly, similar
concepts have been successfully applied to different levels in this
hierarchy. In order to address the role of cellular forces in the
context of connective tissue, whose mechanical properties are
dominated by the extracellular matrix, one can build on traditional
approaches from condensed matter physics for force-generating centers
in a continuum matrix, such as the theory of elastic defects and their
interactions \cite{eshelby_1957,
  eshelby_1959,e:siem68,e:wagn74,e:lau77,e:safr79}. Motivated by
experimental measurements of cellular traction patterns
\cite{dembo_stresses_1999,butler_traction_2002,uss:schw02b}, it has
been suggested that the contractile activity of cells can be modeled
as anisotropic force contraction dipoles
\cite{uss:schw02a,uss:schw02b} and that cell orientation and
positioning can be predicted by minimizing the energy invested by the
cell into straining its environment for a given level of force
generation \cite{uss:bisc03a,uss:bisc04a}. Similar concepts have been
used to predict the contractile action of molecular motors in the
cytoskeleton \cite{silva_active_2011,dasanayake_general_2011}, the
orientation response of single cells to externally applied stress
\cite{de_dynamics_2007}, the collective response of contractile cells
in an elastic medium \cite{zemel_active_2006}, the polarization and
registry of cells as a function of external rigidity
\cite{zemel_optimal_2010,friedrich_nematic_2011,friedrich_how_2012},
and the growth of tissue where dividing cells correspond to force
dipoles \cite{joanny_2010}. Thus the concept of force dipoles is very
general, with applications to molecular, cellular and tissue
scales. However, the details of these different applications strongly
depend on the biological situation of interest.

For epithelial tissue dominated by direct cell-cell contacts, other
approaches are adequate, most prominently vertex models starting from
the fact that cell walls are strongly contractile
\cite{hufnagel_mechanism_2007,farhadifar_influence_2007,rauzi_nature_2008,landsberg_increased_2009,aegerter-wilmsen_exploring_2010,canela-xandri_dynamics_2011,aliee_physical_2012}.
Although this situation is somehow reminiscent of foams, due to the
presence of cell proliferation and death we are dealing with an active
material \cite{c:shra05,basan_homeostatic_2009,joanny_2010}. This
shows again that within the overarching framework of active materials,
different physics concepts have to be used depending on the biological
context.

Because biological systems are very complex, meaningful mathematical
models must be selective and focus on  phenomena that can be treated in a tractable manner in order to  yield physical insight.
The role of forces at the cell-material interface is certainly a
phenomenon which can be only be fully understood with concepts and tools
from physics. For future progress, it is essential to choose the
appropriate parameters and formulate models that are sufficiently simple to
be analysed in detail, but predictive enough to be verified or
falsified by experiments. A theoretical analysis has many
benefits. Apart from providing deeper insight and quantitative
predictions, it usually reveals relations between quantities or
phenomena that would go unnoticed without a theoretical model. For
example, the interplay between cell adhesion and mechanics leads to
interesting predictions regarding the coupling of cell shape and
forces \cite{bar-ziv_pearling_1999,deshpande_bio-chemo-mechanical_2006,
uss:bisc08a,uss:bisc09a,uss:guth12}.
A major focus of this review is to point out the relations
between cell shape, structure, adhesion, and force as they emerge from
our growing physical understanding of the role of physical forces at
the cell-material interface.

This review is organized as follows. We start with a survey of the
relevant soft matter physics that describes and quantifies those parts
of cells that are involved in force transmission. In particular, we
review the properties of liquid crystals, flexible and semi-flexible
chains and gels, and elements of elasticity theory, for both bulk
systems such as elastic solids as well as for finite-sized systems
such as vesicles and capsules.  We then present the minimally required
cell biology background, including a general discussion of the
cytoskeleton and the properties of actin polymers and networks, myosin
molecular motors that endow these networks with active contractility,
and the membrane-based adhesion structures that connect cells to their
environment. The main part of this review then covers recent
developments in the physics of adherent cells. In the spirit of a
multiscale approach, we start on a relatively small scale with simple
models for the physics of adhesion clusters. We then progress to
models for cell shape and structure, which in turn form the basis for
coarse-grained models for entire cells as force dipoles. In
particular, we use this framework to discuss cell response to
mechanical stress as well as actin network polarization and its
dependence on the elasticity of the underlying matrix. Finally we
address the physics of matrix-mediated cell assemblies from the
viewpoint of cellular forces. We close with some conclusions and an
outlook on future perspectives.

\section{Physics background}

\subsection{Soft matter in biological systems}

The present review on physical forces at the cell-material interface
focuses on a view of animal cells as complex, composite, soft
materials comprising fluid membranes that are coupled to two types of
elastic and often contracile polymer networks. Inside the cell, there
exists a highly crosslinked and entangled network of three different
types of polymers (actin filaments, microtubules and intermediate
filaments) collectively called the \textit{cytoskeleton} (CSK). On the
outside, the cell is coupled to another multi-component, gel-like
network (including fibrous protein components such as collagen or
fibronectin) called the \textit{extracellular matrix} (ECM). If
subjected to mechanical forces, the biological material initially
responds like a passive elastic body; thus elasticity theory is an
essential element of the physics of cells and tissues. At longer time
scales, the cell can respond to mechanical perturbations by actively
reorganizing the structure of its CSK (and to a certain extent, its
ECM as well).

Experiments suggest that cells in solution respond elastically up to
times on the order of a few seconds \cite{wottawah_optical_2005}. The
same is true for tissue on a timescale of seconds and minutes
\cite{gonzalez-rodriguez_soft_2012}. The deformation of an elastic
body induces both stress and strain. For example, for a simple, one
dimensional stretch of an elastic slab, the stress $\sigma$ is the
force per area applied to the slab on its top and bottom faces, while
the strain $\epsilon$ is the resulting relative change in length (for
a more detailed introduction to the tensorial theory of elasticity see
below). The simplest constitutive relation that relates stress and
strain is a linear one in which $\sigma = E \epsilon$. The elastic
constant $E$ introduced in this way is known as \textit{Young's
  modulus} and is often called the \textit{stiffness} or
\textit{rigidity}. The larger the Young's modulus, the more stress is
required to stretch the material to the same extent. Because strain
$\epsilon$ is dimensionless, the Young's modulus has the same physical
dimensions as stress $\sigma$, that is $\rm{N / m^2 =
  Pa}$. Physically, this means that the elastic modulus is a measure
of the mechanical energy density of the system. The corresponding
spring constant $k=EA/L_0$ also depends on two geometrical quantities:
the cross-sectional area $A$ and the rest length $L_0$ of the spring.

For much of our discussion of cell elasticity, it is essential to note
that the elastic modulus of a typical tissue cell is in the range of
10 kPa (comparable to very soft cheese or toothpaste). This should be
contrasted with the much higher values of crystal moduli of 100
GPa. With a typical size of the supramolecular assembly of 10~nm,
simple scaling predicts that the typical energy scale for cells is in
the range of $10 \rm{kPa} (10 \rm{nm})^3 = 10^{-20} \rm{J}$, which is
close to the thermal energy scale $k_B T = 4.1\ 10^{-21} \rm{J} =
4.1\ \rm{pN}\ \rm{nm}$ (here we have used $T = 300 \rm{K}$ since most
of biology operates at room or body temperature). Although somehow
simplistic, this argument nevertheless correctly indicates that the
cohesive interactions that stabilize cells are weak. These are mainly
electrostatic attractions between charges or charge distributions
(mainly dipoles) that are screened by water and relatively high salt
concentration (100~mM corresponding to a Debye screening length of
about 1~nm), hydrogen bridges, hydrophobic interactions due to the
special properties of water, and entropic forces such as depletion
interactions, all of which operate on an energy scale of a few $k_B T$
\cite{b:isra11,dill_molecular_2010}.

The relatively weak cohesive energies are also
related to the large length scales that characterize soft matter,
since it is often the energy density (energy per unit volume) that is
relevant.  For example, the large-scale structures of linear
macromolecules (polymers) in solution can be described by disordered,
blob-like structures where the typical blob size can be hundreds of
Angstroms \cite{pggpolymer_1979}.  Water-amphiphile dispersions can
exhibit disordered, sponge-like structures consisting of bilayer
sheets of amphiphilic molecules whose sizes can be a hundred times the
size of an individual molecule
\cite{uss:schw02c,safran_statistical_2003}.  Even those soft materials
that show solid-like elasticity, such as gels \cite{pggpolymer_1979,
  b:boal} or colloidal crystals \cite{pieranski_1983}, have mesh or
lattice constants that are in the range of hundreds to thousands of
Angstrom.  In addition, the overall weak nature of the interactions
({\it e.g.}, gels with dilute crosslinks separated by large distances,
colloidal particles with small surface charges), results in shear
elastic constants that can be many orders of magnitude weaker than
those of hard-matter.  The weak, non-covalent nature of the
interactions in soft matter often compete with the entropy of
the system and leads to large responses and variations in the
structures and phases as the temperature or composition is varied.
The soft matter topics that are most closely related to cellular
forces are liquid crystals, polymers and gels, elasticity and the
adhesion of fluid drops, amphiphilic vesicles and polymerized
capsules.

In practice, the mechanical properties of cells are described by
linear elasticity only in a limited regime, and there exists a hierarchy of
length and energy scales that determine how cells respond to force.
In fact, the viscoelastic response of cells has been measured using
various techniques, in different situations and over a large range of
frequencies \cite{b:fung93}.  These measurements show that cells share
many of the features of the viscoelasticity of {\it{in vitro}}, reconstituted
networks of biopolymers
\cite{bausch_bottom-up_2006,chen_rheology_2010}.

While synthetic soft matter systems are subject and sensitive to
thermal disorder, biological cells exhibit far more noisy behavior
\cite{cellnoise} due to the stochastic nature of many of the
biomolecular processes that take place; in our context such
non-thermal noise occurs mainly in the context of force generation by
molecular motors \cite{b:howa01} and actin polymerization
\cite{mogilner_edge:_2006}. If these processes are correlated only on
molecular length and time scales, we can regard them as \textit{active
  white noise} with regard to modeling cellular behavior on much
longer scales.  In that case, the molecular processes can be
approximated as being delta correlated in both space and time the
results of which \cite{haken} resemble an \textit{effective
  temperature} that determines the width of a Boltzmann-like
distribution. However, this cannot not be generalized to the role of
an effective temperature in a true thermodynamic sense \cite{niryair}.
In particular, the fluctuation-dissipation theorem
\cite{chaikin_principles_2000} is not obeyed \cite{c:mizu07} as it is
in thermal systems.  With these caveats firmly in mind, we shall use
the concept of effective temperature and the resulting Boltzmann
distribution of cellular energies in situations in which the molecular
noise can be regarded as delta-correlated.

\subsection{Liquid crystals}

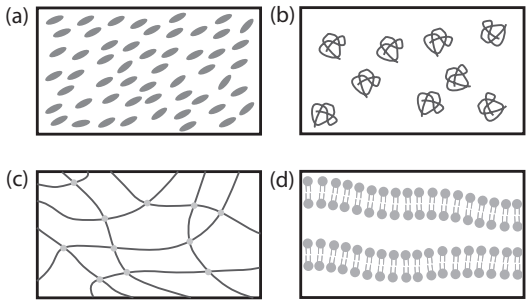
\FIG{Fig01}{fig:SoftMatter}{Passive bulk soft
matter examples that are important model systems
for the understanding of the material properties of cells and tissues.
(a) Liquid crystals often form nematic phases, with no
positional but orientational order. (b) In a dilute
polymer solution, each polymer forms a globule that is well
separated from the other polymers. (c) A cross-linked
polymer gel can behave like an elastic solid but with much weaker
rigidity. (d) Lipids
self-assemble into fluid bilayers, that at high concentration
in turn tend to self-assemble into stacks, the so-called
lamellar phase.}

We begin our review of relevant physical systems with liquid crystals,
which comprise anisotropic ({\it{e.g.}}, rod-like) molecules that can
show orientational (\textit{nematic}) order, but not necessarily
positional (translational) order \cite{pggliquid_1995}.  At lower
temperatures, nematically ordered systems can show a type of
one-dimensional (\textit{smectic}) order in which the molecules form
well-defined layers with the molecular axis oriented parallel to the
layer normal (smectic A).  The layers themselves are fluid with no
translational, in-plane order.  The relevance of liquid crystal
ordering to cells lies in the fact that under external forces (shear
flow or elastic deformation) or in elastic environments of appropriate
rigidity, the polymer networks inside cells can show nematic and even
smectic order; these applications are discussed later on and here we
outline the relevant liquid crystal physics.

The cooperative, orientational interactions between rod-like molecules
can give rise to phase transitions in which they all align in a given
direction, as in \fig{SoftMatter}(a) \cite{pggliquid_1995}.  Such
nematic ordering transitions can arise when the temperature is lowered
in systems governed by microscopic interactions that promote order.
Microscopically, each molecule is characterized by its
thermally fluctuating orientation angle $\theta$ where the $z$ axis that
defines the angle can be defined by convention or by some macroscopic,
symmetry breaking field such as the (non-spherical) shape of the
system.  Because the rod-like molecules have up-down symmetry, the
interaction energy cannot be an odd function of the angle since that
changes sign when the rod is flipped.  Instead, the energies must be
even functions of $\theta$.

These symmetry considerations allow the definition of the local value
of the nematic order parameter, $S_i$, of a given molecule labelled by $i$
as:
\begin{equation}
S_i  = \frac{1}{2}\left(3 \cos{\theta_i}^2 -1 \right)
\label{eq:lcop1}
\end{equation}
A simple mean-field theory for nematic
ordering was formulated by Maier and Saupe \cite{maier_1959}.  The
Hamiltonian of the interacting system of rods in which the energy
depends on the local orientations of nearby molecules is replaced by a
one-body approximation, $U_{ms}$ in which the energy of a given
molecule is proportional to the product of its order parameter with
the thermal average of the average order parameter of the system,
$\langle S \rangle$.  The mean-field nature of this assumption lies in
the fact that the orientations of neighboring, interacting molecules
is approximated by the average order parameter
\begin{equation}
U_{ms}= B  \langle S \rangle \sum_i S_i
\label{eq:lcham1}
\end{equation}
where $B$ is a constant that characterizes the interactions.
The order parameter is determined self-consistently from the
statistical mechanical definition of the thermal average in which
the probability distribution is proportional to the Boltzmann factor,
$\exp[-U_{ms}/k_BT]$:
\begin{equation}
\langle S \rangle = \frac{1}{Z}\int d\Omega \   S_i \ \exp[-B  \langle S \rangle  S_i /k_BT]
\end{equation}
where $\Omega$ is the solid angle and the normalization $Z$
is the average of the $\exp[-U_{ms}/k_BT]$
over all solid angles.

This approximation correctly predicts a first order phase transition
at which the average order parameter $\langle S \rangle$ jumps from a
value of zero to a value of approximately $0.4$ when $B/k_BT =4.6$.
As the temperature is lowered, the order parameter increases till it
reaches its saturation value of unity.  The molecules are still in the
fluid state; there is no translational order, but only orientational
order.

\subsection{Semi-flexible polymers}

Semi-flexible polymers (also known as \textit{worm-like chains})
\cite{stiffchains_1967,marko_1995} are long, one-dimensional chains
of $N$ molecules (monomeric units) whose intermolecular bonds resist
bending; this is in contrast to flexible chains \cite{pggpolymer_1979}
where there is no energetic penalty for bending (at scales that are
comparable to the size of a monomer) and which are completely governed
by entropy. Both types of polymers form globules, as in
\fig{SoftMatter}(b), but with different typical sizes.  The physics of
flexible chains are well known \cite{pggpolymer_1979,doi_1996,
  rubinstein_2003} and their resistance to changes of their ``size''
(end-to-end distance or radius of gyration, $R$) away from their
``random walk'' or Gaussian conformation where $R \sim N^{1/2}$ is
characterized in a mean-field treatment by a free energy per chain
 \begin{equation}
f=\frac{3k_BT}{2}  \ \frac{R^2}{N a^2}
 \label{eq:ideal1}
 \end{equation}
where $N$ is the number of monomers in the chain and $a$ is the monomer size.  Self-avoidance
of the chain due to excluded volume interactions among the monomers
leads to additional interactions and in a mean-field treatment the
scaling of $R$ with $N$ is modified so that $R \sim N^{3/5}$.  Many
biopolymers including DNA and various cytoskeletal filaments such as
actin and microtubules discussed later on, are semi-flexible, and only
bend on length scales of 50 nm (DNA) through micrometers (actin) or even
millimeters (microtubules), while synthetic polymers such as
polystyrene in organic solvents are flexible and easily bend on
nanometric scales.

On a coarse-grained, continuum level, the bending resistance of a
semi-flexible polymer is similar to that of an elastic rod
\cite{b:land70}.  Bending is geometrically characterized by the
curvature of the position vector of the rod, $\vec R(s)$ which is a
function of the monomer distance, $s$, along the contour ($0 \le s \le L$,
where $L$ is the contour length of the rod).  For systems where
positive and negative curvatures are equivalent by symmetry, there can
be no terms in the energy that are linear in curvature, so that in a
small curvature expansion (appropriate when the radius of curvature is
much larger than a monomer size), the energy, $H_b$, is quadratic in
the chain curvature \cite{b:land70}:
 \begin{equation}
 H_b = \frac{\kappa}{2} \int_0^L ds   \left( \frac{d^2 \vec R}{ds^2}  \right)^2
 \label{eq:bend1}
 \end{equation}
 where $\kappa$ is the bending modulus that characterizes the
 elastic resistance to bending.  The lowest energy deformations of the
 rod are the bending modes that do not result in an overall volume
 change of the rod and involve only relative extension and compression
 of its upper and lower surfaces \cite{b:land70}.  For most
 purposes one assumes that the rod is inextensible and neglects any
 stretching of the center of mass distances between molecules.  This
 is expressed by the inextensibility constraint that leaves the rod
 length unchanged:
 \begin{equation}
L =  \int_0^L ds   | \frac{d \vec R}{ds}  |
 \label{eq:bend2}
 \end{equation}
 and is  equivalent to the requirement that the tangent vector given by $d \vec R/ds$ is a unit vector.
 
 In equilibrium, a semi-flexible polymer represented by such a rod
 undergoes thermally driven motion that is resisted by the bending
 energy.  The inextensibility constraint makes this problem difficult
 to treat exactly \cite{stiffchains_1967,rubinstein_2003, marko_1995}.
 For small deformations of a chain oriented in the $z$ direction, one
 can approximate $s \approx z$ and describe the chain position by
 $\vec R = (X(z),Y(z),z)$.  
 The deformations can be resolved into their Fourier components $X(q)=\int dz X(z) e^{i q z}$
 that are the normal modes which diagonalize the bending Hamiltonian.
 Using the equipartition theorem one finds that $\langle |X(q)|^2 \rangle =k_BT/(\kappa \, q^4)$
 and one can show that the tangent vectors $\hat t=d \vec R/dz$ of
 neighboring points are nearly
 equal with:
 \begin{equation}
\langle \left( \hat t(z) - \hat t(0) \right)^2 \rangle \sim k_BT \,
\int dq \frac{\left( 1 - \cos(q z) \right)}
{\kappa q^2} \sim   \frac{k_BT}{\kappa} \, z
 \label{eq:bend4}
 \end{equation}
The tangent correlations diverge as the distance between the points along the rod increases and for large $z$ this invalidates the
approximation of small fluctuations and inextensibility (equivalent to a unit tangent vector).
However, one can find the value of $z$ at which the tangent correlations first  become of order unity;
this defines the persistence length \cite{rubinstein_2003, phillips_physical_2008}, $\zeta$ of the chain and one finds $\zeta \sim \kappa/(k_BT)$.
At scales smaller than the persistence length, the chain shows rigid-rod like behavior with relatively
small bending fluctuations; at longer scales, the fluctuations are large and a random walk (or
excluded volume random walk) picture is more appropriate.

\subsection{Polymer gels}

While single cytoskeletal proteins such as actin filaments or microtubules can be
modeled as semi-flexible polymers, the CSK often contains
crosslinked assemblies (gels) comprising these proteins, as in
\fig{SoftMatter}(c). The assemblies can be network-like
(macroscopically isotropic) or ordered into bundle-like
filaments. Here we review the response of semi-flexible polymers to
applied, static forces that stretch the chains and determine the
regimes in which the chains respond linearly or non-linearly to
applied force. 

For simplicity, we focus on a chain whose projected length is less
than or of the order of its persistence length.  In the context of a
crosslinked gel, the projected length is determined by the distance between the
crosslinks, assuming permanent crosslinks at whose positions the
polymer is rigidly held fixed.  The dissociation of the crosslinks
disrupts the network and can lead to non-elastic ({\it{e.g.}},\ viscous flow)
response to stress; however, we focus on the early time (tens of
seconds and possibly more in strongly adherent cells) behavior where
the network response to force is elastic in nature
\cite{wottawah_optical_2005}.  We consider the elastic response of a
single, semi-flexible polymer.  Naively, one might think that this
response will be typical of a polymer segment in the gel whose
projected length is the {\em average} spacing between crosslinks; the
distribution of crosslinks in the gel implies a distribution of
polymer segment lengths between crosslinks.  This would indeed be true
for affine deformations, in which each chain is stretched in the same
proportions as the macroscopically applied stress or strain.  However,
for large deformations, where the elastic response is highly
non-linear, the distribution of stresses among the chains with varying
segment lengths can be length dependent; the stresses will not be
affine and it is harder to associate the gel with the response of one
chain of {\em average} segment length
\cite{head_deformation_2003,head_distinct_2003,wilhelm_elasticity_2003,heussinger_stiff_2006,heussinger_nonaffine_2007}.

Before treating the case of semi-flexible polymers, we briefly derive
the elastic modulus that characterizes the response of flexible
polymers to applied forces (so-called \textit{rubber elasticity}). The modulus
is completely determined by the changes in the chain entropy that are
due to the applied strain, $\epsilon_i=\lambda_i-1$ $(i=x,y,z)$ that
changes the macroscopic dimensions of the sample from $(L_x, L_y,
L_z)$ to $(\lambda_x L_x, \lambda_y L_y, \lambda_z L_z)$.
Incompressibility of the chains and solvent implies that the volume
must remain unchanged, so that the product $\lambda_x \lambda_y
\lambda_z =1$. The free energy per chain in the unstressed system is
given by Eq.~(\ref{eq:bend1}) and for affine strains where $\vec
R=(X,Y,Z) \rightarrow (\lambda_x X, \lambda_y Y, \lambda_z Z)$, the
free energy per chain becomes:
\begin{equation}
 f=\frac{k_BT}{2} \ \left( \lambda_x^2+\lambda_y^2+\lambda_z^2 - 3 \right)
 \label{eq:affine1}
\end{equation}
We consider a uniaxial deformation in the $x$ direction,
$\lambda_x=\lambda$ and by incompressibility $\lambda_y = \lambda _z
=1/\sqrt{\lambda}$.  The force applied to a single chain is $\partial
f/\partial L_x$ and the stress in the entire system of chains,
$\sigma$, is the total force applied per unit area: $\sigma =\rho k_B
T (\lambda^2-1/\lambda)$, where $\rho$ is the number of chain segments
per unit volume.  For small deformations, $\lambda \approx 1$, an
expansion of the expression for $\sigma$ shows that the stress is
proportional to the product of the strain and $\rho k_BT$, similar to
the pressure of an ideal gas.  For large strains, the stress is
non-linearly related to the strain but this arises from the
incompressibility condition and not from any specific properties of
the chains.  Fluctuations of the crosslinks will further reduce the
strain \cite{rubinstein_2003}.

Semi-flexible chains have a more complex response to applied forces and
one can use the model described above to predict their
stress-dependent elastic modulus. When semi-flexible chains are
stretched near their limit, the additional force to stretch them
further tends to diverge and this results in an elastic modulus that is
intrinsically stress dependent. One considers a Hamiltonian that
includes the bending energy as well as an energy that tends to
equalize the projected length, $L_p$, and contour length,
$L=\int_0^{L_p} dz\,\sqrt{1+  X'(z)^2 + Y'(z)^2}$.
  This arises from a tension (energy per unit length), $\tau$
that   couples to the difference, $L - L_p$.    In the
approximation that the fluctuations are small, one can expand the
square root to obtain:
 \begin{align}
 H_\tau & =\frac{\kappa}{2} \int_0^L  \, dz   \left( X''(z)^2 + Y''(z)^2  \right)^2 \nonumber \\
& + \frac{\tau}{2} \int_0^L  \, dz   \left( X'(z)^2 + Y'(z)^2 \right)^2
 \label{eq:semiflex1}
 \end{align}
Using equipartition of the Fourier modes of the chain fluctuations
\cite{b:land70, safran_statistical_2003}
one can calculate \cite{mackintosh_nato_2006} $\delta \ell$, which is the increase
in the chain extension compared to its zero-tension, fluctuating value:
 \begin{equation}
\delta \ell= \frac{L^2}{6 \zeta} \, \left[ 1 + \frac{3}{ \pi^2 \alpha} -
\frac{3 \coth\left( \pi \sqrt{\alpha}\right)}{ \pi \sqrt{\alpha}} \right]
 \label{eq:semiflex2}
 \end{equation}
 where $\alpha=\tau L^2/\kappa \pi^2$ is a dimensionless
 measure of the applied force and  $\zeta$ is the persistence length
 defined above.
 
 For small forces, $\delta\ell \sim \tau L^4/(\kappa \zeta)$;
 the excess strain, $\delta \ell/L$
 is proportional to the force and the system is harmonic.
 For large forces,  $\delta \ell$ approaches the value for full extension of
 $\delta \ell_0=L^2/(6 \zeta)$ and the difference
 $\delta \ell_0 -  \delta \ell \sim 1/\sqrt{\tau}$.
 This non-linear relationship between extension and applied force
 expresses the fact that as the chain approaches its maximum
 extension, a very large force must be applied.
 The measured elastic constant of the crosslinked, semi-flexible polymer
 gel is then stress dependent as discussed below in the context of actin gels.
 
\subsection{Elements of elasticity}

In the previous discussion, we have employed scalar definitions of the
stress and strain developed in an elastic system that is subject to
applied forces. While liquids and gases also resist compression, they
do not show an elastic response to external forces that act only to
change the shape of the system; such forces (per unit area) that do
not induce any volume change, are called \textit{shear stresses}. The
elastic response (restoring force) to shear stresses are
characteristic of solids.  Crosslinked gels, while being disordered,
are indeed classified as solids since they resist shape changes and
can be described by elasticity theory. This is true when the
crosslinks are permanent, or very long-lived; otherwise, one must deal
with time (or frequency) dependent elastic constants
\cite{b:fung93,b:boal}.  Biological gels are typically not permanently
crosslinked \cite{bauschtrans} and at long times one expects
liquid-like flow instead of an elastic response to shear forces.  This
indeed is the time regime in which the CSK is modeled \cite{
  liverpool_instabilities_2003,kruse_asters_2004,julicher_active_2007,marchetti_soft_2012}
as an \textit{active gel} that flows in response to internal forces
generated by cell activity which is fueled by energy consumption.
Here we restrict our focus to the early-time (tens of seconds)
behavior of the CSK \cite{fabryrev} where the crosslinkers still
maintain the elastic response of the CSK to both internal and external
forces. The elastic approach is also more appropriate for the ECM
which remodels much less than the CSK.

In the presence of external or internal forces that are not part of the elastic network themselves (including thermal forces that
change the positions of the particles), and in a continuum picture,
the material particles that comprise an elastic system are assumed to
be displaced from their equilibrium positions  by a smooth displacement field $\vec u(\vec r)$ where $\vec
r=(r_1,r_2,r_3)=(x,y,z)$.  The elastic energy arises from
interparticle interactions and is thus a function not of $\vec u(\vec
r)$, but of its spatial gradients, that represent changes in the
relative positions of the particles.  This is true in the absence of
any external ``pinning'' forces for which translations of the system
(where $\vec u(\vec r)$ is constant), have no energy cost.  The elastic
energy is thus a function of the strain tensor $u_{ij}$ defined
\cite{b:land70} as
\begin{equation}
u_{ij}=\frac{1}{2} \left( \frac{\partial u_i(\vec r)}{\partial r_j} + \frac{\partial u_j(\vec r)}{\partial r_i}
+ \frac{\partial u_l(\vec r)}{\partial r_i} \frac{\partial u_l(\vec r)}{\partial r_j} \right)
\label{eq:strain1}
\end{equation}
where summation over the repeated index $l$ is implied.  The
non-linear term on the right can be neglected for small strains.  The
local change in a small length element $dx$ is $dx (1+ u_{xx})$ so
that the local volume change (given by the product $dx dy dz$ minus
the initial volume), is determined to first order in the strain by
$tr(u_{ij})=u_{ii}=u_{xx}+u_{yy}+u_{zz}$.  These are coupled to
isotropic compressions or expansions while shear forces that change
the shape of the system couple to the off-diagonal strain components
such as $\partial u_x/\partial y$ that represent changes in the
interparticle spacing in the $x$ direction that vary in the $y$
direction.

Displacing the particles from their equilibrium positions creates
strains that are resisted by internal restoring forces that originate
in the intermolecular interactions (and in the case of polymeric gels,
entropy) that provide shape memory and hence elasticity.  The forces
that arise from the elasticity are described by a stress tensor,
$\sigma_{ij}(\vec r)$.  This is the force per unit area in the $i$
direction that acts on the surfaces whose normal is in the $j$
direction of an infinitesimal volume element. The pressure is the negative of one-third of the trace of
the stress. In the absence of motion, the difference of the stresses on two
surfaces separated by a distance $d \vec r$ is attributed to the
presence of a local force density, $\vec f(\vec r)$ within that volume
element so that \cite{b:land70} in equilibrium, $ f_i(\vec r)=-\sum_j
\, \partial \sigma_{ij}/\partial r_j$. It is important to note that
the force per unit volume, $f_i$, is attributed to forces that are not
included in the system's elastic response and arise either from active
internal elements or from macroscopic forces that act on the system
boundaries. In the absence of such forces, mechanical equilibrium thus
dictates that the divergence of the stress tensor vanishes.

For an isotropic body, rotational symmetry implies that there are two tensor
components that must be considered for the strain and stress: (i) the
trace that describes the local volume change, $u^0(\vec r)=u_{ij}(\vec
r) \delta_{ij}$ or the hydrostatic pressure, $-\sigma^0(\vec r)/3=-\sigma_{ij}(\vec r)
\delta_{ij}/3$ (where one sums over the repeated index) and (ii) the
traceless shear, defined as $u^s_{ij}(\vec r)= u_{ij}(\vec r) - (1/3)
u^0(\vec r) \delta_{ij}$ with a similar expression for the shear
stress.  Since the internal forces that resist deformations can also
include thermal effects at the intra-molecular level (such as changes
in the conformations of polymers in gel networks), one considers the
elastic free energy per unit volume \cite{b:land70}, $f_e$. The
free energy associated with elastic deformations is a scalar and can
be written from the following symmetry considerations. (i) The free energy depends only
on the strains and not on the displacements.  (ii) There is no term
linear in strain since the deformation free energy represents an
expansion about equilibrium where the free energy is minimal.  (iii)
The free energy is a scalar and cannot depend on the coordinate
system.  Since $u^0_{ij} u^s_{ij}=0$, the free energy written up to
quadratic order in the strains can only contain terms with
$(u^0_{ij})^2$ and $u^s_{ij} u^s_{ij}$:
\begin{equation}
f_e=  \frac{K}{2} \left(\sum_i \, u_{ii} \right)^2 + \mu \, \sum_{ij}  \left(\, u_{ij}
- \frac{1}{3} \delta_{ij} \ \sum_l \, u_{ll}   \right)^2
\label{eq:strain2}
\end{equation}
where $u_{ij}$ denotes the local strain, $u_{ij}(\vec r)$.  The first
term accounts for the free energy associated with volume changes and
is proportional to the bulk modulus, $K$, while the second term
accounts for the shear response, proportional to the shear modulus
$\mu$.  These two elastic constants that have the dimensions of energy
per unit volume (the same as pressure, measured in Pa), are material
dependent and can also be expressed (in three-dimensions) by the
Young's modulus $E=9K\mu/(3K+\mu)$ and Poisson ratio
$\nu=(3K-2\mu)/(2(3K+\mu))$. As already mentioned above, the Young's
modulus is the elastic constant that appears naturally for a
one-dimensional stretching experiment. Tensorial elasticity shows that
even in the simplest case of linear isotropic elasticity, two elastic
constants exist, with the Poisson ratio acting as a second elastic
constant that accounts for how different dimensions are coupled to each
other. The Young's modulus can show tremendous variation depending on
the strength of the interparticle interactions and the typical
particle spacing: diamond or carbon sheets have $E \sim \rm{TPa}$, metals
have $E \sim 100 \rm{GPa}$, rubber has $E \sim \rm{MPa}$, while tissue cells
typically have $E \sim 10 \rm{kPa}$.  The large differences between the
rigidities of molecular and cellular systems are mostly determined by
the very different length scales involved: the modulus (with
dimensions of energy per unit volume) scales as the inverse of the
\textit{cube} of the characteristic length that determines the
interactions. Materials whose cohesive energy is due to interatomic or
intermolecular interactions on the nm scale can therefore have elastic
moduli that are 6 orders of magnitude larger than the biopolymer gels
that comprise the CSK or ECM where the crosslink distance can be 100~nm or
more.  For incompressible materials, $K/\mu \rightarrow \infty$ and
$\nu \rightarrow 1/2$, while in the opposite limit of highly
compressible materials, $\nu \rightarrow -1$.  Most biological gels
are fairly incompressible due to the presence of water that
solubilizes the biopolymeric elastic elements, with $\nu$ in the range
of 1/3 to 1/2.

The strains in an elastic material result in forces that tend to
restore the equilibrium, unstrained state. These are most
conveniently given by the stress tensor, $\sigma_{ij}$ (force per
unit area) that is derived from derivative of the free energy with
respect to the strains (analogous to force given by the derivative of
the energy with respect to displacement): $\sigma_{ij}=\partial
f_e/\partial u_{ij}$.  This relationship implies that the elastic
deformation energy per unit volume can also be written:
\begin{equation}
f_e= \frac{1}{2}\, \sum_{ij} \,  \sigma_{ij}\, u_{ij}
\label{eq:stress2}
\end{equation}
Using the expression for the force balance in mechanical
equilibrium, Eq.~\ref{eq:stress2} for the free energy (for
small strains), and the relationship between stress and strain,
one finds:
\begin{equation*}
f_i(\vec r) = - \frac{\partial \sigma_{ij}(\vec r)}{\partial r_j} =
- \tilde E \left[ \frac{\nu}{(1-2\nu)}
\frac{\partial u_{\ell \ell}}{\partial r_i}+ \frac{\partial u_{ij}}{\partial r_j}
\right]
\end{equation*}
\begin{equation}
= - \frac{\tilde E}{2} \left[ \frac{\partial^2 u_{i}}{\partial r_j^2}
+ \frac{1}{(1-2 \nu)} \frac{\partial^2 u_j}{\partial r_i \partial r_j}  \right]
\label{eq:strain5}
\end{equation}
where $\tilde E=E/(1+\nu)$ and a summation is implied by repeated indices.
The second and third equalities in Eq.~(\ref{eq:strain5}) are obtained from the
definitions of the stress and strain tensors.

The solution of such linear differential equations with a source term (the
internal force distribution, $\vec f(\vec r)$) is given by the convolution
of the source (the force at position $\vec r\, '\,$) with the Green's function, $G_{ij}(\vec r, \vec r\,'\,)$
 of the system \cite{arfken_1995}.
In our case this predicts the displacement:
\begin{equation}
u_i(\vec r)=\int d \vec r\, ' \ G_{ij}(\vec r, \vec r\, '\,) f_j(\vec r\, '\,)
\label{eq:strain4}
\end{equation}
The Green's function itself is given by the solution of Eq.~(\ref{eq:strain5}) for
$u_i(\vec r)$  for the case of a delta-function, point force located at $\vec r\, '$.
For an infinite elastic domain, the Greens function depends only
on $\vec R=\vec r - \vec r\, '$ and is written \cite{b:land70}:
\begin{equation}
G_{ij}(\vec R)= \frac{1}{8 \pi \tilde E (1-\nu)R} \left[ (3-4 \nu) \delta_{ij} +
\frac{R_i R_j}{R^2} \right]
\label{eq:strain5Green}
\end{equation}
While the angular dependence is complex and resembles that of an
electric dipole, the distance dependence of $1/R$ is similar to the
potential due to a point charge.  Similar to electrostatics, elastic
stresses and strains due to localized forces are long-ranged. As we
shall see later, this allows cells to communicate with each other and
with the boundaries of their physical environment over relatively
large distances.

\subsection{Adhesion of vesicles and capsules}

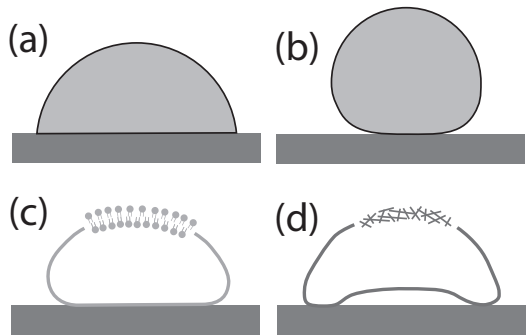
\FIG{Fig02}{fig:CapsuleAdhesion}{Simple soft matter
  models relevant to the passive features of cell adhesion to a flat
  substrate. (a) A liquid droplet adhering to a surface is governed by
  surface tension. (b) A solid elastic sphere gains adhesion energy by
  forming a contact region whose size is determined by the
  balance of the adhesion and shear deformation energies. (c) A closed
  shell of a fluid, lipid bilayer (\textit{vesicle}) is governed by bending
  energy. (d) A polymeric capsule has both bending and stretching
  energy; their interplay can lead to buckling in the contact area.}

Until now we have discussed bulk phases of soft matter and
biomaterials.  We next address finite-sized model systems that can
account for some aspects (mainly passive responses) of cells, namely
fluid droplets, elastic spheres, vesicles and capsules, as in
\fig{CapsuleAdhesion}. We consider the case where these bodies are in
contact with an attractive surface that favors adhesion. While fluid
droplets and elastic spheres are both chemically homogeneous, with the
same chemical species at both the surface and in the bulk, vesicles
and capsules (also known as \textit{polymerized vesicles}) are
characterized by surfaces whose composition differs from that of the
bulk. Vesicles typically consist of fluid, amphiphilic bilayers that
enclose a spherical water core. Although the thermodynamic stable
phase is usually the lamellar phase depicted in \fig{SoftMatter}(d),
vesicles are metastable over very long time scales and ubiquitous in 
biological systems. The bilayers respond to forces that
couple to their curvature (bending response). The
surface of capsules are typically thin polymer films with both bending
and elastic response. Due to their membrane-like nature that is
sensitive to bending and/or elastic forces, the adhesion of vesicles
and capsules are interesting reference cases for the
adhesion of cells.

The interface of a fluid droplet is defined as the region where two
coexisting phases overlap ({\it{e.g.}}, fluid and vapour). The interfacial
energy therefore scales to first order with the product of the
geometrical area and the surface tension $\sigma$
\cite{safran_statistical_2003}. The interfacial Hamiltonian is simply
\begin{equation}
U_i = \sigma \int dA
\label{eq:SurfaceEnergy}
\end{equation}
Variation of this surface functional with a Lagrange parameter $\Delta p$
that enforces the conservation of volume ($\Delta p$ simply corresponds to
the pressure difference between the inside and outside of the sphere)
yields the Laplace law $H = \Delta p / (2 \sigma)$, where $H = ( 1/R_1 + 1/R_2
) / 2$ is the mean curvature of the surface and $R_1$ and $R_2$ are
the two principal radii of curvature \cite{safran_statistical_2003}.
For a free droplet, the solution will be simply a sphere, with the
mean curvature $H = 1/R$ everywhere. For an adherent droplet, the
Laplace law is valid for the free part of the droplet, which therefore
will be a spherical cap of radius $R$, as in
\fig{CapsuleAdhesion}(a). The exact dimensions of this spherical cap
are determined by the overall volume and the contact angle $\theta$,
which in turn is determined by the interfacial energies according to
Young's law:
\begin{equation}
\cos \theta = \frac{\sigma_{SG} - \sigma_{SL}}{\sigma}
\end{equation}
where $\sigma_{SG}$ is the interfacial energy between the substrate and
the gas phase and $\sigma_{SL}$ is the interfacial energy between the substrate
and the liquid phase, respectively.  The contact angle according to
Young's law also determines the direction in which the interface is
pulling as expressed by its surface tension. With a typical contact angle
around 90 degrees, the pulling force is mainly normal to the
substrate. The horizontal component of this pulling force is balanced
by the surface energies associated with adhesion.

A filled elastic sphere of homogeneous composition and with radius $R$
that adheres to a surface, forms a finite-sized contact region of
radius $a$, as in \fig{CapsuleAdhesion}(b). The size of the
adhesion region is determined from the balance of the gain in adhesion energy
per unit area $W$ and the elastic energy penalty from the
deformation that accumulates in the sphere due to the shape change
upon adhesion. For a material that obeys linear elasticity, this
depends on the Young's modulus $E$ and the Poisson ratio $\nu$. The
balance of the adhesion and elastic forces is treated in contact
mechanics and was first solved by Johnson, Kendall and Roberts
(\textit{JKR-theory}) \cite{e:john71,b:john85}, who calculated
\begin{equation}
a^3 = \frac{9 \pi (1-\nu^2)}{2 E} R^2 W
\end{equation}
Thus, the linear dimension of the adhesion area increases with
adhesion energy (due to the gain in adhesion energy) and decreases
with Young's modulus (since it tends to oppose the shape deformation
induced by the attractive adhesion energy). Note that these
calculations assume only normal forces. Contact mechanics predicts
that in order to detach the elastic sphere in the normal direction, a
critical force $F_c = 3 W \pi R / 2$ is required, which surprisingly
depends only on the adhesion energy and is independent of the elastic
constants. For an elastic sphere pushed onto the substrate by a normal
force, one has the Hertzian stress profile $\sigma(r) = \sigma_{0} (1
- (r/a)^2)^{1/2}$, where $r$ is the radial coordinate. In marked
contrast to this, the JKR-solution, which applies to a self-adhered
elastic sphere has an additional contribution $(1 - (r/a)^2)^{-1/2}$
that diverges at the boundary. The localization of the stress to the
boundary make the contact prone to fracture from the periphery due to
crack nucleation.
  
In contrast to droplets and elastic spheres, the interfacial energy of
vesicles and capsules is determined by the force response of the
molecules on the surface. For thin, elastic shells (capsules)
that obey linear, isotropic elasticity with bulk
Young's modulus $E$ and Poisson ratio $\nu$, there are three main
deformation modes: out of plane bending as well as in-plane shear and
stretching. The bending energy reads
\begin{equation}
U_b = 2 \kappa \int dA H^2
\label{eq:BendingEnergy}
\end{equation} 
where $H$ is the mean curvature as above and $\kappa$ is the bending
rigidity which is related to the elastic properties of the material by
$\kappa = E h^3 / 12 (1-\nu^2)$ \cite{b:land70}. A simple material law
for the in-plane contributions is \cite{lim_h._w._stomatocytediscocyteechinocyte_2002}
\begin{equation} \label{eq:RBCthinshell}
U_p = \int dA \left\{ \mu \frac{(\lambda_1 - \lambda_2)^2}{2 \lambda_1 \lambda_2}
+ \frac{K}{2} (\lambda_1 \lambda_2 - 1)^2 \right\}
\end{equation}
where $\mu$ and $K$ are two-dimensional shear and bulk moduli,
respectively, which are related to the three-dimensional moduli by
multiplication by the shell thickness $h$; here $\lambda_i = 1
+ u_{ii}$ are the principal extension ratios.

For vesicles, comprising amphiphilic bilayers that are generally
fluid, the in-plane deformations are not relevant for two reasons.
Due to the fluid nature of the lipid bilayer, the shear modulus
vanishes, and the bulk modulus is so large that the system is
effectively incompressible. Therefore, only the bending energy is
relevant; the form of the bending energy is the same as in
\eq{BendingEnergy}, but the origin of the bending energy depends on the
molecular characteristics; for systems with long chain molecules, the
entropy which is a function of chain length, can play an important
role \cite{safran_curvature_1999}. The typical bending rigidity of
amphiphilic lipids that comprise biological membranes is $\kappa = 20\
k_B T$. A detailed shape analysis of the bending Hamiltonian
\eq{BendingEnergy} and its extensions to account for each of the
monolayers that comprise the bilayer has shown that free vesicles can
adopt a large variety of often surprising shapes
\cite{a:canh70,a:helf73,seifert_shape_1991,miao_budding_1994,seifert_configurations_1997}.
In order to calculate vesicle shape upon adhesion, as in
\fig{CapsuleAdhesion}(c), one must consider the competition of the
bending energy with the adhesion energy where $W$ is the adhesion
energy per unit area
\cite{seifert_adhesion_1990,seifert_configurations_1997}.  For weak
adhesion or small radii of curvature, the bending energy dominates and
the vesicle maintains it spherical shape without deforming to adhere
to the surface. However, in the case of strong adhesion, $W R_0^2 /
\kappa \gg 1$ (where $R_0$ is the equivalent sphere radius defined by
the vesicle volume $V=4 \pi R_0^3 / 3$), the vesicle shape effectively
approaches a spherical cap with a well-defined contact radius. In this
case, the adhesion forces will again be mostly normal and localized to
the rim of the adhesion region.

As we will see later, an important aspect of cell adhesion is that
adhesion molecules are mobile in the lipid bilayers and can form local
clusters. This has indeed been demonstrated experimentally in a
vesicular systems by incorporating such adhesion molecules within the
lipid bilayers \cite{albersdorfer_adhesion-induced_1997}.  Theoretical
models have shown that membrane fluctuations lead to an effective
attractive interaction between the adhesion molecules which can
explain this clustering
\cite{zuckerman_statistical_1995,menes_nonlinear_1997,lipowsky_adhesion_1996,weikl_adhesion-induced_2001,smith_effective_2005,smith_force-induced_2008}
and there is experimental evidence that indeed this mechanism also
operates in biological cells
\cite{delanoe-ayari_membrane_2004}. However, despite the presence of
this local clustering, the contact zone of vesicles that adhere to a
surface through specific adhesion molecules tends to remain rather
homogeneous.

In contrast to vesicle adhesion, capsule adhesion also involves in-plane
elastic energies. It is well known that in particular the
stretching energy cannot be neglected when dealing with the shape of
capsules, because the ratio of stretching and bending energies for
spherical shells scales as $(R/h)^2$ (where $R$ is the radius of
curvature and $h$ is the shell thickness) and is therefore always
large \cite{b:land70}. An important consequence of this fact is that
thin elastic capsules buckle inwards when a critical pressure of $p_c
\sim E (h/R)^2$ is exceeded \cite{b:land70}. In general, the interplay
between stretching and bending (possibly complemented by sheet adhesion to
itself) leads to a very rich
phase diagram of possible shapes \cite{knoche_buckling_2011}. A
rich variety of phenomena also arises for forced crumpling of
planar sheets such as paper or graphene
\cite{lobkovsky_scaling_1995,vliegenthart_forced_2006} or closed
shells such as ping pong balls, fullerenes or virus capsids
\cite{uss:schw00d,vliegenthart_mechanical_2006}. For red blood cells,
one must combine the elasticity of thin shells with the bending
energy; one then finds very good agreement between simulated and
observed shapes, both for free cells
\cite{lim_h._w._stomatocytediscocyteechinocyte_2002} and for cells in
hydrodynamic shear flow \cite{noguchi_shape_2005}.

Because attraction to a flat substrate results in deformations that are
similar to those induced by external pressure or forces, adhesion also
can lead to the inward buckling of an adherent capsule, as in
\fig{CapsuleAdhesion}(d) \cite{fery_mechanical_2007}.  A computer
simulation for spherical shells adhering to a flat substrate has shown
that as the adhesion energy increases, the shell first flattens like
an elastic sphere, then buckles in a radially symmetric manner, and
finally develops a polygonal adhesion region through the formation of
elastic ridges running in parallel to the substrate
\cite{komura_buckling_2005}.  This shows that capsules in adhesion can
develop very inhomogeneous adhesion regions, and suggests that
interfacial stresses will mainly be localized at the rim of the adhesion
area.

Here we focused on the competition of adhesion and deformation
energies in determining the shapes of adhering bodies as relevant
background to understand the specific features of cell adhesion. As we
will see in the next section, however, cells adhesion is characterized
by additional and mainly active features that do not exist in the
passive systems discussed so far. Adherent cells tend to develop very
inhomogeneous contact areas, with adhesion molecules strongly
clustered along the periphery of the adhesion region. In particular,
the inward buckling characteristic for homogeneously adhering capsules
is not observed for cells. The stress localization expected for
capsules is weakened by remodeling processes at the cell periphery,
which are, in turn, closely coupled to the growth and stabilization of
the adhesions. Most importantly, the adhesion structures of cells are
extremely dynamic, with a constant flow of material from the cell
periphery towards the cell center.

\section{Biology Background}

\subsection{Actin cytoskeleton and cell adhesion}

Cells are the smallest units of life and widely vary in their shape,
structure and function
\cite{alberts_molecular_2007,phillips_physical_2008,
  b:boal,b:bray01}. For simplicity, we focus here on animal cells,
thereby excluding {\it{e.g.}}, bacteria, protists and plant cells from our discussion.
Typical cell sizes are of the order of tens of micrometers and there are
roughly $10^{14}$ cells in humans. They can be classified into 200
major cell types ranging from connective tissue cells through
epithelial and muscle cells to nerve cells \cite{alberts_molecular_2007}. All cells in an organism carry
the same genome, but as a result of differentiation, different cell types have
different gene expression patterns, {\it{i.e.}}, different cell types produce different proteins.  If
viewed from the point of view of soft materials, however, all animal cells are
similar, including a spatial organization determined by
lipid bilayers and the polymer networks of the
cytoskeleton.

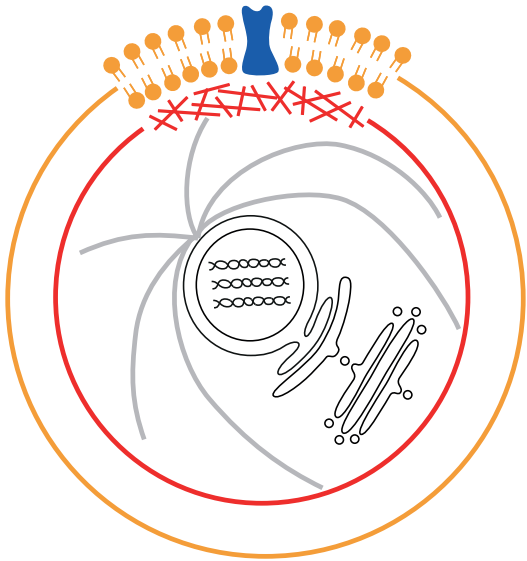
\FIG{Fig03}{fig:CellSchematics}{(Color online) Schematic drawing of
  an animal cell in suspension.  Such a cell is essentially
  round due to its effective surface tension. Important cellular organelles responsible for its internal
  structure and mechanical properties include (1) the plasma membrane
  (orange), a lipid bilayer that envelopes the entire cell and carries
  different proteins, including transmembrane receptors; (2) other
  membrane structures (thin black lines) such as the two membranes around
  the nucleus containing the genes,
  the endoplasmic reticulum, the Golgi apparatus and different kinds
  of vesicles; (3) the actin cortex (red), a thin shell comprising a polymer network
  underlying the plasma membrane; (4) the microtubule system
  (thick grey lines), a system of relatively stiff polymers that radiate outward
  from the microtubule organizing center that is attached to the
  nucleus.}

Fig.~\ref{fig:CellSchematics} shows a schematic representation of the
main structural elements of an animal tissue cell in suspension. The
cell is separated from its surroundings by a \textit{plasma membrane},
which is a bilayer that comprises different lipid molecules and is
enriched by additional components such as cholesterol. The plasma
membrane is fluid in nature (no fixed topological relations of
neighboring molecules, flow under shear deformations) and acts as a
carrier for a large variety of membrane-bound proteins and
sugars. Underneath the plasma membrane is the actin cortex, a
relatively thin (100 nm) dynamic layer of crosslinked actin filaments
whose mechanical properties dominate the elastic response in reaction
to deformations of the cell. The plasma membrane and the actin cortex
are coupled through a variety of linker molecules that are separated
by relatively large distances, so that the membrane between them can
fluctuate relatively freely, leading to the phenomenon of
\textit{membrane flickering}. The \textit{cytoplasm} of the cell
refers to the cellular volume (excluding the nucleus containing the
generic material in the form of DNA) delimited by the plasma
membrane. It contains several organelles important for cell function,
including a variety of additional membrane systems (such as the
\textit{endoplasmic reticulum} and the \textit{Golgi apparatus}) and
polymer networks.  There are three important types of polymer
networks, based on actin filaments, microtubules and intermediate
filaments, respectively
\cite{b:howa01,b:boal,phillips_physical_2008,alberts_molecular_2007}.
Collectively, they are called the \textit{cytoskeleton}. For a cell in
solution, only the microtubule network is well developed in the
cytoplasm.

Animal cells in suspension are usually round as depicted in
\fig{CellSchematics}, indicating an effective surface tension arising
from the combined effect of plasma membrane and actin cortex. The
round shape of a cell changes once it adheres to an external
surface. If a cell encounters an external surface covered with
specific ligand, it undergoes a multi-step process that determines
whether or not it eventually will adhere
\cite{bershadsky_adhesion-dependent_2003,cohen_spatial_2004}. In
general, cells use different mechanisms to avoid non-specific adhesion
({\it{e.g.}}, due to van der Waals forces), including a repulsive
sugar layer anchored in the membrane (\textit{glycocalix}) as well as
the steric (entropic) repulsion due to membrane fluctuations
\cite{safran_statistical_2003}. Adhesion is only induced if it is
promoted by specific molecular signals that are found on the
substrate. The specificity of cell-matrix adhesion is implemented by
transmembrane adhesion receptors (in humans, these are mainly the 24
variants of the integrin family), which bind to complementary ligands
of the extracellular matrix (including collagen, fibronectin,
vitronectin and laminin). Similar to passive vesicles or capsules, the
early stages of cell adhesion and spreading can be strongly determined
by viscoelastic processes, {\it{e.g.}}, the deformation of the rim of
the developing contact region \cite{cuvelier_universal_2007}. Later
stages are more strongly determined by remodeling of the cytoskeleton
and the establishment of localized sites of specific adhesion. During
the remodeling process, the actin system is organized into additional
networks extending throughout the cytoplasm. Because these networks
are crosslinked, the actin cytoskeleton provides the cell with elastic
restoring forces that resist shear deformations and is thus essential
in determining the shape, stability and mechanical response of
cells. While the volume of a cell tends to stay constant during
adhesion and spreading, the surface can increase by up to 50 percent,
which occurs via the flattening of the undulated membrane as well as
by the addition of new lipid material \cite{gauthier_temporary_2011}.

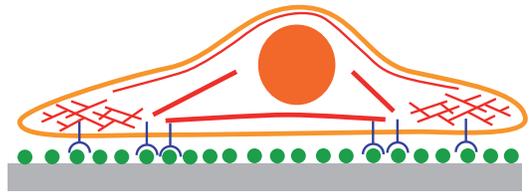
\FIG{Fig04}{fig:CellAdhesionSchematics}{(Color online) Schematic
  drawing of an adherent animal cell. Such a cell typically has the shape of a fried egg,
  with the nuclear region protruding in the middle while the rest of the cell remains relatively flat. As opposed to
  \fig{CellSchematics} for a freely suspended cell, here we do not
  depict the membrane or microtubule systems. In addition to the actin cortex, the actin cytoskeleton
  (red) now forms several additional subsystems that extend throughout the cytoplasm. Here we depict a dendritic
  actin networks that pushes outwards against the plasma membrane
  (\textit{lamellipodium}) and contractile actin filament bundles
  (\textit{stress fibers}) that are anchored to the cellular
  environment through transmembrane receptors (blue) that bind
  extracellullar ligands (green).}

Fig.~\ref{fig:CellAdhesionSchematics} schematically depicts the actin
structures that are typically developed during cell adhesion and
spreading. In contrast to \fig{CellSchematics}, we do not depict the
microtubule system here, because it has only a supportive function in
this context (its main function here is to coordinate processes
involved in active transport and cell migration). The main mechanism
that leads to outward expansion of the plasma membrane and thus to the
development of a contact area with the substrate is the rapid
polymerization of an actin network at the cell periphery
(\textit{lamellipodium}). Lamellipodia grow through the elementary
processes of actin filament polymerization, branching, capping and
crosslinking
\cite{pollard_cellular_2003,pollard_actin_2009,ridley_life_2011},
which have been extensively modeled
\cite{mogilner_edge:_2006,pollard_mathematical_2009}.  The most common
structure of the lamellipodium seems to be a tree-like
(\textit{dendritic}) network with a $\pm 35$ degree orientation
relative to the leading edge of the cell membrane due to the $70$
degree angle in the protein complex Arp2/3 connecting branched-off
daughter filaments to mother filaments \cite{svitkina_arp2/3_1999}.
The exact organization of the lamellipodium varies as a function of
cell type, motility state and external signals
\cite{urban_electron_2010,uss:weic12a}.  One of the most important
aspects of lamellipodia growth is its force-velocity relation, for
which conflicting experimental evidence exists
\cite{marcy_forces_2004,parekh_loading_2005,prass_direct_2006} and
which has been treated by various modeling approaches
\cite{carlsson_growth_2003,lee_force-velocity_2009,uss:weic10b,schreiber_simulation_2010,zimmermann_actin_2012,campas_actin_2012}.

Other types of actin structures that develop in cell adhesion are
bundles and networks that are contractile due to the action of
molecular motors that tend to slide actin filaments relative to each
other. If the filaments are sufficiently anchored to their
surroundings, they can no longer move; thus, instead of motion,
tension is developed in the actin bundles or network by the forces
exerted by the molecular motors. In adhesive cells, this is mainly
achieved by the molecular motor protein myosin II. In contrast to
muscle, where myosin assembles in groups of hundreds of motors, in the
cytoskeleton of non-muscle cells, it organizes into
myosin-minifilaments that typically contain only dozens of
non-skeletal myosins II molecules
\cite{verkhovsky_non-sarcomeric_1993}. The most prominent myosin-based
contractile structures in adhesion-dependent cells are \textit{stress
  fibers} \cite{pellegrin_actin_2007, burridge_tension_2013} shown
schematically as thick red lines in
Fig.~\ref{fig:CellAdhesionSchematics}. One can distinguish different
types of stress fibers \cite{hotulainen_stress_2006}. Dorsal stress
fibers connect to an adhesion site at one end and have their other end
connected to other actin structures in the cell that are far from the
substrate.  Ventral stress fibers are connected to adhesion sites at
both of their ends and thus run parallel to the substrate. In contrast
to dorsal and ventral stress fibers, transverse arcs are usually not
straight, are not connected to adhesion sites and typically extend
parallel to the leading edge. Stress fibers are thought to serve as
the main sources of cellular forces that are exerted on the substrate,
since their endpoints are often found at large adhesion sites that
correlate with large forces \cite{uss:bala01}. Laser cutting
experiments demonstrated that stress fibers are under large tension,
since they retract over large distances when being cut
\cite{kumar_viscoelastic_2006,uss:colo09}. Stress fibers are
distinguished from \textit{retraction fibers}, which are
non-contractile actin bundles that specifically serve to anchor the
cell to the ECM during cell division \cite{thery_experimental_2007}.

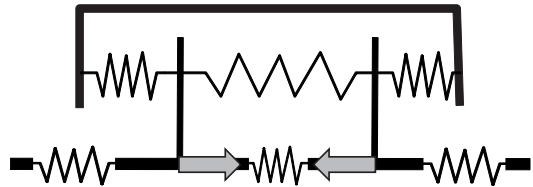
\FIG{Fig05}{fig:ForceBalance}{Scheme for the overall force
  balance in an adherent cell. There are two actin-based processes
  that contribute to force generation at the cell-material
  interface. Contraction by myosin II motors in actin networks and
  bundles corresponds to a stretched spring pulling inwards in the cell center.
  Lamellipodium growth against the membrane corresponds to a
  compressed spring pushing outward at the cell periphery. The entire system is constrained
  by the cell envelope.  Due to the position of the adhesion
  sites, a contractile force dipole emerges as the effective
  traction pattern on the substrate. This leads to deformation of the
  substrate (compression below the cell body, and elongation away from
  the cell).}

The lamellipodium and stress fibers are actin assemblies that create
pushing and pulling forces, respectively; hence, they are the two main
force-generating mechanisms for cells that adhere to flat
substrates. Although its effect is rather indirect, the plasma
membrane plays a very important role in this context. Apart from
acting as host for the transmembrane receptors from the integrin
family, it also controls the polymerization of the lamellipodium and
the contraction of the stress fibers by triggering biochemical signals
that regulate these processes \cite{ridley_life_2011}. Equally
important, the plasma membrane plays an important role in the overall
force balance in the cell, since its tension and curvature elasticity
provide the counterforces to actin-generated forces that tend to
extend and deform the membrane. An imbalance in these forces is
especially important in cell migration
\cite{lauffenburger_cell_1996,fletcher_introduction_2004}.

In \fig{ForceBalance}, we schematically show the overall force balance
in the cell. To first order, the lamellipodium can be depicted as a
compressed spring that pushes outward on the cell membrane and inward
on the focal adhesion. The stress fibers appear as stretched springs
that pull inward on the adhesion. In stationary or slowly migrating
cells the sites of adhesion are typically located in between the
polymerization-dominated lamellipodium and the myosin-dominated
contractile structures that are located closer to the cell body
(\textit{lamella}); thus, both processes effectively lead to
inward-directed forces on the substrate. For a strongly polarized,
stationary cell, the traction force pattern therefore resembles a pair
of oppositely directed forces (pointing from each side of the cell
towards the cell body) of equal magnitude.  As we will see later, this
concept of a \textit{contractile force dipole}
\cite{uss:schw02a,uss:schw02b} is very powerful when describing
cellular forces on a coarse-grained scale. The pulling of the force
dipole on the substrate leads to compression below the cell body and
elongation away from the cell, as schematically depicted by the
springs in the substrate.

The counterforces exerted by the substrate on the cell originate in
the substrate elasticity that resists deformation by the cellular
forces (in physiological tissue, this is the elasticity of the ECM).
The substrate resistance can reorganize the cellular cytoskeleton and
change the size of the adhesive regions.  The feedback between the
cellular and substrate elastic forces means that cellular structure
and function can be very sensitive to the elasticity and in particular
to the rigidity of the substrate
\cite{discher_tissue_2005,uss:schw05}.  For example, cells tend to
migrate from softer to more rigid substrates and to have larger
adhesive regions and overall spread area on more rigid substrates
\cite{c:pelh97,Lo00,engler_substrate_2004,trichet_evidence_2012}. Moreover,
the outside-in forces from the substrate that can modify the
cytoskeletal organization, can also have genetic implications.  In
particular, it was found that skeletal muscle cells differentiate
optimally on substrates with rigidities of 11 kPa
\cite{engler_myotubes_2004} and that stem-cell fate strongly depends
on substrate rigidity \cite{engler_matrix_2006}.

The fluid nature of the plasma membrane means that it is only
indirectly involved in force generation. Transmission of forces and
in particular, the sensitivity to shear, requires a solid-like
structure. In cells, the structural elements that give the cell its
shape integrity and its ability to respond to and to transmit forces
reside in the cytoskeleton.  In addition to this role, the
cytoskeleton is also important in anchoring organelles such as the Golgi
apparatus in their place in the cell, in determining the organized changes that take
place during cell division, in regulating the imbalance of internal
forces that results in cell motion, and in providing a scaffold for
signaling processes inside cells. Since in this review
we focus on force generating processes during cell adhesion, we
will be mainly concerned with the actin cytoskeleton. In cell
adhesion most forces generated in the actin cytoskeleton are balanced
over the sites of adhesion; thus, our second major focus area is the
physics of adhesion sites.

\subsection{Actin filaments and their assemblies}

Most studies of cellular forces have focused on their origin in the
actin cytoskeleton. This motivates our emphasis on the dynamics and
larger-scale structural organization of this important cytoskeletal
component.  Actin (in both monomeric and polymeric forms) comprises
between 5\%-10\% of the protein in eukaryotic cells and is of great
importance in cell structure and motility
\cite{fletcher_cell_2010,stricker_mechanics_2010}.  We begin with a
discussion of the growth of actin polymers.  In contrast to
self-assembling, equilibrium polymerization, these are catalyzed by
the binding of ATP to monomeric (globular) actin (\textit{G-actin}).
While many synthetic polymers are non-polar, actin polymers are chiral
with each macromolecule comprising two helical, interlaced strands of
monomeric subunits.  The two-filament assembly is thus polar so that
the two ends are therefore not equivalent; hence polymerization rates
at one end are not necessarily equal to those at the other.  Actin
polymerization is therefore a polar, energy-consuming, non-equilibrium
process \cite{phillips_physical_2008}.

A dynamical model for the growth of an actin filament takes into
account that the polymer is polar and the dynamics of association of
monomers at the two ends differ. That is, $k_{on}^{+,-}$ and
$k_{off}^{+,-}$ are respectively the rates for monomers to associate
with + (generally growing) or - (generally shrinking) ends and to
dissociate from those ends. For equilibrium polymerization the
association or dissociation energy itself must be the same at either
end since although the monomers are asymmetric at the two ends, the
molecular bonds that are formed are the same.  Hence, by detailed
balance, $k^+_{off}/k^+_{on}= k^-_{off}/k^-_{on}$. Thus, for such
equilibrium polymers one can show that \cite{ phillips_physical_2008}
there is no state in which one end is growing and the other is
shrinking; the polymer either grows or shrinks from both ends --
albeit with different on and off rates for the two ends of polar
chains. However, for the polymerization of actin in cells, detailed
balance does not apply and a richer set of behaviors is found as we
now discuss.

In living systems, polymerization is often a dynamic process that
involves chemical changes that may differ at the two ends of a
polar chain such as actin so that the polymerization and
depolymerization rates differ. The chemical changes are
catalyzed by an input of energy from the conversion of ATP
(adenosine triphosphate with 3 phosphate bonds) to ADP
(adenosine diphosphate with 2 phosphate bonds)
\cite{phillips_physical_2008, alberts_molecular_2007}. This
conversion is known as hydrolysis since  one phosphate group
dissociates from ATP to remain solubilized in water; the
breaking of one of the phosphate bonds   releases about 10-20~
$k_BT$ of energy since the hydration   bonds between ADP and
water and the released phosphate group and water   are
energetically more favorable than the bonds between the
phosphate   bonds in ATP.   The energy released by hydrolysis of
ATP  can be used to modify the conformations of molecules, such
as actin that is bound to ATP in its  lowest energy state.
The resulting conformational changes can result in increased or
decreased bonding of the molecules to other molecules; in the
case of actin, hydrolysis destabilizes polymerization at its
plus end. 

The non-equilibrium nature of actin polymerization in cells is related to the conformational
changes in the monomers that are catalyzed by ATP; G-actin monomers
bound to ATP join the plus end of the actin polymer
\cite{phillips_physical_2008, alberts_molecular_2007}.
Within a time of about 2 seconds, however, ATP is hydrolyzed to form
ADP which reduces the binding strength of the monomers in the chain,
thus destabilizing the polymer.
There is therefore a non-equilibrium competition between growth and
shrinkage of the polymer.
Note that in solution, the G-actin monomers that have dissociated from the chain
can dissociate from ADP and reassociate with ATP to rejoin the polymer; this
turnover makes the process highly dynamic.
Since the actin polymer is polar due to its double helical structure,
the growth and shrinkage at the + and - ends is different, and in principle,
one would need 4 rate constants to describe the on and off rates of
the ATP and ADP bound monomers
at each of the ends.
Filaments elongate about 10 times faster at their + ends compared with their
- ends and this leads to an apparent motion of the + end known as treadmilling
\cite{ phillips_physical_2008}.  Typical values
\cite{b:boal} are $k_{on}^+/k_{on}^- \approx 10$ for ATP-bound actin
and about 70 for the predominant situation of ATP-bound actin at the +
end and ADP-bound actin at the - end.  The ratio $k_{off}^+/k_{off}^-
\approx 5$ for ATP-bound actin at the + end and ADP-bound actin at the
- end, with typical values of $k_{off} \sim 0.3-7.0 \, {\rm sec}^{-1}$
depending on which end is being considered and whether the actin is
ATP or ADP bound.  One can show \cite{ phillips_physical_2008} that
there is a monomer concentration range for which the + ends are
growing while the - ends are shrinking.  Note that the treadmilling
velocity can be finite while the total filament length remains the
same.  Whether the filament can move or not depends on its
environment; for example, treadmilling actin filaments in the vicinity
of the cell membrane have their motion impeded by the restoring forces
(due to surface tension and curvature energy) of the membrane. This
then leads to flow of the actin in the direction opposite to
treadmilling, that is away from the cell membrane (\textit{retrograde
  flow}), as can be measured with speckle fluorescence microscopy
\cite{ponti_two_2004}.

The larger-scale organization of actin can take several forms.
{\it{In vitro}} studies have shown \cite{tempel_1996} that in some
cases alpha-actinin crosslinkers can result in relatively thick actin
bundles; in other cases, a crosslinked, isotropic gel is formed.  The
detailed phase diagram depends on both the actin and crosslinker
concentration \cite{zilman_2005}.  {\it{In~vivo}}, many proteins can
become involved in actin bundling which is utilized by the cell in
maintaining relatively stable \cite{microvili_gov_2006}, finger-like
protrusions called microvilli.  These proteins also participate in
more dynamical protrusions called filopodia \cite{mogilner_2005a} that
exert polymerization forces on the cell membrane and play a role in
cellular motion and shape changes.  Actin bundling is also an
important characteristic of stress fibers
\cite{hotulainen_stress_2006} that typically range over some fraction
of the cell size and provide structural stability to the cell while
transmitting contractile forces to its surroundings.

Due to the dynamics of the crosslinks and the treadmilling of actin,
the cytoskeleton can be remodeled and is therefore not permanently
crosslinked.  However, experiments in which cells are subject to time
varying strains that cause cytoskeleton reorganization, show that the
overall time scale for reassembly and reorientation of stress fibers
can be several hours \cite{brown_tensional_1998,
  wang_specificity_2001, jungbauer_two_2008}.  Entropic fluctuations
occur on a time scale shorter than 0.01~s \cite{Fredberg06}, while on
longer time scales the elastic response to time varying strain has
been characterized as glassy. We first consider the elastic modulus of
actin networks {\it{in vitro}} \cite{b:boal} on time scales shorter
than those at which the shear modulus vanishes due to the crosslink
disconnections \cite{bauschtrans}; on these time scales, the system in
some average sense can be regarded as being permanently crosslinked.

The
simplest model for the elastic constant of a permanently crosslinked
polymeric network predicts a
shear modulus $\mu \sim \rho k_BT$, where $\rho$ is the number
density of crosslinks \cite{rubinstein_2003}. The typical spacing between crosslinks
with a 1:100 ratio of linker to actin monomers, is of the order
of $0.1 \mu{\rm m}$. This yields a shear modulus of about $1
{\rm J/m^3}$ at room temperature which is equivalent to 1~Pa.
The measured value \cite{janmeyactin_1990} in the presence of
the crosslinker ABP (at a ratio of ABP:actin of the order of
1:100)  is about one order of magnitude larger and is sensitive
to the length of the actin segments; the observations also
depend on the history of the sample, since shear can disrupt
actin filaments and give misleadingly low values for the modulus.
Higher values of the modulus than expected from simple
considerations of crosslinked polymer networks can be due to the
stiffening effects of the crosslinks themselves, the
semi-flexible (as opposed to Gaussian) nature of biopolymers
such as actin, and to non-linear shear-stiffening. On the other
hand, the analogy with permanently crosslinked gels must be
reconsidered in light of the finite lifetime of the crosslinks.
Alpha-actinin has a dissociation rate of about $1 {\rm s}^{-1}$
\cite{gardelrev_2008, xu_1998} which may explain why {\it{in vitro}}
experiments using this crosslinker \cite{bauschtrans} in actin gels yield a low
frequency elastic modulus of about $1 \rm{Pa}$ only at the very
highest crosslinker concentrations (alpha-actinin:actin ratios of
1:15). The dissociation rate may be different in different
geometries; in isotropically crosslinked gels, the crosslinkers
can more effectively dissociate compared with their relatively
tighter packing in actin bundles where neighboring filaments are
nearby. The dissociation rate is also strongly temperature
dependent and of course very different for different
crosslinkers \cite{xu_1998}.

In addition to the finite residence time of the crosslinkers at the
network junctions, another important difference between the elastic
modulus of crosslinked biopolymers such as actin and synthetic polymer
gels is the observation that the actin cytoskeleton shows a non-linear
elastic response in a non-perturbative manner
\cite{gardel_2004,storm_2005}. For small stresses, the elastic stress
in actin gels is proportional to the strain, while for larger
stresses, of the order of 0.2 Pa, the effective modulus varies as the
3/2 power of the applied stress due to the entropically dominated
mechanical response of semi-flexible polymers of finite extensibility
as described above \cite{storm_2005}. This entropic nonlinearity is
particularly interesting because the strains may still be relatively
small \cite{gardel_2004, storm_2005} even though the medium responds
very non-linearly; this is quite different from analytical
non-linearities ({\it{e.g,}} due to additional quadratic terms in the
stress-strain relationship or due to geometrical non-linearity) that
arise when the strains become large in non-polymeric systems.
Interestingly, cells can regulate the regime in which they function by
changing their internal stress state through variation of the activity
of molecular motors. Since the cell elastic modulus is of the order of
1-10kPa (reflecting the types of stresses that cells can maintain),
the cytoskeletal elastic response can easily operate in the non-linear
regime.  Finally, we note that many other important biopolymers
\cite{storm_2005,vogel_2009} also show similar non-linear response,
including collagen which is an important part of the ECM.

\subsection{Actomyosin contractility}

In the preceding section, we have summarized the properties of
crosslinked actin gels based on information obtained from {\it{in
    vitro}} experiments.  However, one very important aspect regarding
actin networks and bundles in cells is the fact that these networks
are under tension due to the contractile activity of myosin motors
\cite{b:howa01} (pp.~265-273). Contractile actin networks
\cite{c:mizu07,koenderink_active_2009,silva_active_2011,kohler_structure_2011,murrell_f-actin_2012}
and bundles \cite{thoresen_reconstitution_2011,thoresen_thick_2013}
have recently been reconstituted in biomimetic assays.  In cells, the
myosin motors generate internal forces in the actin network which are
transmitted to its surroundings due to the ``glue'' the cell produces
in the form of proteins that aggregate into focal contacts or focal
complexes \cite{geiger_environmental_2009}. The production of force is
a non-equilibrium process that requires energy input via ATP
hydrolysis that causes conformational changes in the myosin molecular
motors \cite{b:howa01} (pp. 229-238). The internal forces generated by
molecular motors that act upon the crosslinked actin assemblies in
cells distinguishes them from ``dead'', non-active gels and allows
cells to pull on their environment and on each other. Motor activity
also means that the cell can exert forces on itself and this, along
with polymerization of actin, plays an important role in cell
motility. Motors can also influence the conformations of the actin
filaments in a manner that has to do with the motor and motor-actin
dynamics. The stochastic nature of the motor-actin coupling in which
the motor is associated with the actin for a finite time
\cite{b:howa01,b:boal} after which it can detach and diffuse, affects
the fluctuations of the filaments. These motor-driven fluctuations are
distinct from the thermal fluctuations of the actin that are driven by
Brownian motion of its aqueous environment
\cite{c:mizu07,mackintosh_active_2010}. This leads to a breakdown of
the fluctuation-dissipation theorem that relates the thermal
fluctuations of an equilibrium system to its response to a
deterministic force as discussed in the section on the physics
background.

Because stress fibers are an important element of the force-generating
apparatus of cells adhering to flat substrates, a large variety of
models has been developed to describe their physical properties.
Dynamical models show that actin filaments can be sorted by myosin II
motors into a tensile state
\cite{kruse_actively_2000,kruse_self-organization_2003,ziebert_pattern_2004,yoshinaga_polarity_2010,
  stachowiak_self-organization_2012}.  Models for mature fibers are
often motivated by perturbation experiments on stress fibers, such as
studies of contraction dynamics after activation
\cite{peterson_simultaneous_2004} or relaxation dynamics after laser
cutting \cite{kumar_viscoelastic_2006,uss:colo09}. They usually assume
a sarcomeric organization of the stress fiber
\cite{friedrich_plos_2012} and couple elastic, viscous and contractile
elements in a unit cell
\cite{uss:bess07,stachowiak_kinetics_2008,luo_multi-modular_2008,stachowiak_recoil_2009,russell_sarcomere_2009,uss:bess11}.
Recently it has been demonstrated that such sarcomeric models
predict some of the central physical properties of reconstituted contractile actin
bundles (e.g. retraction velocity is proportional to length)
 \cite{thoresen_thick_2013}. On a very coarse-grained scale,
such one-dimensional models can be regarded as more detailed versions
of the force dipole model introduced in \fig{ForceBalance}. In
particular, they usually obey force balance on the substrate by
construction. These types of one-dimensional models can also be used
to predict the cellular response to substrate stiffness
\cite{mitrossilis_single-cell_2009,uss:bess10a,marcq_rigidity_2011,crow_contractile_2012}.

An important reference case for the physics of stress fibers is
sarcomeric muscle, in which actin filaments, passive crosslinkers and
myosin II motors are arranged in a very ordered fashion. The action of
myosin II molecular motors can be modeled either with a generic
two-state theory
\cite{julicher_cooperative_1995,placais_spontaneous_2009} or with more
detailed cross-bridge models that go back to the seminal work of
Huxley
\cite{huxley_muscle_1957,duke_molecular_1999,vilfan_instabilities_2003,uss:erdm12a},
where the cross-bridges refer to the acto-myosin coupling.  This model
accounts for the fact that a myosin II motor loaded with ATP goes
through a cycle where it first binds weakly to the actin
filament. Release of the inorganic phosphate (after hydrolysis of ATP
to ADP) causes the myosin motor to make a powerstroke that creates
force and motion. Finally, after the release of the ADP and binding of
a new ATP-molecule, the myosin II motor unbinds from the filament and
is ready for the next motor cycle. The effective force-velocity
relation has been measured both in the context of muscle
\cite{pate_temperature_1994} and in single molecule experiments
\cite{veigel_load-dependent_2003}.  At vanishing force, the motors
move with an ATP-dependent free velocity of about 1 $\mu$m/s. As the
external counterforce increases, the velocity of the motor along the
filament drops in a hyperbolic manner, until it vanishes at a stall
force of a few pN. This close coupling between force and sliding
velocity was first noted by Hill in 1938, who described it using a
phenomenological equation (called the \textit{Hill equation}), which
can be explained in detail by the cross-bridge models
\cite{b:mahon84,b:howa01}. Alternatively, as we will see later,
Hill-type relations can be used directly as an assumption in
coarse-grained models. In particular, such a model has been used to
argue that rigidity sensing is based on the same principles like
muscle contraction \cite{mitrossilis_single-cell_2009}.

Contractility is also observed in biological systems with no apparent
sarcomeric order.  For example, the contractility of the actin ring
during cell division may rely on depolymerization forces \cite{
  pinto_ring_2012}. It has been shown theoretically that in
one-dimensional actin bundles containing myosin, contractility can
occur even in the absence of spatial organization of the bundle, due
to bundle shortening \cite{kruse_julicher_2000,kruse_julicher_2003}.
The existence of net contractility is related to the assumption that a
myosin motor which binds to or arrives at the plus end (but not at the
minus end) of a filament remains attached for some time.  This model
does not contain additional crosslinkers (similar to Z-bodies found in
sarcomeres) that may tend to associate with only one end of the polar
actin molecules. Another recent suggestion \cite{lenz_2012} for how
contractility can arise in bundles without sarcomeric order is based
on an the asymmetric response of the filaments to longitudinally
applied stresses, {\it{e.g.}}, a tendency to yield under compression
while resisting extension.  Such buckling has been observed in {\it{in
    vitro}} experiments containing actin, smooth muscle myosin and
ATP. We note that additional crosslinkers, such as those found in
sarcomeres and possibly stress fibers, may break the symmetry were not
included. Whether stress fibers, contractile rings, and smooth muscle
fibers are ordered and function in a manner similar to sarcomeres
should be investigated by future experiments.  The sarcomeric order
discussed below in terms of smectic ordering of force dipoles is meant
to apply to nascent muscle cells where striations have indeed been
observed \cite{engler_myotubes_2004, friedrich_striated_2011}.  In a
somewhat similar manner, striations have have also been reported in
studies of non-muscle stress fibers \cite{peterson_simultaneous_2004}.
Here the microscopic, anti-parallel arrangement of adjacent actin
polymers was not directly observed, but the striations measured do
suggest a sarcomeric analogy.

\subsection{Focal adhesions}

Understanding the mechanical response of stationary cells involves
analysis of the internal elastic response of contractile cells as well
as their mechanical coupling to their surroundings.  While the
response of cells to external forces or other mechanical perturbations
can necessitate the disassembly and rebuilding of the actin
cytoskeleton, the stable coupling of the cell to the surrounding
elastic matrix is due to sites of adhesion called \emph{focal
  adhesions} that connect the actin cytoskeleton to transmembrane
adhesion receptors from the integrin family.  These are then
connected, on the extracellular side, to the substrate or
extracellular matrix.

\FIG{Fig06}{fig:LLBoundary}
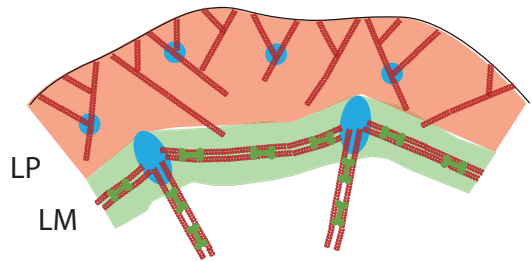
{(Color online) Spatial organization of focal adhesion growth and the actin
  cytoskeleton. The actin lamellipodium (LP) is assembled at the leading
  edge and flows from there towards the cell center. Small adhesions
  are formed along the way and mature into focal adhesion as they move with the actin
  flow.  At the boundary with the myosin-dominated lamella (LM), only a
  few mature focal adhesions persist; those are stabilized by large
  contractile forces which are mainly due to the activity of myosin II minifilaments in stress fibers.}

In contrast to the adhesion of passive vesicles or capsules, the
spatial distribution of the adhesion structure of cells is very
heterogeneous.  It is mainly localized at the cell periphery, because
it is strongly coupled to the growth processes of the
lamellipodium. Fig.~\ref{fig:LLBoundary} depicts the spatial
coordination between the growth of adhesions and the actin
cytoskeleton. Nascent adhesions are initiated close to the leading
edge and then move towards the cell center. This movement is mainly
driven by the flow of actin away from the leading edge
(\textit{retrograde flow}) due to the counterforces exerted on the
polymerizing actin by the membrane. As they move towards the cell
center, the small adhesions either mature into micrometer-sized focal
adhesions, or decay again. This switch typically occurs near the
lamellipodium-lamella boundary, where a more condensed and myosin
II-rich actin network replaces the dendritic network of the
lamellipodium \cite{uss:schw12b, shemesh_physical_2012}. Whether the
adhesion grows and matures or whether it decays is strongly coupled to
the mechanics of the system.  The adhesions are stable only if
sufficient force is exerted upon them and this is not possible on very
soft substrates. This force is mainly applied by contractile stress
fibers and networks connecting them to the focal adhesions in the
lamella, although the force resulting from retrograde flow also might
play an important role. Because focal adhesions are connected to the
matrix, these forces are transmitted to the substrate and can be
measured there with traction force microscopy on flat elastic
substrates
\cite{dembo_stresses_1999,butler_traction_2002,uss:schw02b,uss:saba08a,
  plotnikov_force_2012,legant_multidimensional_2013} or with fields of
elastic pillars \cite{tan_cells_2003,saez_is_2005,
  trichet_evidence_2012}.  These studies have shown that force and
protein assembly are linearly coupled at focal adhesions, resulting in
a constant stress for adhesions of about 5 nN/nm$^2$ = 5 kPa
\cite{uss:bala01,tan_cells_2003}.  For elastic substrates, recently it
has been suggested that this relation only holds for growing adhesions
\cite{stricker_spatiotemporal_2011}. For pillar assays, recently it
has been reported that there exists a constant stress, but that it
depends on extracellular stiffness due to global feedback
\cite{trichet_evidence_2012}. At any rate, the typical stress at
single focal adhesions is close to the value of the physiological
stiffness of matrix and cells; this suggests that these forces are
used for mechanosensing in the physiological environment of the cell.
 
\FIG{Fig07}{fig:FocalAdhesion}
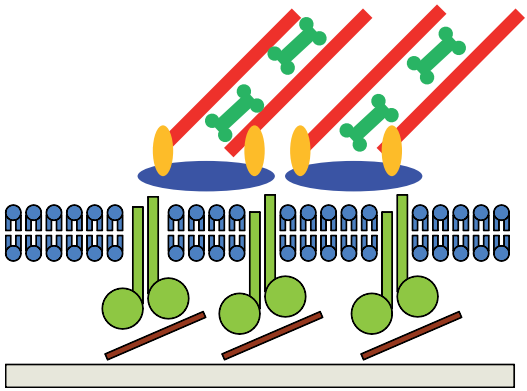
{(Color online) Schematic view of a focal adhesion. The transmembrane
  adhesion receptors from the integrin family (light green, a
  heterodimer with two subunits) bind to the extracellular matrix
  (brown, for example collagen) on the outside and are crosslinked by cytoplasmic proteins
  such as talin (blue) in the inside. Talin binds to actin (red) and this
  binding is further strengthened by proteins such as vinculin
  (orange). The contractility of the actin cytoskeleton is determined by the activity of myosin II
  minifilaments (dark green).}

The detailed molecular organization of focal adhesions is a very
active area of research that is very challenging due to the large
number (more than 150) of different components that are involved
\cite{zaidel-bar_functional_2007, kuo_analysis_2011}. Recent progress
includes the use of electron tomography \cite{patla_dissecting_2010}
and superresolution microscopy \cite{shtengel_interferometric_2009} to
discern focal adhesion structure at scales smaller than the optical
resolution, and the use of high throughput RNA-interference screens to
dissect the regulatory hierarchy of focal adhesions
\cite{prager-khoutorsky_fibroblast_2011}. Kinetic models have been
used to describe the temporal and spatial coordination of the
different components \cite{
  civelekoglu-scholey_model_2005,macdonald_kinetic_2008,uss:hoff13}. Fig.~\ref{fig:FocalAdhesion}
shows a schematic representation of a focal adhesion. In general,
focal adhesions have a layered structure determined by the
two-dimensional nature of the plasma membrane. The transmembrane
adhesion receptors of the integrin family consists of two subunits,
with relatively large headpieces that bind to the matrix and
relatively small cytoplasmic tails. In the absence of special signals,
the integrins have a low affinity for matrix binding. However, due to
inside-out signaling (related to the cytoskeletal forces), the
integrins can become activated and are then primed for matrix
binding. This in turn leads to stabilization of the intracellular part
of the adhesion complex that through the binding of a variety of
cytoplasmic proteins, forms a two-dimensional plaque that reinforces
the attachment of the cytoskeleton and the integrin layer.  One of the
main molecules responsible for cross-linking neighboring integrins is
talin \cite{alberts_molecular_2007} (p.~842) , which extends over 60
nm. Since talin also binds actin, it connects the integrins in the
focal adhesion to the actin cytoskeleton.  As the adhesion matures,
this crosslinking is strengthened by additional proteins such as
vinculin and paxillin, whose recruitment seems to be increased by
force.

Focal adhesions act not only as mechanical linkers that anchor the
cell to its substrate, but also as very prominent signaling centers
that activate biochemical signaling molecules that diffuse into the
cytoplasm and towards the nucleus
\cite{zaidel-bar_functional_2007,vogel_cell_2009}. In our context, the
most important signaling molecules are the small GTPases from the
Rho-family (Rho, Rac and Cdc42), that regulate the assembly and
activity of the actomyosin system. Each of these molecules acts like a
molecular switch which is activated by exchanging GDP by GTP
(analogous to ADP to ATP conversion); the active form then diffuses in
the cytoplasma and activates downstream targets. For example, mature
adhesions are known for Rho-signaling, which upregulates both actin
polymerization (through the formin mDia1) and contractility of
non-muscle myosin II motors (through the Rho-associated kinase,
ROCK). Moreover, many signaling molecules responsible for cell
migration, differentiation and fate are localized to focal adhesions,
most prominently the focal adhesion kinase, FAK, which is known to be
important in many types of cancer \cite{mitra_focal_2005}.

Adhesion proteins localized to focal adhesions coexist with the same
proteins in relatively dilute solution in the cytoplasm or the
membrane. The domains as a whole have relatively long lifetimes (tens
of minutes) which suggests that the two coexisting phases might be at
equilibrium.  However, fluorescence studies
\cite{wolfenson_actomyosin-generated_2011} show that the adhesions
continuously exchange proteins with the cytoplasm even though the
large-scale composition and structure seem to remain
unchanged. Thermodynamic equilibrium would dictate much larger domain
sizes (or no domains at all) and why the adhesions are stable on the
micrometer scale is not obvious \cite{Lenne09}. This puzzle might be
resolved by noting that in addition to diffusion, energy consuming,
active processes may transport free proteins from the dilute phase in
the cytoplasm or membrane to the adhesion sites.  These active
processes involve molecular motor proteins and are highly regulated by
the cell \cite{Kawakami01}. This suggests that non-equilibrium effects
may be important in stabilizing these finite-size domains, similar to
the situation with treadmilling actin filaments.

Naively, one might try to understand cellular adhesions by analogy
with physical adhesion ({\it{e.g.}}, of a synthetic vesicle coated with
ligands that are attracted to an appropriate surface, see the
preceding section).  However, while physical adhesion is a passive process, cell
adhesion involves molecular motors that generate internal stresses. Force
generation consumes ATP and results in the fact that in addition to
passive contacts that result in forces mainly directed in the normal direction, cells
also exert contractile forces that act mainly in the lateral direction, {\it{i.e.}},  parallel
to the substrate. These contractile forces have been
observed in experiments that measure surface deformation
\cite{dembo_stresses_1999,butler_traction_2002,uss:schw02b,tan_cells_2003,uss:saba08a,saez_is_2005}.
Although recent experiments have also provided evidence for vertical
forces \cite{hur_live_2009,delanoe-ayari_4d_2010}, most of these experiments
were done with rather weakly adhering cells, which share some
similarities with the passive reference cases discussed above
(droplets, vesicles, capsules). Thus on a planar substrate it is the actively generated and
often tangential forces that are used by the cell to regulate its response to the
physical environment.  This conclusion is supported by the fact that
use of the myosin blocker blebbistatin not only leads to the
disappearance of these forces, but also eliminates the cellular response to
stiffness. Thus the main challenge is to understand how actively
generated forces allow the cell to probe the physical
properties of its environment.

\section{Physics of cell-matrix adhesions}

\subsection{Physical motivation}

As explained in the previous section, cell adhesion does not occur
homogeneously at the cell-material interface, but instead is
characterized by the local assembly of specific adhesion molecules
into supra-molecular adhesion sites, the so-called \textit{focal
  adhesions}.  For cell adhesion to a flat substrate, these focal
adhesions are mainly situated at the cell periphery.  During the last
decade, it has been shown that the protein assembly that comprises the
focal adhesion is strongly coupled to mechanical force. Experimental
studies inducing changes in mechanical stress at adhesions have used
shear flow \cite{davies1994,Zaidelbar05}, optical tweezers
\cite{choquet_extracellular_1997}, micromanipulators
\cite{riveline_focal_2001,uss:paul08a,heil2010}, laser nano-surgery
\cite{Lele06,kumar_viscoelastic_2006,uss:colo09} or pharmacological
drugs
\cite{chrzanowska-wodnicka_rho-stimulated_1996,kuo_analysis_2011,
  wolfenson_actomyosin-generated_2011} to perturb and hence study the
nature of these contacts.  In all cases, it was observed that focal
adhesions respond to changes in mechanical load by growth, as
evidenced by changes in focal adhesion morphology and size. Quantitative
correlation indicated a linear relation between force and adhesion
size \cite{uss:bala01,tan_cells_2003}, although the history of the
adhesion and global determinants also play an important role
\cite{stricker_spatiotemporal_2011,trichet_evidence_2012}.

From the physical viewpoint, it is interesting to note that tensile
mechanical deformation, such as the shear induced by the tangential
displacement of a pipette, leads to growth of focal adhesions in
the direction of applied force
\cite{riveline_focal_2001}. De-activation of actomyosin contractility
reduces the adhesion size and eventually leads to its complete
disruption \cite{uss:bala01}. Other forces, such as hydrodynamic flow
\cite{Zaidelbar05} or stretching forces applied to the substrate
\cite{Kaunas05}, also cause growth of focal adhesions in the direction
of the force. The anisotropy of focal adhesion growth under force is
characterized both by overall growth of the adhesion (in which the
number of molecules involved increases) and by treadmilling (or
sliding) of the center of mass of the adhesion towards the direction
of the applied force.  These anomalies cannot be explained with
standard models of nucleation and growth of molecules adsorbed from
solution onto surfaces. The role of force in stabilizing and promoting
growth of FA has been discussed from several different points of view
with a focus on predictions of the growth of focal adhesions in the
direction of actomyosin or externally applied forces
\cite{bershadsky_adhesion-mediated_2006}.

However, before discussing the growth response of focal adhesions under load,
it is instructive to consider why this response has evolved
in the first place. From the physical point of view, it seems obvious
that in this way, the system avoids material failure. In general,
failure under mechanical load is a phenomenon relevant to many systems
of practical interest and on widely different scales, from muscle
proteins (on the nanometer scale) through bone on a centimeter scale
to bridges and buildings (on the meter scale) and earthquakes (on the
kilometer scale) \cite{buehler_colloquium:_2010}.   In contrast to the macroscopic systems considered in traditional
fracture mechanics, biomolecular adhesions do not usually
break at sharp, deterministic stability thresholds, but can unbind and
rebind from the surface to which they adhere in a stochastic
manner. This important observation is related to the weak interaction
scales relevant to soft matter compared with hard matter.
Thus, if one considers the stability of such assemblies
(defined by either by their average lifetime or fracture strength),
one must include stochastic effects, in contrast with traditional fracture
mechanics of macroscopic systems where these effects are not relevant.  Before
we address the issue of growth, we therefore first consider the
stability of adhesion clusters under load from the viewpoint of
stochastic processes.

\subsection{Stability of stationary adhesion clusters under force}

\FIG{Fig08.eps}{fig:cartoon_cluster} 
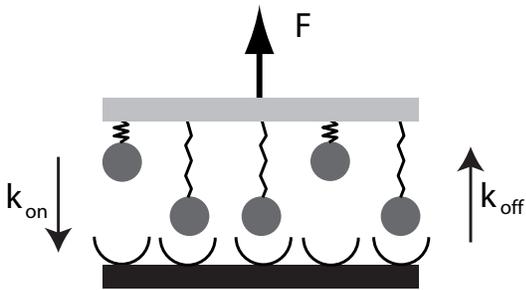
{Minimal model for an adhesion cluster under force: $N_t$
  adhesion receptors are arranged along the membrane. Due to this
  geometry, they share the load.  At any time,
  $N(t)$ bonds are closed and $N_t-N(t)$ bonds are open. The applied
  force $F$ is equally distributed over the closed bonds due to the
  parallel architecture, $F_{bond}=F/N(t)$.  A bond dissociates with a force-dependent
  unbinding rate $k_{off}(F_{bond})$. The molecule can rebind with a constant
  rate $k_{on}$. Adapted from \textcite{uss:erdm04a}.}

We begin our discussion with physical considerations of the stability
of stationary adhesion clusters of constant size. This crucial issue
was first addressed in the seminal work of Bell \cite{c:bell78}.  An
adhesion cluster is modeled as a collection of $N_t$ molecules near an
adhesive surface, of which, at a given time $t$, a number $N(t)$ are
bound and a number $N_t - N(t)$ are unbound, as shown in
\fig{cartoon_cluster}.  Each of the bonds can break with a rupture
rate $k_{off}$ and each of the unbound molecules can bind with a
rebinding rate $k_{on}$. Unbinding is assumed to increase with force
$F$ as $k_{off} = k_0 e^{F / F_0}$, where $F_0$ denotes a
molecular-scale force (typically of the order of pN). While Bell used
this expression as a phenomenological ansatz, it was later motivated by
Kramers theory \cite{hanggi_reaction-rate_1990} for thermally assisted escape from a metastable state
\cite{c:evan97,evans_forces_2007}. Here the basic idea is that force lowers the height
$E_b$ of the transition state barrier. Because the escape rate scales
as $\exp{(-E_b/k_B T)}$, adding a force term to the energy
changes the escape rate by a factor $\exp{(F x_b/k_B T)}$, where $x_b$
is the position of a sharp transition state barrier.  Indeed, this
viewpoint has been impressively verified by dynamic force spectroscopy
\cite{evans_forces_2007}. For the rebinding rate, Bell assumed a
value that is force independent. Note that this assumption does not reflect detailed balance
(in which the ratio $k_{on}/k_{off}$ is exactly equal to a Boltzmann
factor related to the energy difference between the bound and unbound
states). Thus, this model is one that is truly dynamic and can thus
represent {\it{e.g.,}} different routes for binding and unbinding.

For the following, it is helpful to introduce the dimensionless
time $\tau = k_0 t$, force $f=F/F_b$ and rebinding rate $\gamma = k_{on}/k_0$.
Assuming a constant force $f$ equally applied to all bound molecules, the following
rate equation predicts the number of closed bonds:
\begin{equation}
\frac{dN}{d\tau} = - N e^{f / N} + \gamma (N_t - N) 
\label{eq:BellODE}
\end{equation}
While the second term representing rebinding is linear in the number
of bonds, the first term representing forced-unbinding is highly
non-linear and therefore leads to interesting feedback effects. As one
bond opens, the remaining closed bonds must compensate to carry the
additional load. Thus the coupling through force leads to a highly
cooperative system. A bifurcation analysis of its steady state
behaviour shows that the system is unstable (no steady state solution
exists) when the force exceeds a critical value, $f_c$.  This
saddle-node bifurcation occurs when \cite{c:bell78}
\begin{equation} \label{eq:critical_force}
 f_c = N_t \ \rm{pln} ( \gamma / e )
\end{equation}
where the product logarithm $\rm{pln}(a)$ is defined as the solution
of $x e^x = a$. Therefore an adhesion characterized by a finite number
of bonded molecules is only stable up to a critical force $f_c$. For
small rebinding rate $\gamma$, the critical force scales linearly with
$\gamma$. Thus an adhesion cluster is completely unstable if the
rebinding rate is zero; for finite rebinding the cluster stability
({\it{i.e.}}, the number of bonds formed) grows in proportion to the degree
of rebinding. For large rebinding rates, the scaling becomes
logarithmic; that is, once the rebinding rate exceeds the force-free
unbinding rate, very large changes in $\gamma$ are required to change
the cluster stability in a significant manner.

The analysis of Bell immediately shows that due to the finite lifetime 
of single biomolecular bonds, adhesion clusters
can be stable under force only if rebinding takes place.
However, \eq{BellODE} is a mean field description and
does not include fluctuation effects, which are expected
to be highly relevant in the biological context, both for the small
precursors of focal adhesions and for possible subclusters that may exists within focal
adhesions. Moreover, a description based on multiple bonds
is required to treat more detailed
situations of biological interest, as we discuss below.
The natural extension of the mean field approach of Bell
is a one-step master equation in the number $i$ ($0 \le i \le N_t$)
of bonded molecules \cite{uss:erdm04a,uss:erdm04c}. Thus
the probability $p_i(t)$ that $i$ bonds are formed
at time $t$ evolves in time according to
\begin{equation} \label{eq:MasterEquation}
\frac{dp_i}{dt} = r(i+1) p_{i+1} + g(i-1) p_{i-1} - [ r(i) + g(i) ] p_i\ .
\end{equation}
The two positive terms represent the tendency for the number of bonds
in state $i$ to increase due to the dissociation of a formed bonds in
state $i+1$ and the formation of a new bond in state $i-1$,
respectively. The two loss terms represent bond dissociation in state
$i$ (that contributes to state $i-1$) as well as the formation of a
new bond (which changes the state from $i$ to $i+1$), respectively.
The rates corresponding to the Bell model \eq{BellODE} are
\begin{equation} \label{eq:Rates}
r(i) = i e^{f / i},\ g(i) = \gamma (N_t - i)\ .
\end{equation}
As in \eq{critical_force}, the first (rupture) term leads to strong cooperativity between the different bonds.
The Bell equation \eq{BellODE} is recovered from the dynamic,
stochastic model \eq{MasterEquation} if one calculates
the average number of formed bonds, $N = \langle i \rangle$,
in the limit of large system size (Kramers-Moyal expansion).

In contrast to the deterministic equation for the first moment,
\eq{BellODE}, which predicts infinitely long cluster lifetimes below
the stability threshold, \eq{critical_force}, the stochastic model
\eq{MasterEquation} predicts finite lifetimes for any value of the
force. The average lifetime of a cluster with $i=N_t$ bonds at time
$t=0$ can be identified with the mean time $T$ for this cluster to
stochastically evolve to the state $i=0$ (no bound molecules).
For a one-step master equation, this \textit{mean first passage time}
can be easily calculated \cite{b:kamp92}:
\begin{equation} \label{eq:MFPT}
T = \sum_{i = 1}^{N_t} \frac{1}{r(i)} +
\sum_{i = 1}^{N_t - 1} \sum_{j = i+1}^{N_t} 
\frac{\prod_{k = j-i}^{j-1} g(k)}{\prod_{k = j-i}^{j} r(k)}\ .
\end{equation}
The first term in \eq{MFPT} is the result for vanishing rebinding, $\gamma = 0$.
For small force, $f < 1$, it is analogous to the simple case of
proportional (or radioactive) decay (for $\gamma = f = 0$, one
basically deals with the stochastic version of $dN/d\tau = - N$). Using that approximation one finds that the
cluster lifetime scales as $T \approx \ln N_t$; that is, the cluster
size $N_t$ has only a relatively weak effect on the stability of the
cluster, because the bonds dissociate in parallel with no
cooperativity.  Finite force exponentially decreases the lifetime due
to the Bell-equation.  The second term in \eq{MFPT} increases the
lifetime as a polynomial of order $N_t-1$ in $\gamma$ and
approximately exponentially with increasing cluster size $N_t$.  For
$N_t = 2$, \eq{MFPT} reads
\begin{equation} \label{eq:MFPT2}
T = \frac{1}{2} \left( e^{-f/2} + 2 e^{-f} + \gamma e^{-3f/2} \right)\ .
\end{equation}
This simple but instructive formula shows how force $f$ exponentially
suppresses the cluster lifetime, while rebinding $\gamma$ increases it in a polynomial fashion.

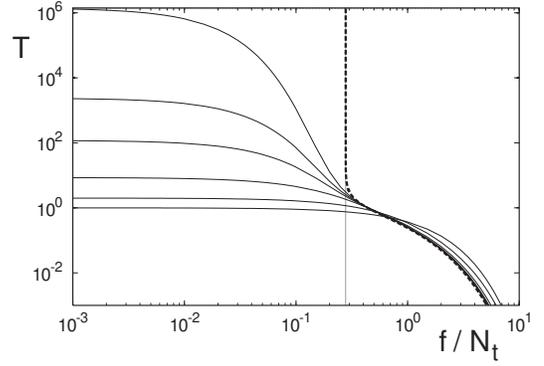
\FIG{Fig09}{fig:MFPT}{Average cluster lifetime $T$ for $N_t = 1, 2, 5, 10, 15$ and $25$
(from bottom to top) as a function of $F / F_b N_t$ for $k_{on} = k_0$
as it follows from \eq{MFPT}.
The vertical line is the critical force predicted by \eq{critical_force}
and the dashed line is the lifetime predicted by \eq{BellODE}.
Adapted from \textcite{uss:erdm04c}.}

In order to study the cluster lifetime as a function of force, in
\fig{MFPT} we plot $T$ as calculated from \eq{MFPT} as a function of
$f/N_t$ for $\gamma = 1$ and different values of the cluster size
$N_t$.  For small force $f < N_t$, the mean lifetime increases
exponentially with the cluster size $N_t$. For large force $f > N_t$,
rebinding becomes irrelevant and all the curves for different values of $N_t$ are very similar. One
very interesting aspect here is the transition region, which is
characterized by strong amplification (a small change in force $f$ has
a strong effect on the mean lifetime $T$).  This transition region
corresponds to the stability threshold from \eq{critical_force}
(vertical line). Above this threshold, a deterministic lifetime can be
defined by numerically solving \eq{BellODE} with $N(0)=N_t$ and solving
for the time at which $N=1$ (this replaces the criterion $i = 0$ from
the stochastic case, which cannot be used in the deterministic case
due to exponential approach of $N = 0$). The result (dashed line)
corresponds well to the mean cluster lifetime from the stochastic
model, but diverges at the threshold.

\eq{MFPT} can be used to make interesting estimates for experimental
situations of interest. For example, in the case of zero rebinding
rate ($\gamma = 0$) and zero applied force ($f=0$), for a single bond
lifetime of one second ($k_0 = 1/s$), a cluster lifetime $T$ of one
minute could only be achieved with for the absurdly large number of
$10^{26}$ bonds; this is because in this case, the cluster lifetime
scales only logarithmically with cluster size.  However, for a finite
rebinding rate of $\gamma = 1$ ($k_{on} = k_0$, still at zero force),
only $N_t = 10$ bonds are necessary, because the lifetime scales
strongly with rebinding rate: $T \sim \gamma^{N_t-1}$.  Increasing the
dimensionless force to $f = 10$ (corresponding to 40 pN for $F_b = 4$
pN) would decrease the lifetime to $T = 0.05$ s, because $T$ is an
exponentially decreasing function of $f$.  To reach a cluster lifetime
of one minute in this case, the cluster size must be increased to 50
or the rebinding rate must be increased by a factor of 10.

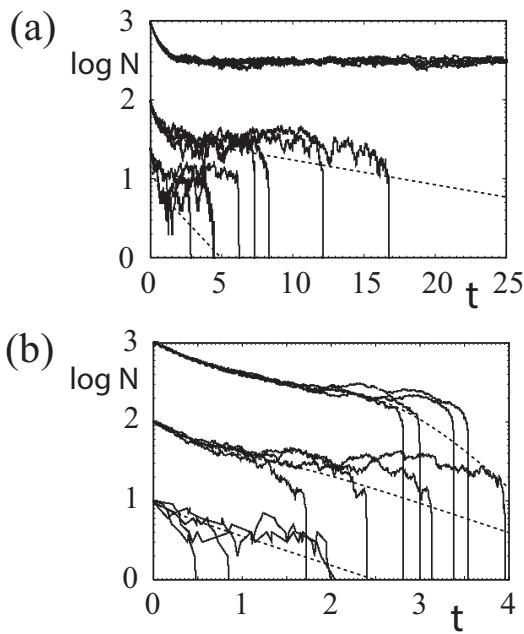
\FIG{Fig10}{fig:Rupture}{Selected trajectories simulated with the
stochastic master equation model. Rebinding rate $\gamma = 1$.
(a) Force below threshold, $f/N_t = 0.25$. Small clusters
($N_t = 10$ and $100$) are unstable due to fluctuations.
(b) Force above threshold, $f/N_t = 0.3$. Now all cluster sizes
are unstable. Lines are first moments. Adapted
from \textcite{uss:erdm04a}.}

One of the strongest advantages of the stochastic model is that it can
be used to simulate single trajectories, which share many
similarities with experimental realizations of individual experiments. 
In \fig{Rupture}
we show selected simulated trajectories for forces below (a)
and above (b) the critical force. We note that failure
is rather abrupt due to cooperative effects: once sufficiently
many bonds have broken, {\it{e.g.}}, due to a fluctuation that leads to a smaller 
number of  bonds, the force on the remaining ones
is so high that rebinding becomes very unlikely and the
cluster fails in a cascade of dissociated bonds that leads
to rupture of the adhesion cluster (\emph{cascading
failure}). Interestingly, this effect is not apparent if one calculates only
the averages, which are shown as lines. We further
note that even below the stability threshold in (a), small clusters
are likely to fail due to the finite probability for a
devastating fluctuation that takes the system to a state with a small numbers of closed bonds.
Above the threshold in (b), clusters of any size are unstable.

The conceptual framework introduced by Bell shows how an adhesion
site can at the same time be highly dynamic and yet be stable up to some maximal force: while
some bonds can dissociate and then rebind, the remaining adhesion 
bonds allow the transfer of force from the cell to the substrate. The
simple model also shows that this mechanism leads to
strong cooperativity, because each bond that forms  or
breaks leads to a fast redistribution of the force, thereby
affecting all the other bonds. The same cooperative mechanism operates in many
other biologically relevant situations, {\it{e.g.}}, during force generation in muscle
\cite{huxley_muscle_1957,duke_molecular_1999}
or during cargo transport by multiple molecular motors
\cite{klumpp_cooperative_2005,guerin_coordination_2010}.
The stochastic extension of the Bell model demonstrates
that adhesions are not only unstable for large applied forces,
but are also unstable at small sizes for which fluctuations to
smaller numbers of closed bonds can be detrimental. 

\subsection{Adhesion between moving surfaces}

Adhesion not only occurs via molecular binding of two stationary
surfaces, but also frequently bridges two surfaces that move relative
to each other. This is especially relevant for matrix adhesion
underneath the cellular lamellipodium, where actin retrograde flow can
transport individual binding molecules from the cell edge towards the
cell body and where nascent adhesions form and mature in the region
between the cell and the substrate. It has been found experimentally
that in this case, a biphasic relation exists between the traction force
on the substrate and the flow velocity in the cell: while the traction
force and flow velocity are linear proportional in the case of mature
adhesions, they are inversely related in the case of fast flow over
nascent adhesions \cite{gardel_traction_2008}.  Experimentally, the
threshold value of the velocity at which this change occurs has been
found to be 10 nm/s and to be insensitive to
various perturbations of the cellular system.  Surprisingly, the
simple conceptual framework introduced above for stationary adhesion
sites can be extended to explain these experimental findings
\cite{srinivasan_binding_2009,sabass_modeling_2010,li_model_2010}.

\FIG{Fig11}{fig:FlowCartoon}
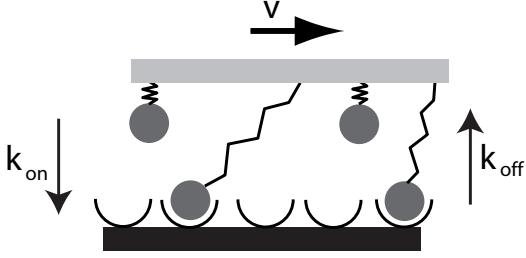
{Minimal model for an adhesion cluster that bridges two
surfaces that move relative to each other with velocity $v$.
Here we discuss this case in the limit of continuous
ligand and receptor coverage. The situation described here is 
very similar to sliding friction as usually studied for macroscopic objects.}

\fig{FlowCartoon} shows the minimal model for the situation of
interest. The upper surface moves with a velocity $v$ relative to the
lower surface. Each bond is modeled as a spring with spring constant
$\kappa$ that immediately gets elongated with velocity $v$ once it is
formed. In contrast to the minimal model for a stationary adhesion
cluster, see \fig{cartoon_cluster}, there are two essential
differences. First, the governing model parameter is the relative
velocity $v$ rather than the external force $F$. This also implies
that different bonds are not coupled by force and a theoretical
description can be constructed using a single bond model. Second, each
bond is characterized by its dynamic length, namely its extension
compared with the case of zero velocity, which we denote by $x$. This is in
addition to the property of the bonds to dynamically form or
dissociate.  We therefore now introduce a probability that depends on
both time and bond elongation, $p_b(x,t)$, that describes the
likelihood that at time $t$ a given bond is closed and has  elongation
$x$.  The complementary probability that the molecule is dissociated
is not related to the elongation $x$ and we denote it by $p_u(t)$. From
normalization we get
\begin{equation}
p_u(t) = 1 - \int_{- \infty}^{+ \infty} dx \, p_b(x,t) = 1 - P_b(t)
\label{eq:FlowNormalization}
\end{equation}
where we have introduced the abbreviation $P_b(t)$ for the overall probability
of a bond to be closed with some elongation $x$.  When we assume harmonic
springs with spring constants $\kappa$, the average traction on the
substrate is determined by the first moment of $p_b(x,t)$:
\begin{equation}
F_T = N_t \kappa \int_{- \infty}^{+ \infty} dx \, x \, p_b(x)\ .
\label{eq:FlowTraction}
\end{equation}

We next consider the evolution equation for $p_b(x,t)$. In contrast to
the equation for the fraction of bound bonds for the stationary
adhesion cluster, \eq{BellODE}, we now have a convective derivative
that accounts for the change in extension due to the fact that the
molecules (one end of which are fixed to the moving surface) are
moving with velocity $v$:
\begin{equation}
\frac{\partial p_b}{\partial t} + v \frac{\partial p_b}{\partial x} = - p_b k_{off} + (1-P_b) k_{on} \delta(x)\ .
\label{eq:FlowODE}
\end{equation}
Here $p_u$ has been replaced by the right hand side of \eq{FlowNormalization}.
The delta-function represents the assumption that a new bond forms
with vanishing elongation $x$.
For the single bond unbinding rate $k_{off}$ and rebinding rate $k_{on}$,
we make assumptions similar to those made for the bonds in
a stationary cluster, namely a Bell model $k_{off} = k_0 e^{r x}$ for the unbinding rate and
a constant rebinding rate, $k_{on} = const$.
Here the reactive compliance $r = \kappa / F_b$ is a typical inverse
length scale of the bond.  With these simple form of the rates, the
steady state with $\partial p_b / \partial t = 0$ can be calcuated
analytically \cite{srinivasan_binding_2009,sabass_modeling_2010}.

\FIG{Fig12}{fig:FlowBiphasicRelation}
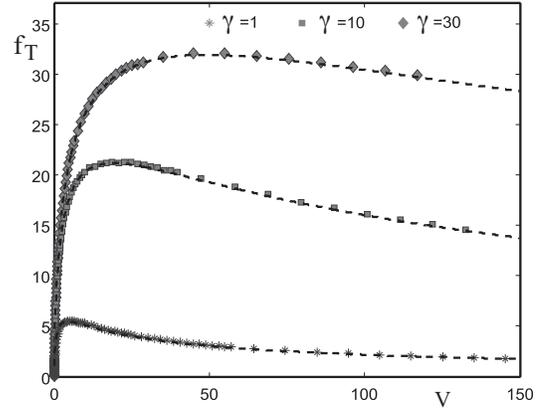
{Biphasic relation between dimensionless flow $V$ and dimensionless traction force $f_t$
predicted by the minimal model for cluster size $N_t = 25$ and for
different values for the rebinding rate. Symbols are the results of
stochastic simulations with $N_t$ single bonds.
Adapted from  \textcite{sabass_modeling_2010}.}

We first note that because bonds form with vanishing elongation and are then 
stretched by the motion of the upper surface (so that the elongation is positive), the probability of negative elongation vanishes,
$p_b(x < 0) = 0$ (the time $t$ does not appear because we consider steady state).
From \eq{FlowODE} we see that for $x=0$ we have
\begin{equation} \label{eq:FlowJump}
p_b(0) = p_0 = (1-P_b) \frac{k_{on}}{v} = (1-P_b) \frac{\gamma r}{V}
\end{equation}
where we have defined a dimensionless velocity $V = r v / k_0$ and
$\gamma = k_{on} / k_0$ is the dimensionless rebinding rate defined above.
For $x > 0$, \eq{FlowODE} is solved by
\begin{equation} \label{eq:FlowSolution}
p_b(x) = p_0 e^{\frac{1}{V} (1-e^{r x})}
\end{equation}
Therefore, the probability for a bond (with probability $p_0$ at
$x=0$) to be elongated by an amount $x$ decays faster than
exponentially as $x$ increases.  After calculating $P_b$ from
\eq{FlowSolution}, one finds the traction force from \eq{FlowTraction}
in dimensionless form:
\begin{equation} \label{eq:FlowBiphasicTraction}
f_T = \frac{r F_T}{\kappa} = N_t \frac{M(1/V)}{E_1(1/V) + (V/\gamma) e^{-1/V}}
\end{equation}
where we have used two special functions, the exponential integral $E_1(x)$
and a Meijer G-function $M(x)$ defined by
\begin{equation}
E_1(x)=\int_0^{\infty} dy e^{-x e^{y}}, \ \
M(x)= \int_0^{\infty} dy y e^{-x e^{y}}
\end{equation}
In \fig{FlowBiphasicRelation}, we plot this result for the traction force
$f_T$ as a function of the velocity $V$ for three different values of the
rebinding rate $\gamma$. One sees that the relation is biphasic: the
traction force first increases linearly with velocity, but then decays
again after going through a maximum value.  The symbols in
\fig{FlowBiphasicRelation} are the results of computer simulations for
stochastic models with the same number of bonds
\cite{sabass_modeling_2010}. The analytical result from
\eq{FlowBiphasicTraction} agrees with the experimental observation
of a biphasic relation in retrograde flow 
\cite{gardel_traction_2008}. The fit of the model to the data can be
improved by making more specific assumptions like catch bonding
\cite{li_model_2010}. For small flow, $V < 1$, the model predicts a
linear relation between flow speed and traction force, as observed for
flow over mature adhesions:
\begin{equation} \label{eq:FlowBiphasicTractionSmallFlow}
F_t = \frac{N_t k_{on}}{k_{on} + k_0} \kappa \frac{v}{k_0}\ .
\end{equation}
Therefore, in this regime the traction force is simply the sum of the
spring forces for the typical extension $x = v / k_0$ which is reached
when a bond is elongated due to a velocity $v$ applied for a time
$1/k_0$. The prefactor represents the equilibrium number of bonds that
were formed, that is the number of springs carrying force.  For large
force, $V > 1$, bond rupture predominates which does not allow
transmission of appreciable levels of force. In that case, the
velocity and traction force are inversely related, as observed for
fast flow over nascent adhesions. The crossover between proportional
and inverse regimes occurs when $V \approx 1$, that is $v = k_0 /
r$. With a typical unstressed unbinding rate of $k_0 = 1 \rm{Hz}$ and a
typical reactive compliance of $r = \kappa / F_b = 0.5 {nm}^{-1}$, this
predicts $v$ = 2 \rm{nm/s}, on the order of the experimentally observed values of
10~nm/s.

\FIG{Fig13}{fig:FlowBistability}
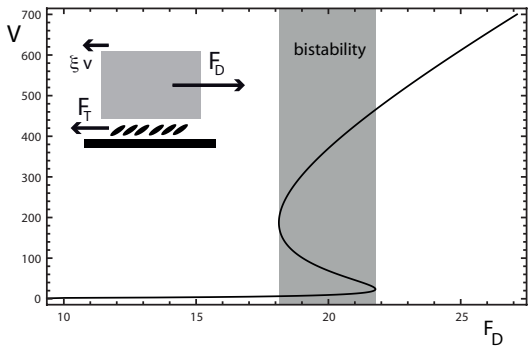
{The non-linear relation between the flow velocity $v$ and the traction 
force $F_T$ from \fig{FlowBiphasicRelation} leads to
a region of bistability for the flow velocity $v$ as a function
of the driving force $F_D$. Parameters $N_t = 25$, $\gamma = 10$,
$\xi k_0 / \kappa = 0.03$.}

The concept of friction has been discussed in the biological context
before \cite{sekimoto_protein_1991,marcy_probing_2007}. Moreover, the
biphasic relation between flow and traction has been noted before in a
non-biological context for sliding friction mediated by discrete
microscopic bonds \cite{c:scha63,filippov_friction_2004}. The large
velocity regime suggests the possibility of an instability: as the
velocity increases, the traction force decreases, thus leading to an
even larger velocity. In order to investigate this point in more
detail, a dynamical model for flow over adhesion sites is required. A
simple model motivated by the typical conditions at the lamellipodium
is to assume that the actin cytoskeleton is driven by a constant
driving force $F_D$ (representing both the push of the polymerizing
actin network away from the leading edge and the pull by the myosin
motors towards the cell body).  This force is balanced by the
frictional force with the substrate and an intra-cellular viscous
force representing dissipative processes in the lamellipodium. Thus
the force balance reads
\begin{equation} \label{eq:FlowBenediktModel}
F_D = F_T(v) + \xi v 
\end{equation}
with $F_T(v)$ from \eq{FlowBiphasicTraction}. In \fig{FlowBistability},
we numerically invert this equation to plot the velocity $v$ as a function of the driving force, $F_D$.
One sees that the non-linear relation from \eq{FlowBiphasicTraction} leads
to a region of bistability:  there is an interval of intermediate 
driving force for which two values of the velocity are stable. In practice,
this will lead to  stochastic switching between periods of slow and
fast flow, a phenomena which is known as \emph{stick-slip motion} 
in sliding friction and which can be easily verified
using stochastic simulations \cite{sabass_modeling_2010}.
Indeed this irregular kind of motion has
been observed for filopodia retraction and has been successfully simulated with a
detailed stochastic model which also included the effect
of stochastic force generation by myosin II motors
\cite{chan_traction_2008}.

\subsection{Load localization and fracture in adhesions}

\FIG{Fig14}{fig:ElasticClusterCartoon}
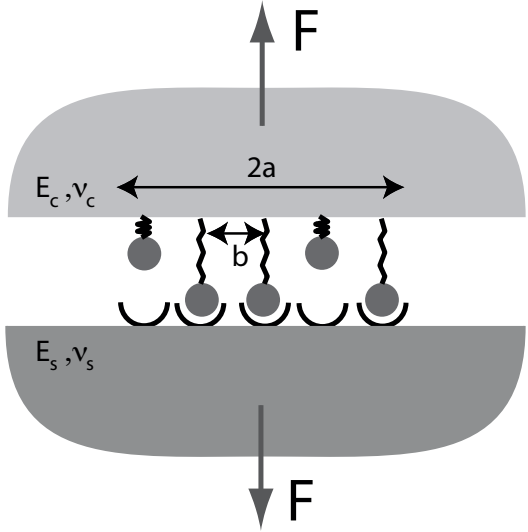
{Model of an adhesion cluster loaded by equal and opposite forces of magnitude $F$, each applied to an 
elastic halfspace. $E_C$ and $\nu_C$ denote the Young's modulus
and Poisson's ratio of the cell, and $E_S$ and $\nu_S$, those of
the substrate. $2 a$ and $b$ are the linear dimension of the adhesion
cluster and the distance between receptor-ligand bonds, respectively.}

The view of focal adhesions as bond clusters is a very flexible
conceptual basis that can be applied to more specific situations of
interest. As an instructive example, we now discuss its extension to
include the role of elasticity of the anchoring bodies. This subject
is very important because cells have been shown to respond very
strongly to changes in cellular and environmental stiffness, mainly
through changes in the stability of their adhesion sites. We consider
the situation depicted in \fig{ElasticClusterCartoon} as analyzed in
\textcite{qian_lifetime_2008}.  A single adhesion site of size $2 a$
is located between two elastic halfspaces, one representing the cell
(C) and the other the substrate (S).  The cell has Young's modulus
$E_C$ and Poisson's ratio $\nu_C$, while the substrate is
characterized by $E_S$ and $\nu_S$.  The two halfspaces are pulled
apart by a pair of equal and opposite forces (a force couple) of
magnitude $F$, acting in the plus and minus $z$-directions.  The
density of ligands in the adhesion is $\rho = 1/b^2$. By considering
cylindrical geometry and a section of thickness $b$ in the
$y$-direction, the model is reduced to one lateral dimension whose
coordinate is denoted by $x$.  The number of bonds in the adhesion is
$N_t = \rho 2a b = 2a / b$.  As above, we ask how the cluster
stability is affected by force, for example by calculating the mean
cluster lifetime $T$ or the critical force $F_c$ as a function of the
model parameters.

In order to treat the mechanical aspects of the model,
we use continuum mechanics. As it is common in contact mechanics,
we define an effective elastic modulus $E^*$ that accounts for both the cell and substrate by 
\begin{equation}
\frac{1}{E^*} = \frac{1-\nu_C^2}{E_C} + \frac{1-\nu_S^2}{E_S}\ .
\end{equation} 
The adhesion cluster is loaded by interfacial stress $\sigma(x)$
that acts in the normal direction; the force determining the 
rupture rate of a single bond located at $x$ would then be $\sigma(x) b^2$.
We first consider the case when all the molecules are bonded to the lower surface. Then $\sigma(x) = \rho \kappa u(x)$,
where $\kappa$ is the spring constant of a single bond as before
and $u(x)$ is the bond extension. From continuum mechanics,
which shows that stress and strain propagate with an elastic 
Green's function that decays in space as $1/r$ \cite{b:land70}, 
the following equation can be derived for $\sigma(x)$:
\begin{equation} \label{eq:GaoStressEquation}
\frac{d\sigma(\hat x)}{d \hat x} = \frac{2 \alpha}{\pi}
\int_{-1}^{1} \frac{\sigma(\hat s)}{\hat x- \hat s} d \hat s\ .
\end{equation} 
Here we have used dimensionless variables and have
defined the \textit{stress concentration index} $\alpha$ as
\begin{equation} \label{eq:GaoStressCondensationIndex}
\alpha = \frac{a \rho \kappa}{E^*}\ .
\end{equation} 
Thus the stress concentration index is linearly proportional 
to the adhesion size, bond stiffness and bond density, and inversely
proportional to the effective elastic modulus.

\eq{GaoStressEquation} can be solved in two limiting cases
of immediate interest. For $\alpha \to 0$, we find that $\sigma(x)$ is a constant independent of $x$.
Thus for rigid surfaces, the elastic model reduces to that of 
stationary adhesions introduced above; in that case, small clusters are the most unstable. For 
$\alpha \to \infty$, \eq{GaoStressEquation} is solved by
\begin{equation}
\sigma(x) = \frac{F}{\pi a b} \frac{1}{\sqrt{1-(x/a)^2}}\ .
\end{equation} 
Thus, the stress distribution at the edge becomes singular, similar to
that of a crack. A numerical solution shows that in general, the
interfacial stress is distributed rather uniformly for $\alpha$
smaller than $0.1$, and becomes localized to the adhesion rim for
$\alpha$ larger than $1$. Because the crack-like distribution will
lead to cascading failure from the rim, large adhesions will be
unstable since they give rise to a large stress concentration index
according to \eq{GaoStressCondensationIndex}.  Thus, both very large
and very small clusters are predicted to be unstable, suggesting that
intermediate cluster sizes have the longest lifetimes.

\subsection{Adsorption kinetics for growing adhesions}

Up to now, our discussion has centered on the physical limits for the stability of
adhesion clusters under conditions in which forces
tend to destabilize the bonds that are formed. We have seen that
force, adhesion cluster size, surface motion and elasticity
define clear limits for the physical stability of
adhesion sites. We now address the question of how biological focal
adhesions (FA) can protect themselves against these limiting
factors. The main mechanism which seems to have evolved
in this regard is growth under force, which we will now discuss.

We begin with a generic treatment of adsorption kinetics that are
governed by the chemical potential differences of molecules in the
solution and those adsorbed to the adhesion on the substrate. If the
FA is under conditions close to equilibrium and if the growth is
reaction-limited as opposed to diffusion-limited, the growth dynamics
of the adhesion depend on the dimensionless chemical potential
difference (in units of $k_BT$) between molecules in solution and
those adsorbed to the FA.  The dimensionless, local (fractional) area
coverage by these molecules is denoted by $\phi(\vec r)$ which is also
time dependent: $d \phi(\vec r)/dt=(1/\tau) \left(\mu_b-\mu_a(\vec r)
\right)$ \cite{andelman1996}, where $t$ is the time variable and
$\tau$ is a characteristic time; since $\phi$ and $\mu$ are
dimensionless each side of the equation scales as the inverse of a
time.
Here $\tau$ is the attempt time for
molecules near the surface to adsorb, and $\mu_b$ and
$\mu_a(\vec r)$ are the dimensionless chemical potentials (in units of $k_BT$)
 of molecules in the bulk cytoplasm and the FA, respectively.
 When $\mu_a < \mu_b$ adsorption is locally favored and
 the local concentration of adsorbed molecules grows in time.
  We classify the chemical potential of molecules in the FA by terms
with different symmetries with respect to the external force and
afterwards discuss the molecular origins of these terms in
different models.  We therefore  write: 
\begin{equation} \label{kineticeq3}
\frac{\partial \phi(\vec r)}{\partial t}=\frac{1}{\tau}
\left(\mu_b - \left( \mu_\ell(\vec f) + \mu_i(\phi(\vec r))+ \mu_f(\vec f,
\phi(\vec r)) \right) \right)
\end{equation} 
The term $\mu_\ell(\vec f)$ depends on the cytoskeletal (CSK) force,
$\vec f$, but not on the local, dimensionless, local area fraction
covered by adhesion molecules, $\phi(\vec r)$.  It originates in those
terms derived from the derivative of the free energy with respect to
$\phi(\vec r)$ that are independent of $\phi$ and are thus not
cooperative in nature. The next term depends on the local
concentration (and hence reflects cooperativity of the adhesion
molecules) but not on the force, while the last term depends on both.
 
The local, chemical interaction (ligand binding) of the adsorbing
molecules with the surface, $\mu_\ell(\vec f)$, includes the
effects of force-induced changes in the single-molecule
conformations.   Since the chemical potential is a scalar
quantity while the force $\vec f$ is a vector,  symmetry dictates that to
quadratic order in the force (which is assumed to be small) it
must have the form $\mu_\ell=\mu_0+\alpha < \vec f \cdot \vec d
>  + \beta f^2$ where $\mu_0$ is force independent and $\alpha$
and $\beta$ are constants.  The angle bracket $<\vec f \cdot
\vec d>$ denotes an average over all the orientations of a
vector that resides in the adsorbed molecule $\vec d$ to which
the force (possibly) couples; this coupling results in a
conformational and hence in an energy change.
If the adsorbing
molecules are  oriented at a fixed angle relative
to the CSK force, this
term will depend only on the magnitude of the CSK force.

The mutual, force independent, interactions of the molecules that
assemble in the FA, are reflected in the term $\mu_i(\phi(\vec r))$
and is derived from the functional derivative of the interaction free
energy with respect to $\phi(\vec r)$ \cite{safran_statistical_2003}.
For convenience, one can consider a Ginzburg-Landau expression for the
free energy for interacting molecules that nucleate a condensed phase
in equilibrium with a ``gas'' (low concentration) phase of adsorbates
on the surface \cite{safran_statistical_2003}. In order to obtain an
analytical solution it is convenient (but not necessary in general) to
focus on values of $\phi$ that are close to the critical value,
$\phi_c$ at which the condensation first occurs as the temperature is
lowered \cite{safran_statistical_2003} as the system parameters
approach a critical value (which for simple systems can be the
critical temperature but for more complex systems can depend on the
interaction energies, effective temperature and other parameters). The
chemical potential, $\mu_i=\delta F_0/\delta \psi$ is derived from a
free energy (in units of $k_BT$) of the form:
\begin{equation}
\label{eq_GLenergy} F_0=\int dx dy \left[
-\frac{1}{2}\epsilon\Psi^2+
\frac{1}{4}c\Psi^4+\frac{1}{2}B\left(\nabla \Psi
\right)^2\right] \end{equation} where $\Psi=\phi - \phi_c$,
$\epsilon$ is the deviation from criticality (for simple systems $\epsilon \sim (T_c-T)/T_c$), $c$ is a number of order unity, and
$B$ is proportional to the attractive interaction between the
molecules \cite{safran_statistical_2003}.

We note that the scalar nature of the chemical potential means
that the lowest order term in $\mu_f$ must be the dot
product of two vectors: the local force and the gradient of the
adhesion molecule density so that $\mu_f$ is proportional to
$\vec f \cdot \nabla \phi(\vec r)$. Such a term cannot be derived from
the direct functional derivative of a free energy that is only a
function of $\vec f$ and  $\psi(\vec r)$ and its derivatives.
However, as discussed in the context of the molecular models,
the adsorption may be coupled to other degrees of freedom in the
system such as  its mechanical properties ({\it{e.g.}}, strain,
anchoring to the substrate).   These can give rise to terms in
the free energy that are manifestly and linearly force dependent
 and that can equilibrate more quickly than  the adsorption.
The
discussion below of the microscopic models demonstrates that this
coupling can effectively yield the term $\vec f \cdot \nabla
\psi(\vec r)$, written here from symmetry
considerations alone.

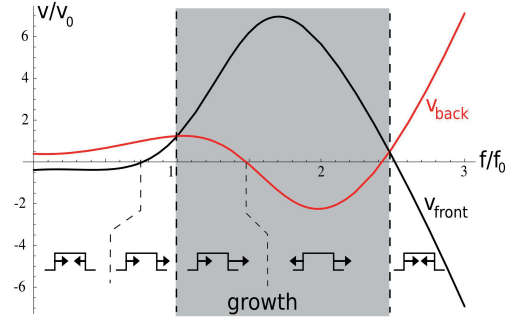
\FIG{Fig15}{fig:alicefig1}{(Color online) Calculated velocities of the front
  (closer to the direction of the pulling force in this figure, to
  the right) and back of the adhesion for a rigid matrix. Courtesy of
  A. Nicolas and adapted from \textcite{Besser06} and
  \textcite{Nicolas08}.}

With these expressions for the various contributions to the
chemical potential, one can analytically solve Eq.
\ref{kineticeq3} in one dimension in terms of $\Psi=\phi-\phi_c$
where $\phi_c$ is the average concentration of the coexisting
high and low density phases on the surface ({\it{i.e.}}, the
adhesion and a low density ``gas'' of isolated, surface adsorbed
molecules).  The protein assembly grows \cite{Besser06} in the
direction of the force with a profile given by:
$\Psi(x,t)=\Psi(x-vt)$ with the growth velocities at front and
back related to the characteristic velocity $v_0=\xi/\tau$ where
$\xi$ is the correlation length that characterizes the interface
width: $\xi^2 \sim B  / \epsilon $. In the limit of small
chemical potential differences between the molecules in solution
and those in the FA, we can calculate the velocities of the front (closer to the pulling direction)
and back of the adhesions as shown pictorially at the bottom of \fig{alicefig1}:
\begin{equation}
\begin{array}{ll} \displaystyle v_{front} & \displaystyle = + v_0
\Delta\mu(f)+ {f} \sigma \\ \displaystyle
v_{back} & \displaystyle = - v_0 \Delta\mu(f)+ {f}
\sigma  \\ \displaystyle v_{tot} & \displaystyle = v_{front}-
v_{back}= v_0 \Delta\mu(f)\\ \end{array}
\end{equation}
where $\Delta\mu(f)=\mu_b-\mu_0 - \alpha <\vec f
\cdot  \vec d>  - \beta f^2$.
The term $f \sigma$ arises from the symmetry-breaking
term in the chemical potential $\vec f \cdot \nabla \psi(\vec
r)$.
One considers the case where $\mu_0 >\mu_b$, $\alpha<0$
and $\beta>0$ (so that   $\Delta\mu(f)>0$), favoring adsorption
for some range of force, but not for forces that are too small
or too large; the quadratic term in $f$ inhibits growth when the
force becomes too large.  Note that the overall growth velocity
of the adhesion depends on the homogeneous activation term but is
independent of the symmetry breaking term $\vec f \cdot \nabla
\psi(\vec r)$. These findings indicate that both force-induced
activation modes are required to explain the anisotropic growth
of focal adhesions: the homogeneous activation term alone could
not explain the anisotropy of the aggregation, whereas the
symmetry breaking term alone could not account for overall
adhesion growth.

The stress  dependence of the velocity is shown in  Fig.
\ref{fig:alicefig1} for a rigid substrate. As expected, this
model accounts for both the growth and sliding of the adhesion,
based on a treadmilling mechanism. This treadmilling mechanism
was recently observed experimentally  \cite{debeer2010}. In
addition, the theory also predicts  a range of stress in which
the adhesion indeed grows. Furthermore a regime where the front
and back edges  move apart is predicted; this was also recently
reported \cite{debeer2010, heil2010}.

\subsection{Force-induced growth of adhesions}

We now discuss two classes of models that motivate the general
treatment presented here from more molecular considerations to explain
the stability and growth of focal adhesions under CSK force: (i)
microscopic models that motivate the nucleation and growth picture
described above and (ii) an analogy to force induced polymerization
that takes into account the imbalance of CSK stresses and adhesion
molecule anchors.  These two approaches differ in several aspects, the
most important of which is whether the symmetry breaking exists
already at the genesis of the adhesion (model (ii)) or whether it is a
spontaneous consequence of force applied to an adhesion (model (i))
with no intrinsic asymmetry in its internal structure. The first class
of models that are based on nucleation and growth is supported by
experiments that suggest that focal adhesion growth occurs primarily
at the front and back of the plaque \cite{debeer2010, heil2010}.  As
explained below, the second class of models would allow for adhesion
molecules to accrue all along the plaque.  On the other hand,
experiments also show that the focal adhesion is not symmetric; the
front is acted upon by the CSK stress fibers while the back is facing
the lamellipodium.  This provides support for the intrinsic asymmetry
of binding and force that is the basis for the second class of
models. All of these models focus on forces that are tangential to the
substrate, appropriate to stress fibers near the basal cell surface.
Adhesion growth induced by forces normal to the surface has been
modeled in \textcite{Walcott_nucleation_2011}.  Their results and
related experiments show that for normal forces, adhesion nucleation
and decay depends sensitively on the substrate stiffness; however, the
growth and decay dynamics themselves are stiffness independent.

(i) {\it{Symmetry breaking due to activation of mechanosensors (within the adhesion) by
force:}} The model introduced by Nicolas and coworkers
\cite{nicolas_cell_2004, Alice06, Besser06, Nicolas08} assumes
that the dynamics of FA is governed by the activation of
mechanosensitive units, that are part of the adhesion.  Since
the molecules in the FA are attached both to the matrix and to
the CSK, they sense variations of mechanical stresses in the
cell as well as  the local, elastic properties of the
extracellular matrix.   The symmetry breaking occurs
spontaneously, without introducing an adhesion geometry that
desymmetrizes the front and back. Instead, the adhesion
treadmills or slides in the direction of the force because the
force itself is a vector that breaks symmetry via the term in
the chemical potential $\vec f \cdot \nabla \psi(\vec r)$ where
it couples to the concentration gradients in an asymmetric
manner at the front and back of the adhesion.

The model includes two modes of deformation \cite{nicolas04}.  (a) The
stretching of the individual molecules of FA assembly by the CSK
force, which corresponds to a shear deformation of the FA and (b) the
effect of compression induced by lateral compressions (due to forces
tangential to the substrate); this is a cooperative effect involving
the displacements of neighboring molecular units
\cite{ali_cooperativity_2011}. The compressive mode is asymmetric in
its response to a local, tangential stress since the back edge of the
FA is expanded while the front edge is compressed. Assuming that
compression induces adhesion growth ({\it{i.e.}}, lowers the
adsorption free energy of molecules in solution that contact the
substrate) and that expansion induces desorption, the front of the
adhesion may grow, while the back may shrink.  This results in a
treadmilling of the adhesion and an apparent motion in the direction
of the force. In contrast to this, the stretching mode is symmetrical
in response to the tangential stress; both the back and the front
edges are activated with identical probabilities. If the stretching
results in conformations that stabilize adsorbed molecules, this will
result in growth of the front of the adhesion in the direction of the
force and in the growth of the back of the adhesion in the opposite
direction; the net result will be an overall growth of the size of the
adhesion.  It should be noted that activation of the adhesion
molecules is expected to require tensile forces. In a picture in which
the mechanosensor is a single molecule, it is not obvious how these
forces can arise from compression at the front edge of the
adhesion. However, if one associates the mechanosensor with a complex
assembly of molecules that include the integrins and the protein
plaque, reorientation and interactions may indeed give rise to
molecular extensions in the direction perpendicular to the compression
{\it{e.g.}}, by the tendency for a molecular assembly to conserve its
volume.  These effects may activate adsorption even in the absence of
externally generated tensile forces.  FA growth as a function of
substrate rigidity can be predicted if one assumes that that the
adsorption is partially controlled by the energy invested by the CSK
in deforming the substrate; this of course depends on the size and
nature of the adhesion.  For soft substrates, the substrate is
deformed over a thickness related to the adhesion size. This limits
the maximal size of the adhesion due to the stresses and strains that
the CSK forces induce in the underlying matrix via the coupling by the
FA \cite{Nicolas08}.

A related point of view was recently presented
by Garikipati and collaborators  \cite{krishna2010} who used a
general thermodynamic argument to motivate the symmetry breaking
term in $\mu_f$. Instead of assuming that the adhesion molecules
are activated to adsorb near regions of compression and desorb
near regions of expansion as in
\textcite{Nicolas08,Alice06}, the ``negative work'' done by the
CSK pulling forces is included in the energetics at the outset. The resulting
energy is decreased when the adhesion moves in the direction of
the force (towards the nucleus); this apparent sliding occurs by
the adsorption of molecules at the edge of the adhesion that is
closest to the nucleus (or direction of applied force) and
desorption at the other end. 
We note that this mechanism can lead to growth all along the
adhesion surface and not just at the edges, as discussed below
in relation to the work of \textcite{Shemesh05pnas}.

Another model that focuses on the integrin binding with the adhesion
has been suggested by Deshpande and coworkers
\cite{deshpande_bio-chemo-mechanical_2006,deshpande_model_2007,pathak_simulation_2008,mcgarry_simulation_2009}
who have developed a thermodynamically motivated computational
approach that has three essential features: (i) coexistence of both
low and high affinity integrins in thermodynamic equilibrium, (ii)
mobility of the low affinity integrins within the plasma membrane, and
(iii) mechanical equilibrium of the contractile forces generated by
the stress fibers -- these forces affect the free energies of the
integrins and give rise to a coupled thermo-mechanical response. An
initial prediction based on this is the correlation of the
distributions of the normalized focal adhesion densities (as
parameterized by the high affinity integrin concentration) and
contours of the stress fiber density. However, this model does not
contain spontaneous symmetry breaking and the growth of adhesions in
the direction of the applied force requires the ad-hoc inclusion of
anisotropic activation signals.

Another recent model of focal adhesion growth focuses on the bond
attachment-reattachment dynamics discussed above and combines this
with considerations related to clustering.  The stochastic elastic
model combines theory and Monte Carlo simulations
\cite{gao_probing_2011} and suggests that FA growth is self-limiting
since growth eventually leads to crack-like delamination failure near
the adhesion edges.  Very soft substrates tend to diminish the
adaptive capability of cells by suppressing bond rebinding
irrespective of the cytoskeleton stiffness, which may prevent
short-lived, small focal contacts from maturing into stable FA.

(ii) {\it{Symmetry breaking due to the geometrical structure of the
    adhesion:}} The model introduced by Kozlov and coworkers
\cite{Shemesh05pnas} uses a general thermodynamic argument to write
$\mu_f=-\vec f \cdot \vec d$ where $\vec f$ is the force applied by
the CSK to one adhesion molecule and $\vec d$ is a particular bond in
the molecule that is stretched by that force.  In the one-dimensional
version of the model described in \textcite{Shemesh05pnas}, the force
and bond are in the same directions. This energy represents the
``negative work'' done by the CSK against the adhesion when an
additional protein is allowed to adsorb. In the absence of this
additional protein, the adhesion would be stretched and deformed by
the CSK force; the presence of an additional protein adsorbed from
solution relaxes this deformation to some extent and thus lowers the
free energy of the adhesion; hence the negative sign in the expression
for $\mu_f$.

To predict the growth of the adhesion in the direction of the
applied force, several assumptions are made.  First, the FA is
assumed to be capable of adsorbing and releasing adhesion
molecules at every point along its surface; this differs from
models of polymerization where the monomer exchange occurs only
at the ends of a linear polymer.  For the FA plaque,
plausibility of this unusual property has some experimental
support \cite{wichert2003}.  However, other studies have
shown that the growth primarily occurs at the two ends of the
adhesion and not uniformly along its surface
\cite{debeer2010, heil2010}. Second, the pulling forces $\vec f$
are assumed to be applied to the FA surface at discrete points
that are  distributed along the FA length with a particular
density. Finally, the FA plaque is taken to be anchored to a
rigid external substrate by discrete linkers spread over the
adhesions length with a density that differs from the density of
the pulling forces. This intrinsic asymmetry in the geometry of
the adhesion leads to its asymmetric growth.

The interplay between the distribution of the pulling forces and
the anchors leads to an inhomogeneous stretching stress within
the FA and, consequently, to an uneven distribution of the
chemical potential $\mu_f$ along the adhesion.  Hence, the
tendency to assemble or disassemble  can vary along the FA
resulting in different regimes of the overall molecular exchange
between the FA and the adhesion molecules in solution.

The model predicts different modes of the FA assembly that are largely
consistent with the experimentally observed FA behavior.  As a
function of the chemical potential difference, the FA can exhibit
unlimited growth, complete disintegration, or reach a stable
force-dependent steady state dimension.  The sliding of the adhesion
in the direction of the force is explained by an explicit
symmetry-breaking introduced in this model due to an imbalance of
local forces and anchors; this would require additional specific
modifications of the generic model and may be due to the presence of
the lamellipodium on the far side (close to the cell membrane) of the
adhesion and the stress fibers on its near side (closer to the
nucleus).  In contrast to other models of FA mechanosensing, the
thermodynamic model does not require any special conformational
changes of proteins upon force application. At the same time, the
model assumes that the FA plaque is an elastic body able to accumulate
mechanical stresses. Moreover, the plaque must possess a mechanism to
accommodate new FA proteins without undergoing stretch-induced
rupture. This may require the presence of delicate molecular
mechanisms having properties similar to those of the members of the
formin protein family \cite{kovat2006}, which are able to maintain a
stable connection to an associated protein complex (the growing ends
of actin filaments) and at the same time, enable insertion of the new
protein monomers into the complex and stabilization of the growing
structure.

\section{Cell shape and forces}

\subsection{Physical motivation}

We now move from the level of cellular adhesions to the level of whole
cells and address the question how the forces in an adherent cell are
balanced over the entire cell. From the preceding section, we have
learned that cells adhere through relatively few but stable sites of
focal adhesion. At the same time, they tend to be very contractile
provided that the environment is sufficiently rigid to balance these
forces. The interplay between contractility and spatially localized
adhesions leads to interesting phenomena which can be understood best
by first considering the shape of cells.

Depending on cell type and environment, cells adopt a large variety
of different shapes \cite{b:bray01,mogilner_shape_2009,
kollmannsberger_physics_2011}. For cells in the human body,
for example, we observe such diverse shapes as the biconcave
disc of the red blood cell, the invaginated shapes of
single fibroblasts in connective tissue, the polygonal
shapes of cells in densely packed epithelial tissue,
and the highly branched networks formed by neurons in the brain.
In fact the term \textit{cell} was coined by Robert Hooke who
in his 1665 book \textit{Micrographia} was the first to report
on the many shapes visible under the microscope.
He chose this term \textit{cell} because of the near rectangular shapes of the building block
of cork reminded him of monk cells in a monastery.
Due to the evolutionary process,
the shape of cells is closely
related to their function. For example, red blood cells
are optimized to squeeze through narrow
capillaries, fibroblasts are sufficiently contractile to
deform and remodel extracellular matrix, epithelial cells
pack tightly to seal certain regions, and
neurons form a highly connected communication
network.

In 1917 D'Arcy Thompson suggested
in his book \textit{On Growth and Form}, that the shapes of
cells and organisms must be closely related to physical
forces \cite{thompson_growth_1992}. The interest in shape somehow
declined in the wake of the molecular revolution in 
biology, but has recently has been re-invigorated by the finding
that cell shape strongly effects cellular
function. By using microcontact printing of adhesive
patterns to constrain cells into pre-defined shapes, it has been
shown that it is not the total amount of adhesive ligand
available to the cell, but rather its spatial distribution that determines
cell fate \cite{chen_geometric_1997}. In fact, relatively 
little ligand is sufficient to ensure cell survival if
it is arranged in such a way as to allow the cell to
spread over a large area on the substrate. In contrast, a cell constrained to occupy
a small area or volume goes into programmed cell
death even if many ligands are available in that small space. More
recently, it has also been shown that stem cell
differentiation depends on cell shape \cite{kilian_geometric_2010}.

\FIG{Fig16}{fig:CellShapes}
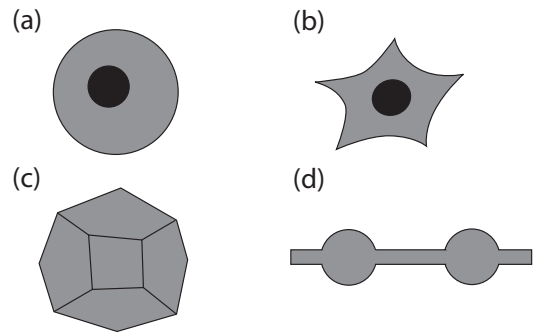
{Cell shapes are often dominated by tension.
(a) Single cells in solution are usually round. (b) Isolated
tissue cells like fibroblasts commonly show an invaginated 
shape between distinct sites of adhesion. (c) Epithelial
cells in closely-packed tissue are usually polygonal,
both in two and three dimensions (here a polyhedron is schematically depicted for the 3d case). (d) Cylindrical cell extensions
such as axons tend to pearl when the tension is increased.}
 
During the last decade, the use of micropatterned 
substrates has developed into a standard technique employed 
to investigate many details of the spatial organization
of cells. For example, it has been shown that the  
spatial coordination of lamellipodia \cite{parker_directional_2002},
stress fibers \cite{thery_cell_2006}, spindle formation
\cite{thery_experimental_2007,fink_external_2011} and cell-cell adhesions
\cite{tseng_spatial_2012} strongly depends on the geometry of the
extracellular environment. Moreover, micropatterned
substrates can also be used to quantitatively analyze
cell shape. It has long been noticed that on flat substrates, most tissue
cell types adopt shapes that are indicative of cell contraction, often
characterized by invaginations between pinning sites
\cite{zand_what_1989}. The tendency to invaginate becomes
even more pronounced when the actin cytoskeleton
is disrupted,
which leads to a ray-like morphology of adhering cells 
\cite{bar-ziv_pearling_1999}. Recently micropatterning
and image processing have been combined
to quantitatively study the relation between cell
shape and adhesive geometry \cite{uss:bisc08a}.

In general, many of the observed cell shapes can be understood with
concepts borrowed from soft matter physics.  The round shape of cells
in solution, the invaginated shapes of single tissue cells on flat
substrates and the foam-like packing of cells in epithelial tissue
point to a strong role of cellular tension, compare
\fig{CellShapes}(a)-(c). Tension also plays an important role in many
dynamical situations, such as the pearling of cell extensions after
changes in pressure \cite{pullarkat_osmotically_2006} or cell
elasticity \cite{bar-ziv_pearling_1999}, as in \fig{CellShapes}(d).
Tensions that contract cell-cell junctions have emerged as a key
factor that determines the dynamics of tissues
\cite{farhadifar_influence_2007,lecuit_cell_2007,paluch_biology_2009,
  aliee_physical_2012}.  However, in order to quantitatively explain
the experimental data for single cells, tension arguments must be
combined with elements of elasticity
\cite{bar-ziv_pearling_1999,uss:bisc08a}. During recent years,
different modeling approaches have been suggested to describe the
interplay between myosin II contractility which is balanced by elastic
forces exerted by the cytoskeleton and the adhesions (that couple to
the substrate).  In the following we discuss and compare some of the
suggested approaches. This then forms a basis that allows us to
consider even more coarse-grained models of cells (cellular force
dipoles) in the next section.

\subsection{Contour-models for cell shape}

Because adhering cells on flat substrates spread to become very thin compared 
with their lateral extensions, it is appropriate to describe them as
approximately two-dimensional objects, see
\fig{ContourModel}(a). The simplest approach is to focus only on
cell shape and to consider only the two-dimensional contour
$\vec{r}(s)$ describing the cell boundary, with $s$ (which has units of length) defined as the  
distance coordinate along the contour. At any point $s$
along the boundary, we define the tangent vector
$\vec{t}(s) = (d \vec{r}(s) / ds) / |d \vec{r}(s) / ds|$ and the
normal vector $\vec{n}(s)$ perpendicular to it. These two 
unit vectors are
connected by the geometrical relation $d\vec{t}(s)/ds =
\vec{n}(s) / R(s)$, where $R(s)$ is the local radius of curvature.  We
first consider the simplest model geometry possible, namely, a contour
which is pulled in towards the cell in the region  
between two adhesion points, as shown in
\fig{ContourModel}(b).  We shall relate this to the forces that arise
from acto-myosin contractility in cells.

\FIG{Fig17}{fig:ContourModel}
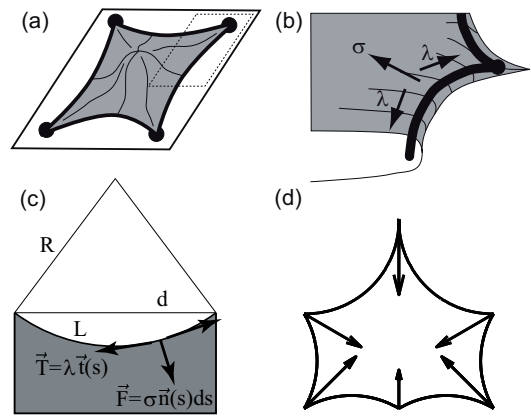
{Simple tension model for cell shape on a flat substrate.
(a) The cell is very flat and therefore effectively two-dimensional.
The outlined region is shown in (b) with more details.
(b)  Along the contour between two neighboring
adhesion sites, surface tension $\sigma$ pulls inward,
while line tension $\lambda$ pulls tangentially.
(c) Along the contour, surface tension and
line tension pull in normal and tangential directions,
respectively. For a circular arc, radius $R$, contour length $L$ and spanning
distance $d$ are geometrically related to each other.
(d) For a house-shaped cell, all arcs have the same
radius, although the spanning distance is larger on the
diagonals. Also shown are the traction forces derived from the shape model.
Adapted from \textcite{uss:bisc09a} and \textcite{uss:guth12}.}

We begin with the most elementary example, the \textit{simple tension model},
in which these forces arise from the energies associated with changes
in the contour length and surface area. In the simple tension model, a
constant surface tension $\sigma$ pulls in the contour (thereby
reducing the surface area) and is balanced by the effects of a
constant line tension $\lambda$ which tends to straighten the contour
(thereby reducing the line length), see \fig{ContourModel}(b). The
surface tension acts on a line element and points in the normal
direction, leading to a force $\vec{F} = \sigma
\vec{n}(s) ds$, while the line tension acts on every point of the contour
and in the tangential direction with a force $\vec{T} = \lambda
\vec{t}(s)$.  This situation is depicted in \fig{ContourModel}(c). The
force balance on a contour element $ds$ then leads to a Laplace law:
\begin{equation} 
\vec{F} = \vec{T}(s+ds) - \vec{T}(s) 
\Rightarrow \sigma\vec{n} = \lambda\frac{d\vec{t}}{ds}= \frac{\lambda}{R}\vec{n}
\Rightarrow R=\frac{\lambda}{\sigma}
\label{eq:LaplaceLaw}
\end{equation}
Note that this result is expected from dimensional analysis. The
simple tension model thus predicts that the contour forms perfect
circular arcs, which indeed is often observed in cell experiments.  In
\fig{ForceFromShape}(a) and (b), this is demonstrated for cells on a
dot micropattern and on a pillar array, respectively. Note that due to
its local nature, the simple tension model does not obey total
momentum conservation that is expected for a closed system like a
single cell.

\FIG{Fig18}{fig:ForceFromShape}
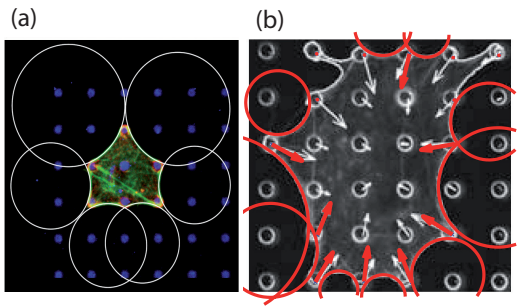
{(Color online) (a) House-shaped cell on an adhesive micropattern created by
  microcontact printing. Note the circular arcs with radii that are
  larger at the two diagonals. From \cite{uss:bisc08a}. (b) Cell on a pillar array that allows a
  simple read-out of local forces from measurements of the deflection
  of each pillar.  Circular arcs describe most of the cell contour,
  but not for example at the upper left corner, where an internal
  fiber distorts the contour. From \textcite{uss:bisc09a}.}

The last equality in \eq{LaplaceLaw} resembles the Laplace law
$R = (2 \sigma) / \Delta p$ for a sphere ({\it{e.g.,}}
a soap bubble) whose surface is contracted by a surface tension $\sigma$
and stabilized by a pressure difference $\Delta p$ (compare section IIF).
We note that in three dimensions, $\lambda$ and $\sigma$ 
(related to line and surface) are replaced by
$\sigma$ and $\Delta p$ (related to surface and
volume), respectively, and that a factor $2$ appears (compare section IIF). These two effects obviously result from the different
dimensionalities. However, a more fundamental difference 
is the fact that while the sphere stabilizes
itself without any need for attachment, the simple
tension model for the invaginated contour only 
makes sense in the presence of the two adhesion sites.
Without the adhesions, both tensions
would work in the same direction and the contour would
simply contract to a point.

In single cells adhered to a flat substrate, 
the two tensions have contributions from
different processes. The surface tension $\sigma$
mainly results from the pull of the
myosin motors in the actin cytoskeleton (including
the actin cortex), but can also have a contribution
from the tension in the plasma membrane. The 
line tension $\lambda$ is expected to 
primarily arise from the elastic pull of the thick and contractile actin
bundles lining the cell periphery. If cells are treated
with pharmacological drugs that disrupt the actin cytoskeleton,
they tend to invaginate more strongly, indicating that
effectively the line tension $\lambda$ is reduced
more than the surface tension $\sigma$
\cite{bar-ziv_pearling_1999,uss:bisc08a}.

Interestingly, very strong invaginations necessarily lead to tube-like
extensions connecting the retracted cell body to the sites of
adhesions (here we assume a three-dimensional viewpoint
again). A cylindrical tube in which the surface tension is the only
relevant force undergoes a Rayleigh-Plateau instability
\cite{safran_statistical_2003}. Thus one also expects pearling of the
cellular tubes. Indeed this is exactly what has been observed
experimentally \cite{bar-ziv_pearling_1999} once the elasticity of the tube is suppressed, similar to the pearling
which can be induced, for example in axons, by changing the osmotic
pressure \cite{pullarkat_osmotically_2006}.

For contractile cells, the force due to the surface tension $\sigma$ is
mainly balanced by the elasticity of the actin cortex underlying the
plasma membrane. Experimentally, it was found that thinner tubes are
unstable whereas thicker ones are not. This can be explained as
follows.  One considers a cylindrical tube with undulations by defining a
local radius that varies along the tube axis in the $z$ direction: $R(z) = R_0 + \Delta R
\cos(2 \pi z/\lambda)$. There are two energy contributions, the change in
surface energy and an elastic energy that includes bending and
stretching energies. Using the constraint of volume conservation
the sum of these two energies reads \cite{bar-ziv_pearling_1999}
\begin{equation}
E_t = \sigma \frac{1}{4} u^2 (k^2-1) + \frac{1}{4} \frac{3 E R_0}{1+\nu} u^2\ . 
\end{equation}
Here $k = 2 \pi R_0 / \lambda$ and $u = \Delta R / R_0$ are
the dimensionless wavenumber and amplitude of a perturbation of wavevector $k$,
respectively, while $E$ and $\nu$ are the Young's modulus and Poisson's ratio of
the cortex, respectively. Because both terms have the same scaling with
$u$, these terms cannot determine the amplitude. However,
the wavelength at which the system is unstable is determined by this expression. The first term can become negative for large
wavelengths, $k < 1$.  In this case, the energy is  negative
if the tension exceeds a critical threshold of
\begin{equation}
 \sigma_c = \frac{3 E R_0}{1+\nu}
\end{equation}
which increases with $R_0$. This makes thicker tubes more stable
than thinner ones, as observed experimentally.

The pearling study indicates that in cellular systems, tension and
elasticity are strongly coupled.  A similar result was obtained
by a quantitative study of cell shape on micropatterned substrates
\cite{uss:bisc08a}. Although this analysis revealed that invaginations
of the cell contour are usually close to circular as predicted by the
Laplace law from \eq{LaplaceLaw}, it also showed that the arc radius $R$
varies with the spanning distance $d$ (defined in Fig.~(\ref{fig:ContourModel})) between the
two neighboring adhesion sites, while the Laplace law would predict a
constant radius independent of spanning distance. Again, this can be
explained by an elastic analysis, the \textit{tension-elasticity
  model} \cite{uss:bisc08a}. The circular nature of the arcs suggests
that a modified Laplace law must hold.  While the surface tension $\sigma$
is expected to be determined mainly by myosin motor activity in the
bulk cytoskeleton and therefore should be the same and constant for
all arcs, the line tension $\lambda$ might be determined locally by
the mechanics of the peripheral bundles. The simplest possible model
is to take into account the fact that the actin cortex, localized near
the line, can behave elastically.  The line tension then has a
contribution from the stretching of this cortex relative to its
relaxed state:
\begin{equation}
\lambda=EA \, \frac{L-L_0}{L_0}
\label{eq:ElasticLineTension}
\end{equation}
where the product of three-dimensional modulus $E$
and cross-sectional area $A$ is the effective one-dimensional modulus of the 
bundle, and $L$ and $L_0$ are its actual stretched length
and relaxed length, respectively. Together with 
the geometrical relation 
\begin{equation}
\sin\left(\frac{L}{2 R}\right)=\frac{d}{2 R}
\end{equation}
between the contour length $L$, radius $R$ and spanning
distance $d$ (compare \fig{ContourModel}(c)),
one obtains a self-consistent equation
for the arc radius $R$:
\begin{equation}
\label{eq:tem}
R=\frac{EA}{\sigma}\left(\frac{2R}{L_0}\arcsin
\left(\frac{d}{2R}\right)-1\right).
\end{equation}
The simplest model assumption for the resting
length is $L_0=d$.  A numerical solution of
\eq{tem} shows that $R$ is a monotonically
increasing function of $d$, as observed experimentally.
For small invaginations, $d/R \ll 1$ and one can
expand \eq{tem} to obtain
\begin{equation}
\label{eq:tem_simple}
R = (EA / 24 \sigma)^{\frac{1}{3}} d^{\frac{2}{3}}.
\end{equation}
Thus $R$ increases with the contour rigidity $EA$ and
spanning distance $d$, but decreases with increasing surface tension
$\sigma$. Note that compared with the Laplace law from \eq{LaplaceLaw}, one still has an inverse relation between radius and surface tension,  but now with a different exponent.

\subsection{Whole-cell models}

As shown in \fig{ForceFromShape}, contour models can also be used to
analyze the shape of whole cells that are characterized by
geometrically prominent features like circular arcs. One model class
that addresses whole cells by construction are cellular Potts models,
which have been successfully used to evaluate and predict cell shape
on dot-like micropatterns \cite{vianay_single_2010}, with an example
shown in \fig{ForceFromShape}(a). In essence, the cellular Potts model
is very similar to the simple tension model, because its main
ingredient is tension at the interface. There are many situations of
interest, however, for which such a simple approach is not sufficient
to account for cell behavior; one example is the case where the actin
cytoskeleton locally reorganizes into contractile bundles (in addition
to those that line the cell periphery). However, a theory based only
on structural elements visible with standard microscopy procedures
might not be sufficient as many observations in single cell
experiments point to the existence of a much finer-scale network of
additional fibers that coexist with the stress fibers.  The natural
theoretical framework for studying this situation is continuum
mechanics. Although traditionally used mainly to address the mechanics
of macroscopic objects like growing tissue
\cite{ambrosi_perspectives_2011}, different continuum mechanics
approaches have recently been developed to describe the shapes and
forces of adherent cells \cite{kollmannsberger_physics_2011}. In
particular, the powerful framework of the finite element method (FEM)
has been adopted for this purpose. A detailed FEM-model integrating
mechanical and biochemical aspects has been developed that is able to
explain many details of cell adhesion
\cite{deshpande_bio-chemo-mechanical_2006,deshpande_model_2007,pathak_simulation_2008,mcgarry_simulation_2009}.
Here, we discuss this model as one representative example that
demonstrates how a detailed whole-cell model can be constructed. Note
that this kind of model ensures total momentum conservation by
construction.

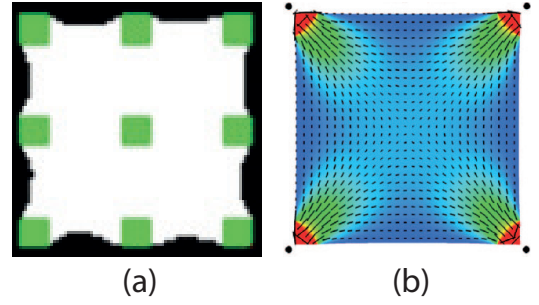
\FIG{Fig19}{fig:FEM} {(Color online) Two model approaches for whole cells. (a)
  Cellular Potts models use spins on a lattice to simulate the contour
  between the cell and its surroundings. From
  \cite{vianay_single_2010}. (b) Finite element methods (FEM) can be
  used to implement any material law of interest. Contractility is
  implemented by thermoelasticity. The vector field represents the
  direction and activation level of stress fibers. From \textcite{deshpande_bio-chemo-mechanical_2006}.}

As before, the nearly flat adherent cell is treated as an effectively
two-dimensional object. In continuum elasticity theory, this
corresponds to a plane stress approximation for thin elastic films in
which the stress is approximately constant throughout the film in z-direction
\cite{b:land70}.  The stress in the cell is assumed to have both
active and passive contributions that are additive :
\begin{equation}
\Sigma_{ij} = \sigma_{ij} + \left( \frac{E \nu}{(1-2\nu) (1+\nu)}
\epsilon_{kk} \delta_{ij} + \frac{E}{(1+\nu)} \epsilon_{ij} \right)
\label{eq:stressdeshpande}
\end{equation}
The first term represents active, contractile stresses $\sigma_{ij}$.
The second term represents the passive elasticity of the
cell, which here is assumed to have a linear elastic response. 
$E$ and $\nu$ are the Young's modulus and Poisson's ratio 
of the cell, respectively, and $\epsilon_{ij}$ is the strain tensor.
If required, this constitutive law can be easily replaced by
a more complicated one, {\it{e.g.}}, the Neo-Hookean model
for non-linear materials.

We now outline how the active stress, $\sigma_{ij}$, can be related to
the kinetics of contractile, acto-myosin stress fibers \cite{deshpande_bio-chemo-mechanical_2006,deshpande_model_2007,pathak_simulation_2008,mcgarry_simulation_2009}.  Because
stress fibers are essentially one-dimensional objects, the
corresponding theory is a scalar one. At any position of the
two-dimensional cellular domain, one assumes a distribution 
of stress fibers to exist, which point in the direction
parametrized by the angle $\phi$. Next, one 
defines a direction-dependent activation level $\eta(\phi)$ for stress
fibers ($0 \le \eta \le 1$) 
where the time derivative of the activation level is determined by the following first-order
kinetics:
\begin{equation}
\dot \eta(\phi) = [1 - \eta(\phi)] \exp(-t/\theta) \frac{\bar{k}_f}{\theta}
- \left[ 1 - \frac{\sigma(\phi)}{\sigma_0(\phi)} \right] \eta(\phi) \frac{\bar{k}_b}{\theta}
\end{equation}
The first term describes stress fiber formation with a dimensionless
rate $\bar{k}_f$. In addition, the model assumes a temporal
decay that accounts for the finite time scale $\theta$ over which a
biochemical signal activates the formation  of stress fibers (for example the influx of
Calcium-ions or the effect of a contractile agent like LPA).  The
second term describes stress fiber dissociation with a dimensionless rate
$\bar{k}_b$. Because the formation term decays in time, the
system will, in principle, eventually relax to vanishing activation. However,
if the system is able to reach maximal stress $\sigma_0$, then
the decay does not take place. Because one typical starts
with the initial conditions $\eta = 0$ and $\sigma = 0$, the system
will develop an appreciable level of stress fiber activation only if
it is able to build up sufficient levels of stress in a certain
direction on the time scale of the decaying activation signal. In the
framework of the FEM, this will depend strongly on the mechanical
boundary conditions, thus making the cell model very sensitive to
external mechanical cues.  Motivated by models for muscle, the tension
$\sigma(\phi)$ in the stress fiber and the rate of strain $\dot
\epsilon(\phi)$ in the fiber are assumed to obey a Hill-like relation (compare section IIIC).
For a fiber of constant length, $\dot \epsilon = 0$, a constant
stall tension is assumed. As the velocity of fiber shortening increases,
$\dot \epsilon < 0$, the tension drops towards zero. For fiber
lengthening, $\dot \epsilon > 0$, the tension remains constant, at the level of the
stall tension. 

In order to connect the tensorial model for passive elasticity and the
scalar model for stress fibers, homogenization techniques are used to
construct the active stress tensor $\sigma_{ij}$ from the scalar
stress $\sigma(\phi)$ and the scalar rate of strain $\dot
\epsilon(\phi)$ from the strain tensor $\epsilon_{ij}$. In
\fig{FEM}(b) a typical outcome is shown for the simulations of a
square cell which adheres at its four corners. One clearly sees that
stress fibers develop in the diagonal directions and along the
boundaries, in agreement with experimental observations. However, the
model does not allow for the crossing of stress fibers in the cell
center due to the averaging procedure for the order parameter
field. Moreover there are clear differences between these simulations
and the contour models discussed above: the cell shape is much less
invaginated and the free boundaries are relatively flat, mainly
because the passive cell elasticity resists compression.

The active stress $\sigma_{ij}$ introduced in \eq{stressdeshpande} can
be implemented in FEM-software by using established routines for
thermal cooling. In general, the analogy between cellular
contractility and thermoelasticity is very instructive and has also
been used to evaluate stresses in cell monolayers. For example, it has
been shown that the proliferation pattern in cell monolayers on
patterned substrates correlates with the stress distribution in the
thin contractile layers \cite{nelson_emergent_2005}. An analytically
solvable thermoelastic model has been used to explain why stress and
strain are localized at the periphery of such monolayers
\cite{uss:edwa11a}. Combining such calculations and experiments, it
has been shown that for larger cell colonies, the traction pattern of
the monolayer is increasingly dominated by the tensional elements
\cite{mertz_scaling_2012}, and that these collective effects disappear
if cell-cell adhesion is disrupted in the monolayer
\cite{mertz_cadherin-based_2013}.

The FEM-model for single cells shows that tension in contractile cells
must be balanced by structural elements that can carry compressive
load, such as microtubules.  Indeed buckling of microtubules has been
observed in many different contexts within cells
\cite{brangwynne_microtubules_2006} and has been shown to occur for
pico-Newton forces in \textit{in vitro} assays
\cite{dogterom_measurement_1997}. This proves that microtubules are
indeed load-carrying elements in the cell. It has been suggested early
on that the balance between contraction in the actomyosin system and
compression of the microtubules is essential for the mechanical
stability of cells and implements an architectural principle known as
\textit{tensegrity}
\cite{ingber_cellular_1993,stamenovic_tensegrity-guided_2009}.  As
cells adhere to substrates, the contractile forces are then
increasingly balanced by sites of adhesion
\cite{stamenovic_cell_2002}.  Because microtubules alone would not be
able to carry the large load developed by adherent cells, the
establishment of large adhesions seems to be a necessary condition for
the development of contractility in adherent cells. In contrast to
FEM-models that couple continuum elasticity to discrete, actively
contractile stress fibers, the tensegrity models consider discrete
structural elements such as compression struts connected by tensed
cables, as a model of single stress fibers
\cite{luo_multi-modular_2008}. Because they model discrete elements
and do not require homogenization, tensegrity models can be compared
more directly with experiments, for example when cutting discrete
elements with lasers
\cite{kumar_viscoelastic_2006,luo_multi-modular_2008}.

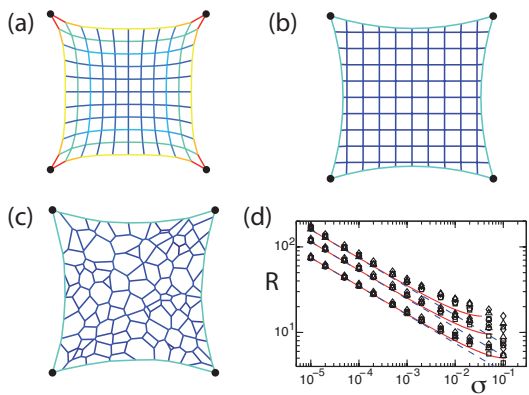
\FIG{Fig20}{fig:ActiveNetworks} {(Color online) (a) A spring network which is
  constrained to be located at the four adhesion points indicated (and
  in which the equilibrium spring length is smaller than the distance
  between neighboring adhesion points divided by the number of springs
  along that line), relaxes to a shape similar to the FEM-model --
  that is, one with relatively flat free edges. Note also the
  variation in force along the boundary indicated by the colors. (b)
  An actively contracting cable network results in circular arcs and
  shows hardly no variation in force along the boundary. (c) The
  results do not depend on network topology, as shown here for a
  disordered network. (d) Arc radius $R$ depends on network tension
  $\sigma$ as predicted by the tension-elasticity model. Adapted rom \textcite{uss:guth12}.}

Given the success of the contour models in explaining the appearance
of circular arcs of adherent cells, it is interesting to ask if
whole-cell models can predict the same shapes.  Recently it has been
suggested that this can indeed be achieved by modelling cells as
actively contracting cable networks
\cite{uss:bisc08a,uss:guth12}. This approach combines
elements from FEM and tensegrity models.  It considers the extreme
case in which contractile forces are balanced by the stretch response
of the cytoskeletal polymers and by forces from the elastic
environment that couple to the cell at the adhesion sites. The model
does not include any compressive response from within the cell or any
area conservation because it is assumed that cytoplasm is not
constrained to the 2d plane; the rest of the 3d cell therefore acts as
a reservoir for adhesion area. A 2d mechanical network is constructed
which consists of a set of nodes locally connected by mechanical
links.  If the links are taken to be linear springs, the network
propagates compression similar to the FEM-model. However, if the links
are taken to be cables, only tension and not compression is
propagated. This representation is appropriate to the polymeric nature
of the cytoskeleton, whose filaments tend to buckle or depolymerize
under compressive strain \cite{coughlin_prestressed_2003}. In
principle, contraction can be modeled by reducing the relaxed length
of the springs or cables. However, this does not represent the
properties of actin bundles contracted by myosin II minifilaments,
which do not have a well-defined reference state, but contract in a
Hill-type fashion until a certain stall force is reached. A simple way
to achieve this feature is to add a pair of constant forces (force dipole) to each
network link, thus creating permanent contraction between two
connected nodes. With these very simple prescriptions, actively
contracting cable networks can be simulated. In
\fig{ActiveNetworks}(a), we demonstrate that contracting spring
networks do not result in the circular arc morphology, but rather show
a flat contour as does the FEM-model.  Circular arcs appear for
actively contracting cable networks, independent of network topology,
compare \fig{ActiveNetworks}(b) and (c). In the second case of a
disordered network, a constant tension per length has been
assumed. Depending on the link density at the boundary, this
translates directly into an effective tension $\sigma$ that acts
within the network.  \fig{ActiveNetworks}(c) demonstrates that the arc
radius $R$ scales with network tension $\sigma$ as predicted by the
tension-elasticity model (TEM), see \eq{tem_simple}. Here different
symbols correspond to different network topologies and the three sets
of curves correspond to three different spanning distances $d$. Solid
lines correspond to the numerical solution of \eq{tem} while dashed
lines are the analytical results from \eq{tem_simple}. Deviations
between computer simulations and TEM occur only at very large
tensions.

The continuum mechanics approaches described here are especially
suited to investigate static situations relevant mature cell adhesion;
however, to treat cell migration, one must consider a dynamically
changing cell shape. A natural framework for this is hydrodynamics in
the overdamped limit, since cellular flows are characterized by very
small Reynolds numbers. For example, the shape of migrating
keratocytes has been investigated by a hydrodynamic model representing
the flux of newly polymerized actin networks
\cite{barnhart_adhesion-dependent_2011}. Here, the shear and
compressive forces in the viscoelastic fluid are balanced by forces
arising from myosin contractility and flow over adhesion
sites. Membrane tension enters as a boundary condition. A similar
approach is taken by active gel theory, which can be considered as a
hydrodynamic theory for polarized active gels
\cite{julicher_active_2007} that is based only on the symmetries and
conservation laws of the system and is independent of any particular
molecular model. Alternative approaches are level set (or phase field)
models \cite{shao_computational_2010,ziebert_model_2012} or models
incorporating discrete elements such as single focal adhesions and
stress fibers
\cite{shemesh_role_2009,shemesh_physical_2012,loosli_cytoskeleton_2010}.

\subsection{Force from shape}

The various elastic forces predicted from the shape models can be verified with experimental measurements,
{\it{e.g.}},  for cells on soft elastic substrates or on pillar arrays. The
simplest possible evaluation can be obtained by comparing measurements with the simple tension
model \cite{uss:bisc09a}. We now discuss the forces one
might expect for different adhesion geometries. Typical
micropatterns can be circular islands, rectangular islands,
islands with concave parts (such as U-, Y- and X-shapes)
or dot patterns. For a circular island of radius $R$,
both tensions pull in the same direction and the
boundary force per length is simply $\sigma + \lambda / R$
(compare the force balance given in \eq{LaplaceLaw}).
For a rectangular island, the inward force per unit
length is simply $\sigma$ along the flat parts of the perimeter,
because here contributions from the curvature and hence from the line tension $\lambda$ vanish. At the corners,
however, the situation is reversed. We can calculate
the corresponding force by approximating the 
corner by an arc with radius $\epsilon$ 
and then taking the limit of a sharp corner:
\begin{equation}
\vec F = \lim_{\epsilon\rightarrow0}\int_{-\frac{\varphi}{2}}^{\frac{\varphi}{2}}
(\sigma+\frac{\lambda}{\epsilon})\vec n(\theta) \epsilon d\theta
= 2 \lambda \cos{\left( \frac{\phi}{2} \right)} \vec n_{b}
\label{eq:corner}
\end{equation}
where $\phi$ is the opening angle, $\varphi = \pi-\phi$ and
$\vec n_{b}$ points in the direction of the bisecting line.
We thus see that the surface tension does not
contribute because it is associated with a line
element. The interpretation of \eq{corner} is
very simple: the force at the corner is simply
the vectorial sum of two forces of magnitude 
$\lambda$ pulling along the two incoming contour
lines. For very small opening angle $\phi$,
these two forces pull in the same direction and
one obtains the maximal value $2 \lambda$.

The same line of reasoning can now also be used to
predict forces for free contours, as they appear
on concave and dot patterns. Again the forces at
the adhesion sites directly depend only on the line
tension $\lambda$. However, now the surface tension
$\sigma$ enters indirectly as it determines the arc
shape and therefore the effective angle of the arc that
pulls on the contact. We consider three
neighboring adhesion sites where the two spanning
distances $d$ are identical and with an opening angle $\phi$. The
force can then be calculated to be: 
\begin{equation}
\vec F = 2 \lambda \left[ \beta \sin \left( \frac{\phi}{2} \right)
+ \sqrt{1-\beta^2} \cos \left( \frac{\phi}{2} \right) \right] \vec n_{b}
\label{eq:arc}
\end{equation}
where $\beta = \sigma d / 2 \lambda$ can be interpreted as a
dimensionless measure of  the strength of the inward pull or of the
dimensionless spanning distance $d$. In the limit
$\beta = 0$, the contour becomes straight and
we recover the result from \eq{corner} for
pinned straight edges. For finite values of $\beta$,
however, the edge is curved and the spanning distance $d$ and surface tension
$\sigma$ enter through the arc shape. The larger the spanning distance
$d$, the larger the surface tension $\sigma$ or the more
acute the opening angle $\phi$, the steeper the inward pull and the
closer the force comes to its maximal value $2 \lambda$.
At the critical parameter value $\beta_c = \sin(\phi/2)$,
the two arcs actually touch each other and pearling is
expected to occur as explained above.

These results suggest a simple procedure to estimate forces from
shape, see \fig{ForceFromShape}(b). Using pillar assays or
micropatterned elastic substrates, one could look for images of cells
in which two circular arcs meet at the same adhesion point. In this
case, the traction force at this adhesion point is the vectorial sum
of the two arc forces. The direction of each of these forces follows
from fitting a circle to the arc; the force magnitude is simply the
line tension $\lambda_i$. Because the same surface tension acts on
both arcs, from the Laplace law \eq{LaplaceLaw} we have $R_1/R_2 =
\lambda_1/\lambda_2$.  Therefore, one only needs to calibrate the
force at one adhesion to obtain the force of the others from
geometrical considerations. Applying this procedure to an experiment
with a pillar array (\fig{ForceFromShape}(b)) resulted in a value of
$\sigma =$ 2 nN/$\mu$m. This tension value is higher than the lysis
tension of lipid membranes and presumably corresponds to the actin
cortical tension generated by myosin motors. Interestingly, a very
similar value has been reported for the effective tension in a cell
monolayer \cite{mertz_scaling_2012}.  The procedure outlined here also
shows that forces measured on the substrate might be substantially
smaller than the forces that act within cells, because it is only the
vectorial sum of the internal force which is transmitted to the
substrate, as in the example treated here of the sum of two forces
from two adjacent arcs.

\section{Active response of cells}

\subsection{Mechanical response of force dipoles}

In this section we focus on active cell mechanics due to the presence
of acto-myosin force dipoles; since cellular contractility is due to
ATP-dependent conformational changes of myosin, these non-equilibrium
processes are denoted as active. The concept of force dipoles is
useful at multiple scales.  At the level of the entire cell, the
overall force balance in the cell suggests a coarse grained picture in
which a contractile cell is modeled as a pair of equal and opposite
forces (\textit{contraction force dipole}), as shown in
\fig{ForceBalance} and suggested experimentally
\cite{uss:schw02b}. Indeed each of the whole-cell models discussed in
section V suggest such an approach. The same argument also applies
within a cell to individual molecular force generators in the actin
cytoskeleton (CSK) of tissue cells, namely myosin II minifilaments,
see \fig{FocalAdhesion}. In either of these two scenarios, all
internally generated stresses must balance due to momentum
conservation and thus the force monopole term in a force dipolar
expansion is expected to vanish, leaving the force dipole term as the
leading contribution.  The overall force exerted by the cell depends
on the arrangement of its internal force dipoles which in turn is a
function of cell shape as discussed in section V.  Transmission of
these internal forces to the cellular environment occurs via the
adhesions discussed in section IV.

The stress generated by cellular force dipoles is balanced by the
elastic restoring force of the medium (the CSK in the case of
acto-myosin minifilaments together with the matrix/substrate or the
matrix/substrate alone in the case of a coarse-grained force dipole
that represents the entire cell).  This is illustrated by the cartoon
in Fig.~\ref{fig:CellSprings1} for the case of a cell in an elastic
medium (in Fig.~\ref{fig:ForceBalance}, we presented a more detailed
cartoon that included representations of the internal structure of the
cell; we now focus on the effective force dipole arising from this
configuration of forces). The situation depicted in
Fig.~\ref{fig:CellSprings1} has been analyzed with calculations
carried out at varying levels of detail {\it{e.g.}}, in
\textcite{uss:schw06a,uss:schw07a,mitrossilis_single-cell_2009,zemel_optimal_2010,walcott_mechanical_2010,marcq_rigidity_2011}.

In general, these studies have validated the following simple physical
picture \cite{uss:schw06a}. By measuring how much force or work it
requires to achieve a certain deformation, the cell can sense the
stiffness of its environment. Because cells themselves are soft
objects, the cartoon also shows that the cell tends to deform not only
the environment, but also itself. For two springs in series, the
inverse of the effective spring constant is the sum of the inverse
spring constants. Therefore, if the environment is very stiff, the
cell only deforms itself and cannot sense its surroundings. On the
other hand, if the environment is very soft, the cell can easily
deform it, but does not build up much force and therefore does not
gain much information; in particular, any positive feedback triggered
by mechanosensors will not work well. Therefore the best working or
set point for a cell seems to be a situation in which the two
stiffnesses of cell and environment are nearly matched. Indeed it was
found experimentally that cells tend to match their stiffness to that 
of the environment \cite{c:solo07}.

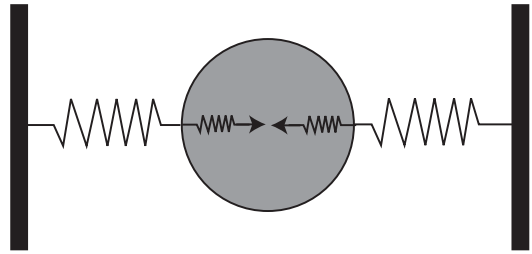
\FIG{Fig21}{fig:CellSprings1} {Two-spring model for
  cell-substrate interactions.  In addition to springs that
  characterize the cellular and substrate deformations, the cell
  exerts active, contractile forces shown by the double arrows. This
  simple cartoon shows that cells can measure the stiffness of their
  environment, which is, however, convoluted with their own
  stiffness. In contrast to \fig{ForceBalance}, here we do not
  consider compression of the substrate, because we consider only the
  far field.}

In this section, we discuss how active force generators such as myosin
II minifilaments in the actin cytoskeleton or entire contractile cells
in an elastic matrix interact with their mechanical environment.  The
deformations induced in the matrix by cell activity allow us to deduce
effective interactions between the force dipoles themselves via their
mutual effects on the elastic environment.  In the case of individual
acto-myosin minifilaments modeled as force dipoles, the elastic
environment includes the CSK of the cell itself (in addition to its
surroundings), while in the case of entire, contractile cells modeled
as force dipoles, the elastic environment is the matrix or substrate.
The elastic properties of the environment can be easily controlled in
experiments so that predictions and measurements of their role in
modulating cellular force dipole assemblies provide insight into these
fundamental processes.

In order to obtain analytical insight, linear elasticity
theory is applied to an isotropic medium (or substrate) to model the
mechanical properties of cells and their environment
\cite{b:land70}, although later on, we briefly comment on
possible extensions to more detailed models for the elasticity of
biomaterials, including non-linear elasticity. Alternatively one could
also employ more microscopic models for the propagation of stress and
strain in polymer networks
\cite{head_deformation_2003,head_distinct_2003,wilhelm_elasticity_2003,heussinger_stiff_2006,heussinger_nonaffine_2007},
but this may preclude the insight gained by the use of elastic theory
generalized to include force dipoles. In the framework employed here,
cytoskeletal force generators ({\it{e.g.}}, acto-myosin minifilaments or
entire cells) are viewed as active, elastic inclusions in (or on) a
homogeneous and isotropic medium.  This suggests the use of Eshelby's
theory of elastic inclusions, that was originally developed in the
context of materials science \cite{eshelby_1957, eshelby_1959}.  In
the limit that the inclusions are much smaller than the length scales
of interest, one can use the theory of elastic point defects, which
also originated in materials science, and has been used extensively to
model the elastic interactions of hydrogen in metal
\cite{e:siem68,e:wagn74}.

There are a few important assumptions made in the following treatment
of force dipoles and cellular/matrix elasticity: (i) The build-up of
force occurs on short enough time scales (10s of seconds) for which
the CSK responds elastically to internal forces and does not flow.
This was discussed previously in a comparison of elastic versus
flowing gel models of the CSK. Even if the matrix flows on larger time
scales, the cell will react on a shorter time scale by building up new
forces via a remodeled CSK. Elastic relaxation after pharmacological
inhibition of actomyosin contractility, laser cutting or disruption of
adhesion (e.g. by trypsination) proves that adherent cells and matrix
are continuously under elastic stress. (ii) The matrix-induced forces
that act to organize the cytoskeleton arise from the potential energy
that accounts for the deformation of the matrix by the dipoles.  In
this sense, matrix can refer either to those parts of the CSK not
included in the acto-myosin dipoles themselves and/or to the
surrounding matrix or substrate.  However, these forces are taken to
determine only the organization of the dipoles ({\it{e.g.}}, nematic
or smectic order) but not their existence or magnitude which depends
in a more complex manner on non-equilibrium cellular activity. These
same forces would be present and act on artificial force dipoles were
they present in the CSK or matrix/substrate. (iii) The time scale on
which elastic signals are propagated in the CSK or matrix/substrate is
faster than internal relaxation times due to dissipation arising from
internal viscosity or fluid flow; for experimental measurements of
these time scales see Fig. 3 of \textcite{fabryrev}.  This allows us
to predict force dipole organization by elastic forces from energetic
or force balance arguments.  We recognize that this is a crude
approximation that must be tested under various circumstances; a
dynamical theory of elastic signal propagation in the CSK or
matrix/substrate is a topic of current research.  (iv) The theories
below focus on cytoskeletal force generation and predict dipole
arrangements for a fixed cell shape.  Experiments that measure the
response of cells to time varying stresses show
\cite{faust_cyclic_2011} that cell shape {\it{follows}} cytoskeletal
reorientation by several hours.

\subsection{Force dipoles and their interactions}

A simplified model of a contractile actomyosin unit (or in
a coarse gained picture, an entire, polarized cell) is that of a
source of two equal and opposite forces, $\vec F$, separated by a
nanoscale (or, for cells, micrometer scale) distance, $\vec d$.  By
analogy with electrostatics, these are termed \textit{force dipoles}.
The dyadic product of the force and the distance defines a local
elastic dipole: $P_{ij}= d_i F_j$.  In contrast to electric dipoles
that are vectors given by product of the scalar charge and the
distance, the elastic dipole is a tensor. In a continuum
representation, valid for scales much larger than $|d|$, one therefore
considers a coarse-grained force density $f(\vec r)$  (which is not the same as the local force $\vec F$)
whose average in some small
volume vanishes (since the forces are equal and opposite) but whose
first moment with $\vec r$ is finite. The force density is
written as the sum of two, localized force distributions with opposite
signs whose centers are separated by a distance $\vec d$.  One expands
these distributions for small values of $\vec d$ relative to the
distance $\vec r$ at which the strains and stresses are measured and
finds that the net force is related to derivatives of the localized
force distributions. Using this approximation for the local force
density and defining the local dipole tensor density, $p_{ij}$ (local
force dipole tensor per unit volume), one can show that:
\begin{equation}
\partial p_{ij}(\vec r)/\partial r_j=-f_i(\vec r)
\label{eq:stress3}
\end{equation}
Using this relationship between the force density and the
divergence of the dipole density in Eq.~\ref{eq:strain4} and
performing an integration by  parts (with the assumption that
the surface terms vanish or are accounted for explicitly), shows
that the strain is related to the dipole density by:
\begin{equation}
u_{ij}(\vec r)=\int d \vec r \, ' \ G_{il,k'j}(\vec r, \vec r \, ') p_{kl}(\vec r \, ')
\label{eq:strain5GG}
\end{equation}
where the two indices in $G$ after the comma indicate derivatives
with respect to $r \, '_k$ and $r_j$ respectively.
Eqs.~\ref{eq:stress2},  \ref{eq:stress3}  and
\ref{eq:strain5GG} and further partial integrations demonstrate
that the deformation energy of the medium acted upon by
localized elastic dipoles can be written as an effective
interaction of those dipoles.   This is important since if the
dipoles are free to arrange themselves to minimize the
deformation energy ({\it{e.g.}}, if the cell activity tends to
minimize the energy expended in deforming the CSK or the
matrix), their spatial arrangement can be deduced from    their
interactions, perhaps also accounting for noise which can in
some cases be  modeled as an effective temperature
as discussed in the section on physics background.

For translationally invariant systems where the Green's function
depends on the difference $\vec r - \vec r'$, the total
deformation energy of the medium is:
\begin{equation}
F_e=\frac{1}{2} \int d \vec r  d \vec r \, ' \, p_{ij}(\vec r) \, G_{il,k'j}(\vec r-\vec r \, ') \,
p_{kl}(\vec r \, ')
\label{eq:elasticenergy4}
\end{equation}
In this case, one can also write the interaction energy in
terms of the Fourier transforms of the dipole density and the
Greens' function:
\begin{equation}
F_e=\frac{1}{2} \int d \vec q    \, p_{ij}(\vec q) \,
G_{il,kj}(\vec q) \, p_{kl}(-\vec q)
\label{eq:elasticenergy4ft} \end{equation}
where
$G_{il,kj}(\vec q)= G_0 (q_k q_j/q^2)
\left( \delta_{il} + q_i q_l/(2q^2(1-\nu)) \right)$
follows from the exact (\textit{Kelvin}) solution for a full elastic space
and $G_0$ is a constant proportional to $1/E$.
In the presence of an externally imposed
strain, $u^e_{ij}(\vec r)$ (in addition to the strains induced
by the internal force dipoles), one can use Eq.~\ref{eq:strain2}
to show \cite{uss:bisc04a} that the interaction energy of the
dipoles with the external strain is:
\begin{equation}
F_e= \int d \vec r  \, p_{ij}(\vec r) \,  u^e_{ij}(\vec r)
\label{eq:elasticenergy5}
\end{equation}

The theory described here is relatively simple to use for the case of
dipoles in an infinite medium. It can be applied to an entire
contractile cell that is placed in one region of a much larger elastic
environment (as a model of an infinite medium).  However, when the
force dipole concept is used to account for the interactions of
acto-myosin minifilaments within one cell which itself is an elastic
medium, the situation is more complex.  The cell itself can be
situated in (or on) an elastic matrix (substrate) whose rigidity can
be different from that of the cell.  The resulting Green's function
will then depend on the cell shape and both the cell and matrix
elastic constants.  The same complications can also arise in the case
of entire contractile cells modeled as effective force dipoles
\cite{uss:schw02a, uss:bisc04a, uss:bisc03a,zemel_active_2006} that
are placed in an elastic matrix which itself is surrounded by another
material with different elastic properties.  This may be applicable to
models of cells in tissues that are contained within the extracellular
matrix that itself is coupled to another elastic medium in which the
tissue resides. In this case as well, it is necessary to consider the
shape and elastic boundary conditions in order to predict the
interactions among the cells that lead to their self-assembled
structures \cite{zemel_2007}.

The real-space solution of the elastic deformation of such a composite
medium was considered by Eshelby \cite{eshelby_1957, eshelby_1959} who
calculated the strain inside of an ellipsoidal inclusion embedded in a
three-dimensional elastic matrix.  Although these inclusions do not
actively produce forces, they can exert stresses on their surroundings
as the temperature is changed and the inclusion thermally expands or
contracts in a manner that is larger or smaller than that of the
matrix.  The Eshelby results can be mapped to the strain experienced
either by an entire contractile cell modeled in a coarse grained
manner or by individual acto-myosin force dipoles \cite{zemel_2007,
  zemel_optimal_2010} within a cell embedded in a three-dimensional
matrix.  Once the local strain due to a given dipole distribution is
known, the dipole interactions are given by the product of the local
dipole density and the local strain.  Later in this review, we
consider specific predictions for the orientational and spatial
organization of force dipoles in cellular systems and summarize the
results obtained from Eshelby theory.  However, this theory with its
focus on the real-space boundary conditions is complex
\cite{eshelby_1957, eshelby_1959, mura_1991} and is also specific to
systems of ellipsoidal inclusions whose dimensionality is the same as
that of the surrounding matrix.  This is not quite the case studied
experimentally where well-spread (nearly two-dimensional) cells are
plated on semi-infinite, elastic substrates, although recent
experiments \cite{florian_2012} show that the behavior of thick
substrates and of 3d surroundings are very similar.  Thus, to provide
a simpler and more intuitive theory of deformation induced
interactions of elastic dipoles that are relevant to cells on
semi-infinite substrates we review here a simplified model
\cite{friedrich_how_2012} that can be solved using Fourier methods.

\subsection{Elastic model for cells on a substrate}

\FIG{Fig22}{fig:cellsubstrate1}
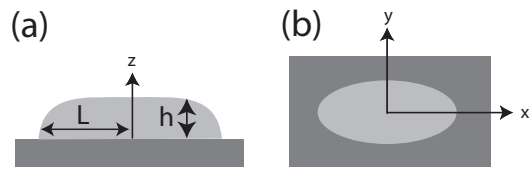
{Simple model of a contractile and widely spread cell on an semi-infinite elastic substrate.}

The particular scenario that we focus upon here is that of acto-myosin
minifibrils modeled as force dipoles within a cell that is spread and adhered to an
elastic substrate.  The theory predicts the conditions under which the
elastic interactions of these dipoles with both the cellular CSK and
the substrate mediate dipole-dipole interactions that tend to orient
the minifilaments.  In this simplified model, the spread cell adheres
to the upper free surface of the substrate, which is taken to lie in
the $z=0$ plane, of a semi-infinite ($z<0$) substrate, as in
Fig. \ref{fig:cellsubstrate1} \cite{friedrich_how_2012}.  The thickness
of the cell may vary as a function of position and is denoted by $h
S(x,y)$. Here, $h$ denotes the height of the contractile region of the cell, while the function
$S(x,y)$ specifies variations of cell thickness within the cell, and
is zero outside the cell. Thus, the dimensionless function $S(x,y)$
characterizes the shape of the cell. The integral $\int
dx\,dy\,S(x,y)=A$ defines a weighted cell area $A$ (which is related
to the volume of the cell by $A=V/h$) and a characteristic
length-scale $L=A^{1/2}$ of the spread cell. For simplicity, the
discussion is restricted to cell shapes that are symmetric with
respect to each of the $x$- and the $y$-axis. In this case, the
Fourier transform $S(\vec q)=(2\pi)^{-1} \int dxdy\,S(x,y)\,\exp(-i
q_x x-i q_y y)$ of the shape function is real and has typical
dimensions $A$ or $L^2$.  There are three important moments of the
cell shape function defined by:
\begin{equation}
\label{eq_moments}
\mathcal{J}_n=\frac{1}{L} \int\! d^2\vec q\, q\, |S(\vec q)|^2 \exp ( 2 i n w(\vec q) )
\end{equation}
where $w(\vec q)=\tan^{-1}(q_y/q_x)$ and $n=0,1,2$.
These moments are dimensionless and depend only on the shape of the cell.
They play a role similar to that of the shape-dependent
depolarization  factors in electrostatics \cite{bellegia_2006}
or the Eshelby tensor in Eshelby's theory of elastic inclusions
\cite{eshelby_1957, mura_1991, bellegia_2006}.
For cells with
mirror-symmetry (considered here for simplicity),  the
moments $\mathcal{J}_n$ are real. $\mathcal{J}_1$ characterizes
cell shape anisotropy and is zero for radially symmetric cells.

The discrete actomyosin contractile elements are
characterized, in a coarse-grained  description, by
a bulk force dipole density $p_{ij}(\vec r)$ with units of energy per
unit volume.
A mean-field approximation
regards these dipoles as being uniformly distributed within
the cell with a mean force-dipole density $\ol{p}_{ij}$.
The
upper surface of the cell is stress-free, and for a cell
whose thickness is much smaller than its lateral extent,
the stresses and force dipoles with components in the
$z$-direction can be neglected (for details see
\textcite{friedrich_how_2012}).
To highlight the symmetries of the
problem, one can decompose the mean dipole density tensor
$\ol{p}_{ij}$ into an isotropic part that is analogous to a
hydrostatic pressure, $\bar P_0=\ol{p}_{xx}+\ol{p}_{yy}$, and
two invariants that characterize pure shear stresses, $\bar
P_1=\ol{p}_{xx}-\ol{p}_{yy}$ and $\bar P_2=\ol{p}_{xy}$. In
matrix notation, $\ol{p}_{ij} = \bar P_0 \E_0 + \bar P_1 \E_1 +
\bar P_2 \E_2$ with respect to a convenient basis of the space
of symmetric rank-2 tensors,
\begin{equation}
\E_0=
\frac{1}{2}
\begin{pmatrix}
1 & 0 \\
0 & 1\\
\end{pmatrix},\quad
\E_1=
\frac{1}{2}
\begin{pmatrix}
1 & 0 \\
0 & -1\\
\end{pmatrix},\quad
\E_2=\begin{pmatrix}
0 & 1 \\
1 & 0\\
\end{pmatrix}
\end{equation}
By virtue of the Stokes theorem, each force dipole density
$\bar P_k E_k$ is equivalent to a set of surface forces $f_k=\bar P_k E_k \cdot \vec n$
that act at the boundaries of a cellular volume element.

\textit{Cellular strain for soft substrates: }
The active dipolar stresses contract and elastically deform the
cytoskeleton.
For a spread cell whose thickness is much smaller  than its
extent, $h\ll L$, the local displacement, $\vec u(x,y)$ has only
$x$ and $y$ components.
The corresponding strain matrix $u_{ij}=(\partial_i u_j + \partial_j u_i)/2$ can be written as the
superposition of a homogeneous dilation $U_0 \E_0$
where $U_{0}=u_{xx}+u_{yy}$ is the trace of the strain matrix,
and the traceless strain matrix $u_{ij}-U_0 \E_0=U_1 \E_1 + U_2 \E_2$,
which characterizes pure shear strain without area change
with two measures of shear strain
$U_1=u_{xx}-u_{yy}$ and
$U_2=u_{xy}$.
The geometry of the problem implies that
$U_1 \E_1$ and $U_2 \E_2$ are symmetric and anti-symmetric with
respect to a reflection about a coordinate axis, respectively.
The elastic deformation energy of the cellular domain is thus
written:
\begin{equation}
\label{cell1}
F_c=h \int dx dy  \, S(x,y)  \left[
\frac{K_c}{2} U_0^2 +
\frac{\mu_c}{2} \left(  U_1^2 +  4 U_2^2 \right) .
\right]
\end{equation}
If the cell were not  coupled to the substrate (or if the substrate had vanishing rigidity),
the cell would not be subject to restoring forces from the substrate
and the cell boundary would be stress-free.
In this case, the only source of cellular elastic stress
$\sigma^{(c)}_{ij}$ would be the forces exerted by the dipoles
and thus $\sigma^{(c)}_{ij}=p_{ij}$.
Alternatively, one can solve for the resulting minimal strain by
minimizing a Legendre transform that includes the work done by the dipole (so that one now can minimize the transformed free energy with respect to the strains for a given ? but arbitrary -- dipole arrangement), $G=F_c+F_d$ of the free energy
 where
\begin{equation}
F_d=- h \int\! dxdy\, S(x,y) \left( U_0 \bar P_0 + U_1 \bar P_1 + 4 U_2 \bar P_2 \right)/2.
\end{equation}
The convexity of the free energy dictates that cellular strain is
constant throughout the cellular domain
and moreover is independent of cell shape since the cell boundaries are stress-free so that
$U_i \equiv \bar U_i = \bar P_i / (2 B_i)$
where $i=1,2,3$ and $\vec B_i=(K_c,\mu_c,\mu_c)$.
Note that, irrespective of any anisotropy of cell shape,
the strain components depend only on force dipole components of the same symmetry type.
Actomyosin production that
controls the strength of microscopic, contractile force dipoles, but not their orientation,
induces only an isotropic (negative) ``hydrostatic pressure'' $\bar P_0 \E_0$
and thus a homogenous dilation $\bar U_0 \E_0$, but no shear.
Thus, in the absence of other factors that break the system symmetry,
a ``floating cell'', decoupled from its substrate,
does not feel anisotropic mechanical guidance cues,
which could drive nematic ordering of force dipoles ({\it{e.g.}}, alignment along one of the cellular axes).
The conclusion is that in the limit of a substrate with zero stiffness,
the actomyosin network will remain symmetric,
not withstanding the fact that the cell shape may be asymmetric.
This situation changes fundamentally, once
the elastic deformations of the cell and the substrate are coupled.

Due to the coupling of the CSK forces to the focal adhesions, active cell
contractility induces substrate deformations, $\vec v(x,y,z)$. The substrate surface
strain at $z=0$ is decomposed into its symmetry components
$v_{ij}(x,y,z=0) = V_0 \E_0 + V_1 \E_1 + V_2 \E_2$.
For simplicity \cite{friedrich_how_2012} one can take the average strain inside the cell to be equal to the
average substrate strain underneath the cell:
$\ol{V}_k =\ol{U}_k$.

\textit{Substrate elastic energy: }
The substrate deformations induced by its coupling to the cell can now be included.
The elastic energy of the substrate is written in terms
of the Fourier transforms of the substrate displacements $\vec v(x,y,z)$
derived in  \textcite{Alice06}. 
This energy is proportional to $\mu_m$, the shear modulus of the substrate,
which is for simplicity taken to be incompressible.
One next expresses the coupling condition in Fourier space and
determines the strains by minimizing \cite{friedrich_how_2012} the
Legendre transform $G=F_c + F_m + F_d$ of the free energy of both the
cell and substrate subject to the coupling condition
$\ol{V}_k =\ol{U}_k$.  The cellular strain components, $U_i$ are then
found as a function of the dipole components $P_i$: $U_i= A_{ij} P_j$.
The coupling coefficients $A_{ij}$ are functions of the cell-shape
moments (Eq. \ref{eq_moments}) and of the cell and substrate elastic
moduli \cite{friedrich_how_2012}.  

Of particular interest is the fact that
there are off-diagonal terms in this relationship.   This means,
for example, that a homogeneous and isotropic dipole distribution
characterized by a non-zero value of $P_0$ can induce a shear
strain such as $U_1$. The coefficient that quantifies  this
symmetry breaking, $\mathcal{A}_{01}$  depends on both the cell
shape and matrix rigidity. In particular, it is proportional to
$\mathcal{J}_1$ (which vanishes for cells with circular cross
sections) and vanishes when the substrate modulus is either very
small or very large.

The elastic energy is the product of the local strain
and local force dipole density and can be written in terms
of effective interactions of the force dipoles:
\begin{equation}
\label{eq:freequadp}
F_i = (L^2h/4) \left( \mathcal{A}_{00} \bar P_0^2 +
2\mathcal{A}_{01} \bar P_0 \bar P_1 + \mathcal{A}_{11} \bar
P_1^2 + 4 \mathcal{A}_{22} \bar P_2^2 \right). \end{equation}
The terms proportional to $P_0^2$ and $P_1^2$ are
the ``self-energies'' of the isotropic and nematic components
of the force dipole respectively; they represent
the energy to deform the cell itself together with the elastic
substrate \cite{fernandez_compaction_2009}.
 In addition, the  shear stress induced by the isotropic
dipole component couples
the $\bar P_1$ nematic dipole component to the isotropic
contractility, $\bar P_0$, for cell shapes that are anisotropic
for which $\mathcal{J}_1$ is non-zero. This implies that the effects of cell
shape on elastic interactions can induce nematic order of the cellular force dipoles even if
local cell activity results only in isotropic contractility.

The coupling of the isotropic component of the dipole tensor to the cellular shear allows
the cytoskeletal shear to present mechanical guidance cues for the polarization and alignment of
cytoskeletal structures and eventually of cellular traction
forces. Initially isotropic cytoskeletal contractility can
result from a local regulation of myosin activity
within the cell, that may tune $\bar P_0$ to a set
value $\bar P_0^\ast$. Considering $\bar P_1$ as an effective
degree of freedom, the total elastic energy of cell and
substrate, Eq.~\ref{eq:freequadp}, is minimized when
$\bar P_1$ is non-zero, which corresponds to anisotropic cellular
contractility.

Physical insight into the coupling of symmetry modes can be obtained
considering the limiting cases of very soft and very stiff
substrates. Since the strain propagates
\cite{friedrich_how_2012,banerjee_contractile_2012} into the substrate
a distance of order of the cell extent, $L$, (but only a distance of
order $h$ -- the cell thickness -- within the cell) the effective
Young's modulus of the substrate is given by the product of its Young's modulus $E_m$ and
a factor of $L/h$ where $h$ is the cell thickness: $\Em=E_m L/h$. This
predicts that both the stiffness ratio as well as the cell geometry
(height and lateral extent) will determine cytoskeletal organization
and ordering, which can be tested by changing both substrate rigidity
and cell volume \cite{ming_2012}.

In the limit of a very soft substrate, $\Em \ll E_c$, the
situation is that of an isolated cell; cellular strain
components couple only to force dipole components of the same
symmetry type. Cell activity that results in locally isotropic
contractile dipoles cannot give rise to nematic order. In the
limit of a very stiff substrate, the cellular strain scales as
$1/\Em$. For isotropic, cellular contractility with only $\bar
P_0\neq 0$: $\bar U_0 = \mathcal{J}_0 \bar P_0 / 8 \Em$, $\bar
U_1 = \mathcal{J}_1 \bar P_0 / 8 \Em$, and $\bar U_2=0$. Thus,
while the symmetric part of the cell contractility, $P_0$, does
induce shear strain due to the shape anisotropy, this shear
strain attenuates as $\Em\rightarrow \infty$. Similar
conclusions are reached for the induction of nematic order
(non-zero values of $\bar  P_1$) by locally isotropic
contractility. If, however, substrate stiffness and cellular
stiffness match, $\Em\sim K_c,\mu_c$, the symmetric shear
component $\bar U_1$ is induced by the symmetric dipole
component: $ \bar U_1 \sim \mathcal{J}_1 \bar P_0 . $
This shear
is a result of anisotropic, cell-shape dependent, elastic
restoring forces from the substrate. For an asymmetric cell
shape that is elongated in the direction of the $x$-axis,
$\mathcal{J}_1<0$ and  isotropic contractility with $\bar P_0<0$
induces cellular shear strain $\bar U_1>0$, which is expansive in
the $x$-direction (and compressive along the $y$-direction). This
causes nematic ordering of the dipoles themselves along the
$x$-axis, characterized by negative values of $\bar P_1$.

\subsection{Cell polarization guided by substrate rigidity}

These intuitive results were used in a more formal theoretical model
to predict that cells with anisotropic shapes on substrates of
intermediate rigidity will spontaneously show CSK nematic order
\cite{zemel_optimal_2010,friedrich_nematic_2011}.  A phenomenological model
couched in terms of active gel theory was presented in
\textcite{banerjee_substrate_2011}.  The case of stem cells is
particularly applicable since at early times, the CSK is not yet well
formed and oriented and one can study the genesis of CSK formation and
orientation starting with relatively short actomyosin minifilaments
that can indeed be modeled as force dipoles contained within the cell.
The paper by \cite{zemel_optimal_2010} uses the real-space Eshelby
formalism \cite{eshelby_1957, eshelby_1959} for an ellipsoidal
inclusion to calculate the real space strains inside a cell contained
in an elastic medium of the same dimensionality (either 2d or 3d). The
shear strains induced by the medium are non-zero for cells that are
not circular (2d) or spherical (3d) and are predicted to give rise to
orientational order of the internal force dipoles (short actomyosin
minifilaments). The nematic order, as expressed by $\bar P_1$ is
related to the shear strain by a susceptibility whose form in the
limit of large noise (expressed as an effective temperature) was
discussed in \textcite{zemel_active_2006}. The work of
\textcite{friedrich_nematic_2011} provided a more general statistical mechanical
basis for the nematic ordering. An energy similar to
Eq. \ref{eq:freequadp} was obtained using Eshelby theory; this was
used as a Hamiltonian in a Maier-Saupe \cite{maier_1959} theory as
described above in Eq. \ref{eq:lcham1}.  This self-consistently
predicts the nematic order parameter, $\bar P_1$ as a function of the
cell and matrix elastic constants, cell shape, and the noise (modeled
as an effective temperature). In both models, if the noise is
moderately large, the nematic order of the CSK is maximal in some
optimal range of substrate rigidity and is small for very small or
very large rigidities.

The dependence on the boundary conditions ({\it{i.e.}}, the
global cell shape and substrate rigidity) highlights the
importance of the long-range elastic interactions, in contrast
to the general situation for nematic ordering in molecular
systems where the interactions are short range. The Mair-Saupe
theory predicts that for small noise (or large values of $\bar
P_0$), the nematic order may increase monotonically as a function
of substrate rigidity due to the increasing importance of
short-range interactions such as the excluded volume of the
dipoles themselves.

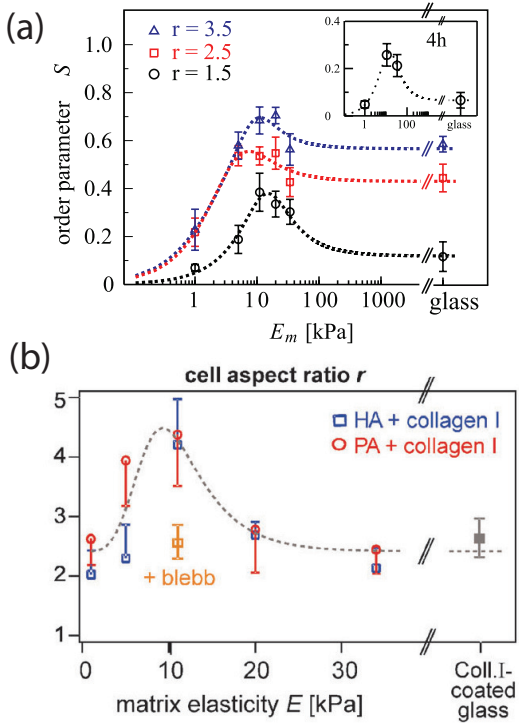
\FIG{Fig23}{fig:discher1}{(Color online) (a) The experimental values of the
  stress-fiber order parameter, $S \sim \bar P_1$ (which indicates the
  extent to which the nascent stress fibers are aligned along the long
  axis of the cell) for three groups of stem cells (of aspect ratios
  1.5, 2.5, 3.5) as a function of the Young's modulus of the matrix,
  $E_m$.  For the smallest aspect ratio, the order is clearly maximal
  for $E_m \sim 11{\rm{kPa}}$; for the other aspect ratios this trend
  may also be obeyed although it is less clear. The fit is motivated
  by the theory described in the text, see
  \textcite{zemel_optimal_2010}. From
  \textcite{zemel_optimal_2010}. (b) Aspect ratio of stem cells as a
  function of substrate elasticity.  The point marked blebb refers to
  cells where myosin activity has been suppressed by treatment with
  the drug blebbistatin; this shows that the peak observed for
  untreated cells is related to cell contractility. From
  \textcite{florian_2012}.}

Experiments were carried out \cite{zemel_optimal_2010} to
systematically analyse the alignment of stress fibers in human
mesenchymal stem cells as a function of the cell shape and the
rigidity of the environment.  Cells were cultured on substrates of
varying stiffness and sorted by their aspect ratio.  A quantitative
analysis of stress-fiber polarization in cells was obtained by
staining for both actin and non-muscle myosin IIa and applying a
segmentation algorithm to map their spatial organization in the cell.
Both the magnitude of the dipoles, as measured by the number of
actomyosin minifilaments \cite{zemel_2010b} and their orientation were
measured.  The results are shown in Fig.~\ref{fig:discher1}.  They
suggest a generic mechanical coupling between the cell shape, the
rigidity of the surroundings and the organization of stress fiber in
the cytoskeleton of stem cells, again pointing to the role of
long-range interactions.  This identifies a mechanical property of
cells -- stress-fiber polarization -- that is maximized at an optimal
substrate rigidity, analogous to the optimal rigidity found in
stem-cell differentiation (for example, to muscle cells)
\cite{engler_matrix_2006}.  The fact that the CSK is maximally
polarized for substrate rigidities of about 10kPa may help explain why
stem-cell differentiation into muscle cells occurs optimally in this
same rigidity range.  Stem cells on such substrates are muscle-like in
their CSK structure and the resulting contractile forces; the latter
may play a role in nuclear deformations resulting in gene expression
that is muscle-like for precisely such contractile cells.  We note
that the time scales observed for stress fiber development
\cite{zemel_optimal_2010} and orientation (1-24 hours) and for the
genetic changes in the cell \cite{engler_matrix_2006} (about a week)
are quite different. In addition, recent experiments have shown that
differentiation may also depend on the mechanics of the ligand
molecule and not only on the bulk rigidity of the substrate
\cite{trappmann_extracellular-matrix_2012}.

The ordering of the cytoskeleton on substrates of different rigidities
eventually affects the overall mechanical response of the cell.
Recent studies \cite{janmey_mechanisms_2011} have shown that cell
cortical stiffness increases as a function of both substrate stiffness
and spread area.  For soft substrates, the influence of substrate
stiffness on cell cortical stiffness is more prominent than that of
cell shape, since increasing adherent area does not lead to cell
stiffening.  On the other hand, for cells constrained to a small area,
cell shape effects are more dominant than substrate stiffness, since
increasing substrate stiffness no longer affects cell stiffness.

\subsection{Single cell response to rigidity gradients}

The previous section looked ``inside the cell'' and considered
the elastic interactions and orientational ordering of short
actomyosin minifilaments modeled as force dipoles that are internal
to the cell. This is relevant to stem cell development at
relatively early times (1-24 hours  in which the stress fibers
do not yet span the entire cell). Mature cells such as fibroblasts or
muscle cells have long and well-ordered stress fibers
\cite{hotulainen_stress_2006,thery_cell_2006} and in some cases, the cell
can be represented by a single, anisotropic force dipole
\cite{uss:schw02a, uss:bisc03a,
uss:bisc04a, pompe_dissecting_2009}.
In the following we discuss the response of an entire, polarized cell
(modeled in a coarse-grained approximation
as a single, anisotropic force dipole) to rigidity gradients 
and in the next section, its response to dynamically applied
stress.

Cell spreading, alignment and locomotion are controlled by both
biochemical activity within the cell as well as by the rigidity of the
substrate on which the cell is plated \cite{c:pelh97, Lo00,
  discher_tissue_2005, Saez07, c:solo07, wong_2009}.  In general,
these activities are enhanced on more rigid substrates.  In addition,
the forces that even static cells exert on substrates have been shown
to increase with substrate rigidity \cite{choquet_extracellular_1997,
  c:yeun05, saez_is_2005, zemel_2010b}.  Moreover, very recent studies
indicate that substrate viscoelasticity also plays a role in stem cell
morphology and proliferation \cite{justin_2011}.  Active gel theory
has been used to model isotropic rigidity sensing in
\textcite{marcq_rigidity_2011}.  Here we review how models of
contractility based on force dipoles \cite{uss:bisc04a, Nicolas06,zemel_2010b} can provide insight into these
observations.

Contractile cells are pre-programmed to exert force on their
surroundings. It has been argued that the experimental observation
that most cells types prefer stiff over soft substrates can be
described by the assumption that cells effectively minimize the
elastic energy invested in deforming the matrix \cite{Nicolas06,
  uss:bisc03a, uss:bisc04a}. Neural cells, that prefer soft over stiff
substrates, are exceptions to this rule of thumb \cite{
  janmey_hard_2009}.  This minimization can be the result of evolution
in producing optimized biological systems \cite{tlusty_2010}, since
the energy that the cell invests in deforming its surrounding is not
directly useful to the cell.  Alternatively, one can think of this
approach as a convenient framework for analytical
progress. Considering the cell as a uniform distribution of dipoles
(that can either be ordered or random in their orientation), one can
use Eq.~\ref{eq:freequadp}.  The dipole densities $P_0$ and $P_1$ are
proportional to the number of dipoles and the cell volume while for
fairly rigid substrates $\mathcal A_{00}$ and $\mathcal A_{11}$ scale
inversely with $\bar \mu_m =\mu_m L/h$.
Assuming that the cell optimizes its activity to avoid
investing energy in substrate deformations, this predicts that cells
will favor and spread optimally on rigid substrates.

This tendency of the cell to prefer rigid substrates is particularly
important for cells on substrates with rigidity gradients or boundary
regions \cite{Allioux-guerin08, wong_2009, Lo00,ladoux_physically_2012}. The limiting cases
of a cell on a substrate with a given rigidity near a boundary of a
substrate with a much larger or smaller rigidity, can be understood by
optimizing the deformation energy of a single force dipole in a medium
with either clamped or free boundaries respectively. The corresponding
elastic problem takes into account these boundary conditions using the
technique of ``image dipoles'' \cite{uss:bisc04a}. As shown there, the
preferred cell orientation close to the surface, as predicted by the
configurations of minimal deformation energy, are parallel and
perpendicular to the boundary line for free and clamped boundaries,
respectively.

This leads to the prediction that cells preferentially locomote
towards a clamped boundary, but tend to migrate away from a free
boundary.  One may think of a clamped (free) surface as the interface
between the substrate on which the cell is placed and an imaginary
medium of infinite (vanishing) rigidity, which effectively rigidifies
(softens) the boundary region.  Thus for clamped (free) boundary
conditions, the cell senses maximal stiffness in the direction normal
to (parallel to) the boundary line.  The cell exerts force on the more
rigid medium.  For free boundaries, the substrate is more rigid and
the cell orients parallel to the boundary to maximize the deformation
of the substrate.  Near a clamped boundary, there is less deformation
if the cell orients perpendicular to the boundary line. Indeed such
behavior has been observed experimentally, {\it{e.g.}}, for cells
close to the boundary between soft and rigid regions of a soft
substrate \cite{Lo00}. The tendency to migrate towards stiffer regions
has been termed \textit{durotaxis}. On a more microscopic level, this
can be understood from the preferred growth of focal adhesions on more
rigid substrates (see section IV).

\subsection{Dynamical response of cells to mechanical stress}

Cells in tissues respond to a variety of mechanical forces
that influence their behavior and alignment such as
gravity, muscle tension, blood pressure as well as from local active
tractions of nearby cells \cite{chen_2008, ingber_2003a}.
The forces that act on cells can be static as well as time
varying, {\it{e.g.}}, continuous loading occurs during
development of long bone growth while cyclic loading occurs due
to periodic blood pressure variations.
These mechanical signals
typically induce an active reorganization of the cell
cytoskeleton and readjustment of the contractile forces exerted
by the cells \cite{Fredberg06, Stamen2007}.
The active nature of
this mechanotransduction is demonstrated by the fact that it
often vanishes when actin-myosin contractility is inhibited
\cite{zhao_2007}.
 
The response to mechanical stress is demonstrated by cells that
actively reorient and align themselves in preferred directions.  It is
interesting to note that, while in some studies cells were shown to
align parallel to the direction of a static or quasi-static stress
field \cite{Collins00, Eastwood98, brown_tensional_1998,
  Vandenburgh90}, other experiments find that cells remain randomly
oriented \cite{jungbauer_two_2008}.  On the other hand, when subject
to dynamically oscillating stress and strain fields (designed
originally to study the effects of heart beat and blood pressure),
cells tend to orient away (nearly, but not exactly perpendicularly)
from the stress direction \cite{Hayakaya01, Kurpunski06, Shirinsky89,
  wang_specificity_2001, Wang00, jungbauer_two_2008,
  faust_cyclic_2011}.  As discussed below, this reorientation does not
occur as a rigid body motion of the cell, but rather involves the
disassembly and then re-assembly of the CSK in directions determined
by the applied stress; it may also involve rotations of the stress
fibers \cite{deibler_actin_2011}. In some of the experiments on static
stress \cite{brown_tensional_1998}, cells were placed in 3d collagen
matrices and it is not clear whether remodeling of the matrix
\cite{takakuda_1996, fernandez_compaction_2009} by the stress
contributes to cellular orientation or if the orientation is solely a
result of CSK reorganization within the cell in response to stretch.
The experiments in Fig. 5 of \textcite{Eastwood98} do indicate random
collagen alignment even under tension. In general, the roles of
passive (CSK elasticity) and active (actomyosin contractility) forces
in determining cell response to applied stress have yet to be fully
elucidated \cite{genin_2008}. In the following, we first review some
recent mechanobiological measurements that provide new insights for
understanding the response of cells to applied stress.  We then
summarize several theoretical approaches that quantify these ideas.

{\it{CSK disassembly and reassembly: }} Recent experimental
studies have shown that cell stretch induces CSK
fluidization \cite{fredberg_2005, fredberg_2010, Fredberg06,
fredberg_2009, fredberg_2007}
 which occurs
through direct physical effects of physical forces upon weak
cytoskeletal crosslinks.
CSK fluidization is typified by marked
decreases of CSK stiffness, CSK tension, and cellular traction
forces, and marked increases in the rate of CSK remodeling
dynamics \cite{fredberg_2009, fredberg_2007}
and is accompanied by extremely rapid
disassembly of actin bundles.
These effects depend \cite{fredberg_2009, fredberg_2007} on the
load, loading frequency and on the magnitude of the
pre-stretch actomyosin contractility.

To restore homeostasis in the cell ({\it{i.e.}}, a fixed level
of contractility),
CSK fluidization is  immediately succeeded by CSK
reassembly, a signaling driven response that restores
molecular interactions that were disrupted by fluidization
\cite{fredberg_2007}.
CSK reassembly  results in
gradual increases of CSK stiffness, CSK tension, and cellular
traction forces, and gradual decreases in the rate of CSK
remodeling dynamics.
These are driven by slow reassembly that
acts predominantly on those spatial sites where traction forces
were markedly reduced by CSK fluidization \cite{fredberg_2009,
fredberg_2007}.
These processes govern the response of cells to
applied stress in which the reorientation is a result of the
CSK fluidization and reassembly.

{\it{CSK stiffness changes in response to applied stress: }}
While it is clear that the CSK reorganizes in response
to applied stretch, it is also important to know whether
cell contractility and stiffness is increased or decreased
during stress application.
Experiments on fibroblasts in three-dimensional, collagen
gels showed that  overall,  cells  reduce
their contractility during the stretch-relax cycles
\cite{brown_tensional_1998}.
This led those authors to suggest that cells have
a homeostatic (or set-point) contractility that is
reduced when the surrounding medium is stretched.
The dynamics of this process were investigated in more detail
in \textcite{genin_2008} who showed that
when stretched for several minutes, contractile
fibroblasts initially diminished the mechanical tractions they
exert on their environment through depolymerization of actin
filaments.
The cells then restored tissue tension and rebuilt
actin stress fibers through staged Ca dependent processes that
consisted of a rapid phase that ended less than a minute after
stretching, a plateau of inactivity, and a final gradual phase
that required several minutes to complete.
Active contractile
forces during recovery scaled with the degree of rebuilding of
the actin cytoskeleton.
The final cell stress following
a stretch exceeds the pre-stretch value; this is in contrast
to the results reported by  \textcite{brown_tensional_1998}.
However, the observations of cellular ensembles might
not be indicative of a ``typical'' cell; the highly repeatable
ensemble behaviors may  represent a diversity of responses at the
level of individual cells

Trepat and coworkers \cite{trepat_2004} developed an experimental
system to subject adherent cells to a global stretch while
simultaneously measuring the local complex shear modulus
($G^*=G+iG''$) of the cells.  They used this system to study the
viscoelasticity of alveolar epithelial cells in response to stepwise
stretch and found that with increasing levels of stepwise stress, both
$G'$ (elastic response) and $G''$ (viscous response) increased.  These
findings indicate that the cytoskeletal response shows a non-linear
elastic response characteristic of strain-stiffening and that
intracellular dissipation also increases with increasing cytoskeletal
tension.  In addition, they found that the ratio $G''/G'$ decreased
with stretch, consistent with an increase in the elastic rigidity of
the CSK, corresponding to reassembly.  In a later study, these authors
observed that when the cytoskeleton was contracted with thrombin
before application of a stepwise stretch, the strain-stiffening
response was abrogated \cite{trepat_2006}, which suggests that the
strain-stiffening regime is restricted to a range of cytoskeletal
tension.  The same experimental setup was used to test the
viscoelastic response of a broad variety of cell types that were
subject to a transient application of stretch-unstretch
\cite{fredberg_2007}.  Contrary to the case of a stepwise stretch, a
transient stretch that returns to zero strain caused a sharp drop in
both $G'$ and $G''$ and a sudden increase in the ratio $G''/G'$.
Thus, while a stepwise stretch induces CSK rigidification, a transient
stretch induces cell softening and fluidization.  To test whether
stretch-induced cell stiffening and softening were associated with
changes in cytoskeletal tension, Gavara and coworkers developed a
system to map traction forces at the cell-substrate interface during
application of stretch \cite{navajas_2008}.  They observed that
cytoskeletal tension increased with application of a stepwise stretch,
but decreased below baseline levels upon stretch removal.  Analysis of
traction maps before, during, and after stretch indicated that the
regions of higher traction force application were those that exhibited
a larger relative drop of traction after stress cessation, suggesting
that those cellular structures subjected to a higher tension are
disrupted by stretch.  Taken together, these findings point to the
existence of two different mechanisms by which cells respond to
stretch.  Strain-stiffening during a stepwise stretch is likely to
arise from non-linear stretching of single cytoskeletal filaments.  On
the other hand, strain-softening after a transient stretch is probably
caused by inelastic unbinding or unfolding of cytoskeletal crosslinks
and actomyosin crossbridges. In response to a constant stepwise
stretch, filament stretching appears to dominate over inelastic
unbinding and unfolding of crosslinks and crossbridges. After stretch
cessation, however, the contribution of filament stretching becomes
negligible and the effect of inelastic unbinding and unfolding
dominates.

{\it{CSK and cellular reorientation in response to cyclic
stretch: }}
In the introduction to this section, we mentioned several
studies that showed that the cells orient away from the stress
direction of cyclically applied stress.
Recent experiments have provided quantitative measures of these
effects. Experiments described in \textcite{jungbauer_two_2008, deibler_actin_2011}
investigated the dynamic reorientation of rat embryonic and
human fibroblast cells over a range of stretching frequencies
from 0.0001 to 20 s$^{-1}$ and strain amplitudes from 1\% to 15\%.
Their measurements  show that the mean cell orientation
changes exponentially in time with a frequency-dependent characteristic
time from 1 h to 5 h.
At subconfluent cell densities (at which the cells are not yet close packed), this
characteristic time for reorientation shows two characteristic
regimes as a function of frequency. For frequencies below 1
${\rm{s}}^{-1}$, the characteristic time decreases with a power law as the
frequency increases.
For frequencies above 1 ${\rm{s}}^{-1}$, it
saturates at a constant value.
In addition, a minimum threshold
frequency was found below which no significant cell reorientation
occurs.
The results suggest a saturation of molecular
mechanisms of the mechanotransduction response machinery for
subconfluent cells within the frequency regime studied.
One possible interpretation of these two time scales is given in the
theoretical model described in the next section.
Interestingly, recent work \cite{spatz_2011} by these researchers
showed that the time scale is correlated with the amount of
actin in the cell; aged cells, with less actin show
faster reorganization in response to uniaxial tensile stress
compared with younger cells which contain more actin and are
elastically more rigid.
In addition, other biochemical changes can modify the response
time of the cytoskeleton and thereby control its orientation
in response to cyclically varying stress \cite{hoffman_dynamic_2011}.

To control the strains both parallel and perpendicular to the stress
directions, the researchers in \textcite{faust_cyclic_2011} used
elastomeric chambers that were specifically designed and characterized
to distinguish between zero strain and minimal stress directions and
to allow accurate theoretical modeling. Reorientation was only induced
when the applied stretch exceeded a specific amplitude, suggesting a
non-linear response. However, on very soft substrates no
mechanoresponse occurs even for high strain. This suggests an
explanation for the necessity of rather stiff environmental conditions
to induce cellular reorientation in mammalian tissues.  For all
stretch amplitudes, the angular distributions of reoriented cells
could be modeled as discussed in the next section. Cyclic stretch
increases the number of stress fibers and the coupling to
adhesions. Changes in the cell shape follow the cytoskeletal
reorientation with a significant temporal delay; this indicates that
cell reorientation and shape is induced by CSK reassembly in response
to stretch. In the frequency range studied of 10-50 mHz, the stress
induces cell reorientation (after about 16 hours) in the direction of
zero strain.  A recent study by Livne and Geiger (unpublished)
analyzed the reorientation dynamics of cyclically stretched cells,
over a wide range stretch configurations, and observed a systematic
deviation between the measured cell and stress fiber orientation and
the zero strain prediction (up to ~10 degrees). To address this
discrepancy, a novel model which shifts the focus of the reorientation
process to the FAs was developed.

{\it{Theory of cell response to applied stress: }} Models of cell
response to applied stress are motivated by the questions of why
stress fibers or cells orient nearly (but not always exactly)
perpendicular to the direction of the applied stress.  Macromolecular
or biochemical models of cellular orientation and stress-fiber
rearrangement in response to applied forces have been discussed in
\textcite{mogilner_2005a, wei_2008, kaunas2009, pirentis_2009,
  pirentis_2011} while a more generic theoretical approach is given by
\textcite{SamBiophysJ08, de_dynamics_2007, de_dynamical_2008,
  safran_nonlinear_2009}.  We first review more molecularly-based
models that focus on the role of the stress fibers and then present a
more phenomenological and general approach that in principle coarse
grains over both stress fiber and focal adhesion response..

{\it{Molecularly-based models: }} The work of \textcite{wei_2008}
predicts the orientation of stress fibers in response to cyclic
stretch based on a biochemical-mechanical model that relates the
contraction and extension rate sensitivity of the stress fibers to the
magnitude and frequency of the applied stress.  These kinetics depend
on a biochemical activation signal -- the tension-dependent fiber
dissociation rate -- and the rate of force generation by myosin II
motors.  This assumes that the stress fibers are intact throughout the
application of dynamically varying strain.  Experiments
\cite{fredberg_2005, Fredberg06, fredberg_2007} show
that cells respond to mechanical stress via an initial, fast (sec
timescale) fluidization of the stress fibers that then reassemble and
reorganize.  Motivated by this, \textcite{pirentis_2009,
  pirentis_2011} focus on a mathematical model that simulates the
effects of fluidization and reassembly driven rigidification
\cite{fredberg_2009, fredberg_2010} on cytoskeletal contractile
stress.  They show how these phenomena affect cytoskeletal realignment
in response to pure uniaxial stretching of the substrate.  The model
comprises individual elastic stress fibers anchored at the endpoints
to an elastic substrate and predicts that in response to repeated
stretch/unstretch cycles, stress fibers tend to realign in the
direction perpendicular to stretching.  The authors conclude that
relaxation of cytoskeletal contractile stress by means of fluidization
and subsequent stress recovery by means of CSK reassembly may play a
key role in reorganization of cytoskeletal stress fibers in response
to uniaxial stretching of the substrate.

A somewhat more general approach to the mechanical response of stress
fibers that was taken by Kaunas and colleagues \cite{kaunas2009,
  kaunas_2011} who developed a model that tracks the fate of
individual, stretched stress fibers based on the hypothesis that
stress fibers have an optimal prestrain due to actomyosin
contractility \cite{lu_2008}; perturbing the strain from that optimal
value promotes stress fiber disassembly.  Motivated by experimental
evidence of stress fiber viscoelastic properties, stress fibers are
assumed to relax at a rate proportional to the perturbation in fiber
stretch away from this optimum.  The dynamic turnover of stress fibers
was described using a stochastic approach with the probability per
unit time of stress fiber disassembly expressed as a constant plus a
term quadratic in the deviation of the strain from its optimal value.
The disassembly of a stress fiber is assumed to be immediately
followed by the assembly of a new stress fiber at its optimal stretch
and oriented in a randomly chosen direction.  Model parameters were
determined by fitting experimentally measured time courses of stress
fiber alignment performed at different rates of strain ({\it{i.e.}},
0.01 to 1 Hz).  The model predicts that reorganization of the stress
fibers is determined by the competition between the rates of stress
fiber assembly and load-dependent disassembly.  The stress fibers
preferentially disassemble in the direction of stretch, while stress
fibers reassembling in stochastically chosen directions gradually
accumulate about the direction of least perturbation in fiber tension.
At low strain rates, the stress fibers are predicted to align with
random orientations with respect to the applied stress direction.
While this has been reported in some cases for very slow cyclic
stress, other experiments report alignment in the stress direction as
discussed above.  Recent studies \cite{tondon_dependence_2012} using
non-sinusoidal waveforms show that the stress fiber reorientation is
most sensitive to the rate of lengthening; this provides support for
the role of stretch of the actin filaments in cell reorientation under
stress.

{\it{Coarse-grained models based on force dipoles: }} One goal of this more phenomenological approach
\cite{SamBiophysJ08, de_dynamics_2007, de_dynamical_2008,
  safran_nonlinear_2009} is to explain the observed frequency
dependence of cell orientation mentioned above.  Another, is to
understand why the characteristic time for the cell to reach its
steady-state orientation, $\tau_c \sim 10^3-10^4 $ seconds, is
strongly frequency dependent for stretch frequencies smaller than
about 1 Hz while at higher frequencies, $\tau_c$ is frequency
independent.  The experiments were conducted on anisotropic cells such
as fibroblasts so the theory focuses on needle-like cells in which the
entire cell is modeled in a coarse grained approximation as a single
force dipole; for needle-like cells, the dipole component $P_2=0$.
The dipoles can then be characterized by their magnitude $P_0 \equiv
P<0$ (to signify contraction) and direction, $\theta=\arctan \left[
  (P_0-P_1)/(P_0+P_1) \right] =\arctan \left[p_{yy}/p_{xx}\right]$,
relative to the external stress.

It has been suggested \cite{brown_tensional_1998} that cells
actively  adjust their contractility by reorganizing the FA and
stress fibers to maintain an optimal (or set-point) value  of the
stress  or strain $U^{\star}$  in the
adjacent matrix  \cite{SamBiophysJ08, de_dynamics_2007, de_dynamical_2008, safran_nonlinear_2009}.
   This translates via elastic theory, into an optimal
value of the cellular dipole $P^{\star}>0$.
The stresses are converted to energy units by multiplying by the cell
volume and the externally applied stretch is denoted as $P_a(t)>0$.
In the presence of such time-dependent stretch
that acts at an angle $\theta$ relative to the cell axis,
it is assumed that the
homeostatic, set-point {\it{total}} local stress in the matrix
is achieved when the cellular force dipole obeys \cite{safran_nonlinear_2009}:
\begin{equation}
P=-P^{\star}+ \alpha_0 P_a(t)   \left(\phi - \phi_1 \right)
\label{pstar1}
\end{equation}
where $\phi=\cos^2 \theta$.
Two limiting cases are where:  (i) the cellular dipole
is controlled by the  matrix stress
where $\phi_1=0$
(ii) the dipole is controlled by the matrix strain and
$\phi_1=\cos^2 \theta_0 \equiv \phi_0$, where  $\theta_0$ is the
zero strain direction given  by $\cos^2 \theta_0=\nu/(1+\nu)$.
In
general, $\alpha_0$ can be either positive or negative
corresponding to matrix stretch that causes either a decrease or
an increase in the cytoskeletal forces respectively.

 Deviations  from the
set-point result in   internal forces within
the cell that  reestablish the optimal stress
condition. These forces can be derived from derivatives of an
effective, harmonic ``free energy'' (more precisely, a cost
function whose minimum represents the optimization of the
cellular activity) due to cell activity, $F_a$, that includes
the active processes within the cell  that establish  cellular
response to its {\it local} environment. \begin{eqnarray}
F_a={1\over 2}\chi  \left( -P + \alpha_0  P_a(t) \left( \phi -
\phi_1 \right) - P^{\star} \right)^2 \label{cell_activity}
\end{eqnarray} where $\chi P^{\star 2}$ (with units of energy)
is a measure of cell activity that establishes the set-point.

In addition to the cell activity, the model also includes the effect
of mechanical matrix forces, Eq.~\ref{eq:elasticenergy5} that yields
an energy, $F_e$ proportional to the product of $P_a(t)$ and $P$.  The
goal is to solve for the dipole magnitude and direction in the
presence of a time-varying stress: $P_a(t)=P_a (1- \cos \omega_a t)$,
where $P_a>0$ for stretch.
In general, the dynamics of the cytoskeleton are governed by complex,
viscoelastic processes that also involve liquification and reassembly
of the stress fibers \cite{Fredberg06}.  In a coarse grained picture,
one can write relaxational equations for the dipole magnitude and
direction, that are governed by the derivatives of $F=F_a+F_e$:
\begin{equation}
\frac{d p(t)}{d t} = - \frac{1}{\tau_p}  f_p \quad    \quad
\frac{d \theta(t)}{d t} = - \frac{1}{\tau_\theta}
f_\theta
\label{dynamic2}
\end{equation}
where $p=P/P^*$, the dimensionless, effective free energy $f=F/(\chi
P^{*2})$, and $f_p=\partial f/\partial p$ and $f_\theta=\partial
f/\partial \theta$.  Noise terms modeled as a dimensionless, effective
temperature, $T_s$, can also be included in this formalism
\cite{safran_nonlinear_2009}; note the caveats on the use of effective
temperature and Boltzmann distributions discussed in the section on
the physics background.

Based on experiments \cite{Fredberg06, genin_2008,
navajas_2008}, it has been suggested that the liquification and
repolymerization of the actin stress fibers after stretch is applied, occurs on
a short time scale on the order of several seconds, while the
correlated reorientation occurs on much longer time scales (on
the order of many minutes) \cite{brown_tensional_1998, Eastwood98,
jungbauer_two_2008}.
It is thus assumed that $\tau_p \ll \tau_{\theta}$: the time
scale associated with changes in the magnitude of the dipole is
much faster than that associated with the dynamics of its highly
correlated reorientation. In this approximation, the dipole
magnitude reaches a steady-state value in a short time; this
value may be time dependent and oscillatory due to the cyclic
nature of the applied, time dependent stress.

One therefore first solves for the dipole magnitude, $p(t)$, treating
the slowly-varying dipole orientation, $\phi(t)$, as a constant; for
details, see \textcite{safran_nonlinear_2009}.  The average value of
$\phi=\cos^2 \theta$ is calculated as a function of the frequency
depends on the effective temperatures for the cases of both stress and
strain as set-points.  At high frequencies and low effective
temperatures, the average angle is nearly~perpendicular (or in the
zero-strain direction, $\theta_0$, for cells whose set-point is
determined by matrix strain), due to the dynamical frustration of the
cell which is unable to adjust its force dipole to the time-dependent
matrix stresses. At very low frequencies, the average angle is nearly
parallel and for both the case of stress and strain as set-points,
consistent with some of the experiments \cite{brown_tensional_1998}.
At higher effective temperatures, the orientation distribution is
random and the average value of $\phi$ (in two-dimensions) is 1/2 for
all frequencies.  At intermediate temperatures, one finds the
interesting possibility of nearly~perpendicular orientation for high
frequencies, but nearly random orientation for low frequencies (for
details, see \textcite{safran_nonlinear_2009}.  Biochemical changes
can modify the response time of the cytoskeleton and thereby change
cell orientation from nearly~perpendicular -- when the CSK cannot
follow the applied, cyclic stress -- to parallel -- when CSK
remodeling time scales are short enough \cite{hoffman_dynamic_2011}.
In addition, experiments on 3d matrices \cite{tissue_rev_2012}
indicate parallel orientation of stress fibers even at relatively high
frequencies.  The systematic understanding of when cells respond by
orienting parallel compared to the relatively well-studied response of
cells on relatively stiff substrates to cyclic stretch is a challenge
that has yet to be met.

The dynamical theory is also used to calculate
\cite{safran_nonlinear_2009} the characteristic time, $\tau_c$, for
a cell to attain its steady-state orientation. At high frequencies,
$\tau_c$ is frequency independent, while at low frequencies, $\tau_c
\sim 1/\omega^2$; in both regimes $\tau_c$ depends on the amplitude of
the applied stress and this is related to the fact that the
zero-strain or zero-stress direction orientation occurs only when the
applied stress exceeds a threshold value
\cite{de_dynamics_2007,de_dynamical_2008, safran_nonlinear_2009}. Both
the predicted frequency and amplitude dependence are in qualitative
agreement with experiments \cite{jungbauer_two_2008}.

\section{Cell assemblies}

\subsection{Matrix-mediated cell interactions}

After treating the response of isolated cells to changes in their
elastic environment, we now discuss the elastic responses of and
interactions in ensembles of cells. We restrict the analysis to the
effects of elastic interactions on the relative positions and
orientations of cells.  Moreover, we focus on the case in which cells
are well separated and interact with each other only via the matrix
and not through direct cell-cell interactions.  This is fundamentally
different when modeling growing epithelial tissue or tumors, when
cell-cell interactions dominate. A popular model system for eipthelial
tissue formation is the Drosophila wing disc, which often is treated
using vertex models
\cite{hufnagel_mechanism_2007,farhadifar_influence_2007,rauzi_nature_2008,landsberg_increased_2009,aegerter-wilmsen_exploring_2010,canela-xandri_dynamics_2011,aliee_physical_2012}.
In our focus here on matrix mediated interactions, we do not discuss
the effect of cell proliferation and cell death, that also leads to
interesting features in cell assembly
\cite{c:shra05,basan_homeostatic_2009,joanny_2010}.

An early study that highlighted the effect of elastic substrate
deformations in modulating the relative positions of cells placed far
apart was presented in \textcite{Korff99}. They observed that
capillary-like structures formed by two, initially separated groups of
cells were associated with tensional remodeling of the collagen matrix
and directional sprouting of the outgrowing capillaries towards each
other.  These experiments presented evidence that tensional forces on
a fibrillar, extracellular matrix such as type I collagen, but not
fibrin, are sufficient to guide the directional outgrowth of
endothelial cells.  More recently \textcite{c:rein08,king_2010} used
matrices of varying stiffness and measurements of endothelial cell
migration and traction stresses, to show how cells can detect and
respond to substrate strains created by the traction stresses of a
neighboring cell; they demonstrated that this response is dependent on
matrix stiffness.  Other studies suggest that on some matrices, cells
can sense each other (most probably via elastic deformations) at
distances on the order of 400$\mu$m \cite{janmey_plos_2009}.  Recently
it was also reported that cardiac cells can synchronize their beating
through substrate deformations \cite{Tang_how_2011}. The various
experiments imply that matrix mechanics can foster tissue formation by
correlating the relative motions or even internal dynamics of cells,
thereby promoting the formation of cell-cell contacts. The theoretical
studies below model an entire cell as a single, usually anisotropic,
force dipole.  Interactions among cells are taken into account for
several simple geometries.  A simple analogy to dielectric media with
predictions of the elastic susceptibilities and ``dielectric
constants'' of force dipole assemblies was presented in
\textcite{zemel_active_2006} where the effective elastic constants of
materials containing force dipoles are calculated as a function of the
dipole density. The results, valid in the relatively dilute limit,
indicate an effective stiffening of the material due to the alignment
of the contractile dipoles parallel to the applied stretch.  It
remains to be seen how to take this analogy further to include
dynamical, tensorial and non-local spatial effects as well as the
development of a theory that is valid for both high and low force
dipole concentrations.  In addition, one must decide whether the
dipoles are translationally mobile (as is the case for counterion
screening in electrostatics) or only orientationally mobile (as
assumed in \textcite{zemel_active_2006}).

\subsection{Elastic interactions of force dipoles}

The anisotropic and long-range nature of the interactions of cell
dipoles leads to a rich variety of self-assembled
structures \cite{uss:bisc03a,uss:bisc04a,uss:bisc05,uss:bisc06a}.
Monte Carlo simulations of these dipolar interactions in the presence
of noise, modeled as an effective temperature, predicted cellular
structure formation on elastic substrates as a function of the cell
density and Poisson ratio of the substrate. One interesting situation
considered was that of an infinitely extended string of aligned force
dipoles spaced at equal distances, $a$ \cite{uss:bisc05}.  An
additional dipole is placed at a horizontal distance $x$ and with a
vertical offset $y$. Despite the long-ranged character of the elastic
dipole interaction, the nearby dipoles in the string screen each
other's strain fields; thus, the effective interaction between an
infinite string and a single dipole (or a second string) is short
ranged and decays as an exponential function of $x/a$.  The magnitude
of the interaction depends strongly on the Poisson ratio of the
substrate.  The results suggest that long-ranged effects do not
dominate structure formation at particle densities sufficiently large
as to allow formation of strings of aligned dipoles.

\FIG{Fig24}{fig:ilka1}
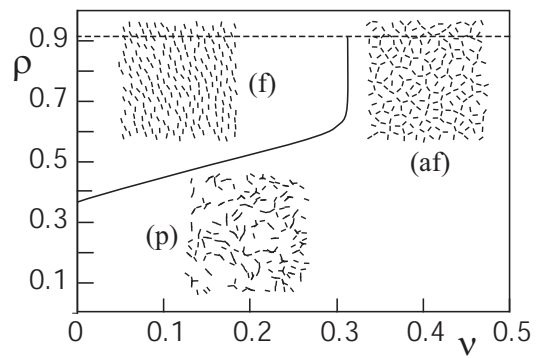
{Phase diagram for positionally disordered cells. At low
values of the scaled cell density, $\rho$, an orientationally
disordered (paraelastic) phase (p) prevails. At high cell
density, orientational order sets in, with a nematic string like
(ferroelastic) phase (f) at low values of Poisson ratio, $\nu$,
and a isotropic ringlike (antiferroelastic) phase (af) at large
values.  From \textcite{uss:bisc06a}.}

The orientational interactions of dipoles at random spatial positions (but constrained to obey excluded
volume) were considered; the elastic energies and noise were used to
equilibrate the dipolar directions \cite{uss:bisc06a}.  At low density, the
simulations show many short strings with few correlations
(paraelastic phase) among them.  At high density and small values of the Poisson
ratio, spontaneous polarization occurs (ferroelastic or nematic), that results in a unidirectional contraction of the substrate.
At large values of both
the cell density and the Poisson ratio, the system becomes
macroscopically isotropic again, with a local structure which is
ringlike rather than string like (antiferroelastic).  The predicted
structures are shown in Fig.~\ref{fig:ilka1}.

\subsection{Myofibril registry modulated by substrate
elasticity}

We now summarize a model that shows that elastic interactions can tune
the registry of long actomyosin fibers whose nematic (orientational)
order is already well established \cite{friedrich_striated_2011}.  In a
variety of cell types, various types of actomyosin bundles 
exhibit periodic internal structure with alternating localization of
myosin filaments and the actin crosslinker $\alpha$-actinin.  Examples
include striated stress fibers in fibroblasts and striated stress
fiber-like actomyosin bundles in some developing muscle
cells \citep{pellegrin_actin_2007,russell_sarcomere_2009,hotulainen_stress_2006,Rhee:1994}.
The striated architecture of these fibers is similar to
the sarcomeric architecture of myofibrils in striated muscle, but is
much less regular.  In both adherent, non-muscle cells as well as in
developing striated muscle cells, the striations of neighboring, but
distinct fibers are often in registry, {\it{i.e.}}, the positions of the
respective $\alpha$-actinin and myosin bands match, see
Fig.~\ref{fig:registry1}.  This inter-fiber registry of striated
fibers represents a further state of cytoskeletal order, which might
be termed ``smectic order'' using liquid crystal terminology.

Experiments on cultured cells plated on flexible substrates have
shown that substrate stiffness is one factor that can regulate
cytoskeletal order in general, and myofibril assembly in particular
\cite{engler_myotubes_2004, Engler:2008, Jacot:2008,Serena:2010,majkut_cardiomyocytes_2012}.
Relative sliding of striated actomyosin bundles
into registry was previously reported in
\textcite{McKenna:1986}. In \textcite{engler_myotubes_2004} the
amount of striated myosin (which serves as a measure of
myofibril condensation) depended on the stiffness of the matrix
upon which various cells were cultured, with a pronounced
maximum at an optimal stiffness of about $E_m\approx 10\,\rm
kPa$. Interestingly, this value is close to the longitudinal
stiffness of relaxed muscle.

The striated fibers are under constant tension due to the activity of
myosin filaments that link actin filaments of opposite polarity, see
figure \ref{fig:registry1}.  These actomyosin contractile forces
strain the $\alpha$-actinin-rich crosslinking regions 
(termed Z-bodies) of premyofibrils and nascent myofibrils in
developing muscle cells.  Because the crosslinking regions can be
mechanically connected to the substrate by means of adhesive contacts,
the tension generated in them may be transmitted to the substrate.  Thus,
the substrate underneath a striated fiber is strained with regions of
expansion below the crosslinking bands and regions of compression in
between.  The strain fields induced by a single bundle of actomyosin
propagate laterally towards its neighbors, inducing an effective
elastic interaction between the fibers; this biases the spatial
re-organization of fibers to favor registry, that results in
smectic ordering of the crosslinkers and the myosin in neighboring
bundles.

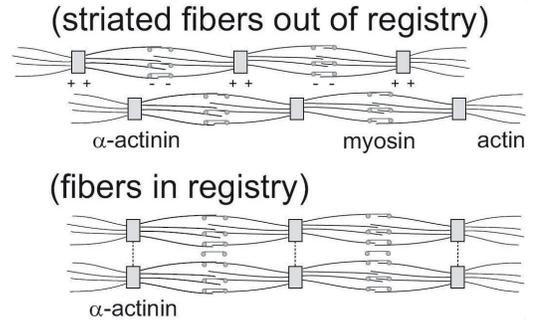
\FIG{Fig25}{fig:registry1} {Schematic view of two striated
  fibers. Striated stress fiber-like acto-myosin fibers form close to
  the cell-substrate interface of adherent, non-muscle cells. Each
  fiber is a bundle of aligned actin filaments that has a sarcomeric
  sub-architecture: Z-bodies (containing alpha-actinin) that crosslink
  actin filament barbed ends alternate with regions rich in myosin II
  in a periodic fashion.  Striated fibers can slide past each other
  until their periodic structures are in phase. Adapted from
  \textcite{friedrich_striated_2011}.}

A minimal model for this effect \cite{friedrich_striated_2011} considers
the cell-substrate interface as the $xy$-plane
with a single contractile fiber parallel to the $x$-axis.
 The forces transmitted by the fiber onto
the substrate can be effectively described by a dipole
distribution (with units of energy per unit area) $\Pi_{ij}(x,y)=
\rho(x)\delta(y)\,\delta_{ix}\delta_{jx}$ of force dipoles that
are localized to the adhesive contacts whose lateral extension
are on the order of $100\,\rm nm$ and are thus much smaller than the
spacing $a\approx 1\,\mu\rm m$ of Z-bodies.  Here, the force
dipole density (with units of energy per unit length) $\rho(x)$
is a periodic function of $x$ due to the sarcomeric ({\it{i.e.}}, 
periodic) architecture of a single striated fiber. For
simplicity, the analysis focuses on  the principal Fourier mode:
$\rho(x)=\rho_0+\rho_1\cos(2\pi x/a)$ where $a$ corresponds to
the sarcomeric periodicity of the striated fiber.

The strain field $u_{ij}(x,y)$ at the surface of the substrate with
Young's modulus $E_m$ (and located at $z=0$) that is induced by this
periodic dipole ``string'' can be found from the Green's function of
Eq.~\ref{eq:elasticenergy4}.  The parallel strain component
$u_{11}(x,y)$ can be written as a product of a ``lateral propagation
factor'' $\Phi$ that characterizes the propagation of strain in the
lateral $y$ direction and a harmonic modulation in the $x$ direction
along the striated fiber
\begin{equation}
\label{eq_u11}
u_{11}(x,y)=\Phi\left(|y|/a,\nu\right) \, \frac{2\rho_1}{E_m a^2}\, \cos(2\pi x/a)
\end{equation}
Thus the strain field $u_{11}$ is periodic in the $x$-direction with period $a$
reflecting the periodicity of the striated fiber.  The factor $\Phi$
characterizes the propagation of strain in the lateral direction away from
the centerline of the fiber; this depends on the distance from the
fiber as well as the Poisson ratio.  The interaction between two such
strings of dipoles is described by the elastic interaction energy,
which is the local product of the dipole and the strain.
The energy which a given fiber (string of dipoles), must invest
in order to deform the substrate is the sum of a ``self-energy'' of
the first dipolar string, $W_{\rm self}=\int d^2\x\, \Pi_{ij}^{(1)}
u_{ij}^{(1)}$, which accounts for the substrate deformation energy in
the absence of the second string of dipoles, and an interaction term
$W_{\rm int}=\int d^2\x\, \Pi_{ij}^{(1)} u_{ij}^{(2)}$. The term
$W_{\rm int}$ characterizes an effective, substrate-mediated
interaction between the two contractile fibers and can guide their
spatial reorganization.

Inserting the specific strain field induced by a single striated
fiber, Eq.~(\ref{eq_u11}), into the general formula for elastic
interactions,  yields the elastic interaction energy
between the two fibers  (per mini-sarcomere) as a function of
the phase shift $\Delta x$ and the separation of their
centerlines, $d$:
\begin{equation}
\label{eq_W_int}
W_{\rm interaction} = \Phi(d/a,\nu)\, \frac{\rho_1^2}{a E_m}\, \cos(2\pi \Delta x/a)
\end{equation}
Here $W^\ast =\rho_1^2/(a E_m) \approx 10^{-18}{\rm J} \approx
250\,k_B T$ sets a typical energy of the elastic interactions.
Registry of fibers with $\Delta x=0$
is favoured for inter-bundle spacings where the propagation
factor, $\Phi<0$.

For incompressible substrates with Poisson ratio close to
$\nu=1/2$, such as those used in experiments
\citep{engler_myotubes_2004, buxboim_matrix_2010}, it can be shown \cite{ friedrich_striated_2011} that
the sign of the prefactor $\Phi$ of the elastic interaction
energy is negative provided that the lateral fiber spacing is larger
than some  threshold $d/a>d^\ast/a\approx0.247$. Hence, elastic
interactions favor a configuration where neighboring  fibers are
in registry with $\Delta x=0$.
The opposite trend is found when $\nu
\approx 0$ \citep{uss:bisc05,friedrich_striated_2011}.
It is therefore possible that elastic interactions also set a preferred
lateral spacing of striated fibers.
Additionally, steric interactions may prevent neighboring fibers from
getting too close and could enforce the condition $d>d^\ast$.

We previously discussed experiments and theory that demonstrated
that cells tend to prefer rigid substrates where the elastic
deformation energy cost is minimal.
This also determines the optimal value of the cellular dipole
magnitude in a simple model that balances the ``cell activity''
(described in Eq.~\ref{cell_activity} with zero applied field,
$P_a=0$) with the elastic energy cost of deforming the
substrate (analogous to the expression in Eq.~\ref{eq:freequadp})
\cite{de_dynamics_2007,
safran_nonlinear_2009,friedrich_how_2012}.   In terms of the present
model, this determines the optimal amplitude of the dipole
string, $\rho_1^\ast$, from the force balance given by
minimization with respect to the dipole amplitude, $\rho_1$, of
the sum of the two energies (per unit length of the bundle): the
activity optimization,  $ W_{\rm active} =  \chi (\rho_1 -
\rho_1^\ast)^2/2  $  and the deformation energy,  $ W_{\rm
deform} \sim \mathcal  \rho_1^2 / ( 2 a^2 E_m )$ where $E_m$ is
the Young's modulus of the substrate
\cite{friedrich_striated_2011}. This predicts that $ \rho_1 =
\rho_1^\ast \, E_m/(E_m + E_m^\ast) $ where $E_m^\ast = 1/(
a^2 \chi)$. The set-point value $\rho_1^\ast$ corresponds to
the amplitude $\rho_1$ of the dipole density on very stiff
substrates with $E_m \gg E_m^\ast$. On soft substrates with $E_m
\ll E_m^\ast$, however, $\rho_1$ can be considerably smaller than
$\rho_1^\ast$.

Using this expression for the saturation of  $\rho_1$ on
substrate stiffness further predicts that the
registry force between two parallel striated fibers becomes a
non-monotonic function  of $E_m$ with a Lorentzian form and has
maximal magnitude for $E_m=E_m^\ast f_{\rm reg} \sim - {1}/{E_m} \rho_1^2 \sim - {E_m}/ {(E^\ast_m+E_m)^2}$. Here, it is assumed that the
lateral spacing $d$ of the fibers is larger than the critical
distance $d^\ast$ and independent of substrate stiffness.

The theory was compared with recent experimental studies of the
inter-fiber registry in human mesenchymal stem cells that were plated
on polymeric gels of different stiffness (ranging from 0.3 kPa to 40
kPa) \cite{friedrich_striated_2011,majkut_cardiomyocytes_2012}.  Well
established, inter-fiber registry of adjacent striated fibers was
observed primarily for cells that were cultured on 10 kPa gels as
opposed to softer or more rigid substrates.  Myosin bands
perpendicular to the axis of nematic fiber organization were clearly
visible and most likely connect neighboring actomyosin bundles in
registry.  Out of approximately 20 cells examined per gel, roughly
30\%-50\% exhibited aligned, striated fibers.  The guidance mechanism
for the registry of striated fibers by elastic interactions due to
their elastic interactions predicts maximal registry at an ``optimal''
value of the substrate rigidity and represents a plausible mechanism
for the establishment of inter-fiber registry observed in the
experiments.  Further experiments are needed to resolve the extent of
striations within one acto-myosin bundle from the registry of
striations in neighboring bundles; the theory presented here addresses
the question of registry among bundles.  It assumes that each bundle
is well ordered; a possible mechanism for the development of such
order was recently suggested in \textcite{friedrich_plos_2012}.

\section{Conclusions and outlook}

Research at the interface of physics and biology is an exciting
adventure that has led and is still leading in several different
directions. In this review with a theoretical focus, we have
appropriately modified soft matter physics approaches to analyze how
adherent cells mechanically interact with their environment through
forces at the cell-material interface. Although passive soft matter
systems such as droplets, fully elastic particles, vesicles and
polymeric capsules are important reference cases for the adhesion of
single cells, our discussion has shown that the main feature missing
from such theoretical frameworks are active processes. In the context
of force generation and sensing of adherent cells, the most prominent
active processes are the polymerization of lamellipodia at the cell
edge and the myosin II generated tensions in the actin cytoskeleton,
including the contractile bundles (stress fibers) and networks that
form during mature adhesion. These cytoskeletal processes are closely
integrated with the dynamics of spatially localized sites of focal
adhesions. Together, this system allows cells to sense and react to
the mechanical properties of their environment. Our review shows that
the active and dynamic nature of cellular systems must be addressed on
many different scales, from the modeling of nanometer-scale molecular
association and dissociation events in adhesion clusters, to force
generation in large supramolecular complexes and the shape and
effective force balance at the 10 micrometer scale for animal cells. A
further level of cooperativity arises if one considers the tissue
scale at which cells can be further abstracted as discrete particles
or defects. In the future, it is hoped that these different approaches
will converge into a systems-level understanding of cellular systems
that not only includes genetic and biochemical aspects (which were not
the focus of this review), but also structural and mechanical ones;
the latter are at least equally influential as biochemistry and genetics
for the interactions of cells with their environment.

An important aspect of this review is to point out in which regard
concepts from physics can be used to improve our understanding of
cellular systems.  By focusing on the physical constraints posed by
the overall force balance in single myosin-minifilaments and cells, we
arrived at the notion of force dipoles, which turns out to be a very
powerful concept to rationalize many important aspects of the
interactions of adherent cells with their physical
environment. Motivated by pioneering experiments with adhesive
micropatterns \cite{chen_geometric_1997} and soft elastic
substrates \cite{c:pelh97}, during the last two decades or so,
a growing body of research has addressed the physical understanding of
how cells sense and respond to the physical properties of their
surroundings, including adhesive geometry, topography and
stiffness \cite{geiger_environmental_2009}. The generic nature of the
experimental observations (including the essential role of active
contractility) suggests that measurements that focus on mesoscale
(tens of nm to micrometers) behavior along with ``coarse grained" models
that capture the physics with only a small number of molecular
parameters, can provide insight into
the generic aspects of cell mechanosensitivity. In particular, we
discussed models for the observed force dependence of the initial
stages of cell adhesion in terms of either polymer-like elasticity or
nucleation and growth. The genesis of the CSK in stem cells and its
dependence on the rigidity of its elastic environment can be
understood in terms of models that focus on the interactions of
actomyosin force dipoles (within the cell) through the elastic
deformations they induce in the cytoskeleton and the substrate.  These
deformations are long range and the ordering that develops in the CSK
is therefore dependent on global boundary conditions such as the cell
shape and the substrate rigidity.  This can be demonstrated either by
a rigorous treatment of the elasticity (that extends known results for
passive inclusions to the case of active contractile elements) or by a
simplified version based on the approximation of the cell as a thin,
actively contractile film coupled to an elastic substrate.  In the
latter theory, the shape dependence enters via moments of the Fourier
transform of the lateral spatial dependence of the cell height. The
response of cells to time varying, externally applied stresses can be
understood in terms of either a specific elastic response of stress
fibers or in terms of a generic theory that treats the entire cell as
an elastic dipole that exerts forces on an elastic substrate. When the dipole dynamics (the formation and
orientation of actomyosin bundles and their adhesions) are fast enough
to follow the applied field, the cell is predicted to align parallel
to the stress direction.  However, if the applied strains or stresses
vary too rapidly, the cell cannot adjust and orients its CSK in the
zero strain or zero stress directions.

Despite some successes of these models in understanding and in some
cases, predicting the experimental findings, many questions remain
unresolved. Regarding the relation between focal adhesions, actin
cytoskeleton and rigidity sensing, recent experimental progress has
posed new challenges to theory. Quantitative studies with elastic
substrates have shown that the size and traction force of focal
adhesions can be very variable, depending on the history and internal
structure of the adhesion (including a possible templating effect for
growth by the actin cytoskeleton)
\cite{stricker_spatiotemporal_2011}. Studies of cell forces with
microplates \cite{mitrossilis_real-time_2010} and pillar assays
\cite{trichet_evidence_2012} have suggested that rigidity sensing is a
more global process than formerly appreciated; however, models
integrating focal adhesion dynamics over entire cells present a great
challenge. Finally RNA-interference studies have revealed the
regulatory complexity of rigidity sensing
\cite{prager-khoutorsky_fibroblast_2011}, but a theoretical framework
to integrate the biochemical, genetic and mechanical features of focal
adhesions on a systems level is still missing.

Another important challenge is improving our understanding of cell
behavior in three dimensions. The physiological environment of tissue
cells in three dimensions is a viscoelastic porous matrix and it is
thus not surprising that cell behavior in three dimensions tends to be
different from the one on flat culture dishes
\cite{cukierman_taking_2001,baker_deconstructing_2012}. Surprisingly,
however, if one cultures cells in open three-dimensional scaffolds,
many of the features known from two-dimensional scaffolds seem to be
conserved (in particular arc-like stress fibers and focal adhesions)
\cite{klein_twocomponent_2011,klein_elastic_2010}. Recent experiments
with three-dimensional hydrogels have shown that the dependence of
cytoskeletal orientation on the matrix rigidity is similar in both two
and three dimensions \cite{florian_2012}. In the future, a careful
quantitative comparision should be made of those factors that are
substantially different in various experimental assays.

From the mechanical point of view, many of the theoretical models
described here treat the CSK and the matrix as linear elastic
materials in which the stresses are proportional to the strains.
However, biopolymers that are important in either the cytoskeleton or
the extracellular matrix often show interesting non-linear responses
to applied forces \cite{gardel_2004, storm_2005, vogel_2009} as
discussed above.  Generalizing the theory of stress generation,
response and interactions of elastic dipoles to non-linear elastic
environments, either within the CSK itself or via the coupling of
actomyosin forces to non-linear substrates by focal adhesions, is thus
an important future goal.  Some experiments report that cells sense
each other (most probably via elastic deformations) at distances on
the order of 400$\mu$m \cite{janmey_plos_2009} on non-linear, elastic
substrates.  In addition, those observations report that cell
spreading becomes independent of the (small-stress) elastic modulus,
suggesting that a mechanical ``tug-of-war'' persists until neither the
cell nor the non-linear substrate can increase its resistance
\cite{janmey_plos_2009}.  A recent theory
\cite{shokef_scaling_2012,shokef_erratum_2012} of the deformations
induced by force dipoles in non-linear elastic media predicts a
linear-type response in the far-field regime, but with an amplitude
that is magnified by the non-linearities important in the near-field
where the stresses are large.  The predicted amplification can be
quite large even for modest forces applied by the dipole.  This can
also modify the interactions between dipoles.  The theory suggests
further quantitative measurements of the long-range effects reported in
\textcite{janmey_plos_2009} along with corresponding theoretical
calculations of the interactions of force dipoles in non-linear
elastic medium.  An experimental hint of some non-linear effects was
presented in \textcite{pompe_dissecting_2009} where a non-quadratic
dependence of the deformation energy on the cellular force dipole
moment was reported; linear elasticity would predict a quadratic
dependence as in Eqs.~\ref{eq:elasticenergy4} and \ref{eq:freequadp}.

Apart from making use of non-linear elasticity, another interesting
avenue is the development of models for non-traditional mechanics,
such as the actively contracting cable networks discussed in
section~V \cite{uss:bisc08a,uss:guth12}. By focusing on two essential
physical aspects of biological materials, namely the asymmetric
mechanical response of filaments and the generation of tension by
molecular motors, these models capture some of the essential physics
but are still relatively easy to handle. This will allow the ideas to
be used in new ways for detailed comparison with experiments on
micropatterned and elastic substrates. Interestingly, these models
also demonstrate a close relation between elasticity and
tension \cite{uss:bisc08a,uss:guth12,uss:edwa11a,mertz_scaling_2012}, which recently has
been confirmed by experiments on cell layers \cite{mertz_scaling_2012}.

Although here we focused on the physical aspects of cellular systems,
it is worth noting that some of the questions addressed in this
framework come quite close to central questions currently studied in
biology, for example stem cell differentiation and
development. Experiments that report genetic effects of substrate
rigidity and their implications for stem cell
differentiation \cite{engler_matrix_2006} are based on observations performed
on the scale of several days while those that report the physical
effects of CSK nematic order in response to rigidity
changes \cite{zemel_optimal_2010} are based on observations performed
on the scale of hours.  Are these two effects related and is CSK
nematic order in stem cells and its optimization on substrates of
particular rigidities a precursor of differential of stem cells into
muscle cells?  While it is true that muscle cells show highly
developed nematic order of actomyosin bundles, it is not yet clear
that the early-time development of nematic order in the same rigidity
range triggers stem cell differentiation into muscle.  Further
experiments and models that explore how CSK stresses translate into
nuclear stresses and possibly chromosomal rearrangements
\cite{ingber_2009, shiva_2008, keng_2011, shiva_2011} are needed
before conclusions can be drawn. 

Another, related area are the effects of elastic interactions, substrate
rigidity, and applied stresses on development. Understanding the role
of elastic stresses on development involves not only an interplay of
genetic expression controlled by CSK and nuclear deformations within a
single cell, but also the interactions of many developing cells via
both chemical signals and elastic stresses.  The spatial development
of ``order'' as evidenced by differentiation within a developing
tissue will be influenced by both the diffusion of signaling
morphogens \cite{naama_2008} as well as by the long-range elastic
interactions explored here in simpler contexts.  The connection to
morphogen diffusion requires an understanding of the dynamical elastic
interactions of cells and this may involve both their elastic (``speed
of sound'') and viscous (damping) dynamics as well as a complete
theory that may bridge the elastic nature of adherent cells to
active-gel theories of cytoskeletal flow and cell
motility \cite{liverpool_instabilities_2003,kruse_asters_2004,julicher_active_2007,marchetti_soft_2012}.

\section{Acknowledgements}

The authors are grateful to
L. Addadi,
N. Balaban,
M. Bastmeyer,
A. Bershadsky,
A. Besser,
D. Ben-Yaakov,
I. Bischofs,
A. Brown,
A. Buxboim,
K. Dasbiswas,
R. De,
D. Discher,
C. Dunlop,
T. Erdmann,
J. Fredberg,
B. Friedrich,
F. Frischknecht,
H. Gao,
K. Garikpati,
M. Gardel,
B. Geiger,
G. Genin,
P. Guthardt Torres,
B. Hoffmann,
R. Kaunas,
R. Kemkemer,
M. Kozlov,
E. Langbeheim, 
R. Merkel,
D. Navajas,
A. Nicolas,
R. McMeeking,
F. Rehfeldt,
D. Riveline,
B. Sabass,
Y. Shokef, 
J. Spatz,
X. Trepat,
C. Waterman,
J. Weichsel, 
Y. Yuval and
A. Zemel 
for useful discussions and comments.
USS is a member of the Heidelberg cluster
of excellence CellNetworks and acknowledges support by the BMBF-project
MechanoSys and the EU-project MEHTRICS.
SAS thanks the Israel Science Foundation, the Minerva Foundation, the
Kimmel Stem Cell Institute and the U.S.-Israel Binational Science Foundation for its support. 


%

\end{document}